\documentclass[a4paper,11pt,twoside]{article}
\usepackage[english]{babel}
\usepackage[utf8]{inputenc}

\usepackage{amsmath,amssymb,amsthm,mathtools}
\usepackage[all]{xy}
\usepackage{enumerate}
\usepackage{mathrsfs}
\usepackage[hmarginratio=1:1]{geometry}
\usepackage[pdftex]{graphicx}
\usepackage{epstopdf}
\usepackage{psfrag}
\usepackage{moreverb}
\usepackage{dsfont}

\usepackage{fancyhdr}
\usepackage[titletoc,title,header]{appendix}
\usepackage{relsize}

\usepackage{enumitem}
\usepackage{pdfpages}

\usepackage{tabularx}
\usepackage{hyperref}
\usepackage[all]{hypcap}
\usepackage{xifthen}
\usepackage[strict]{changepage}
\usepackage{authblk}
\usepackage{needspace}

\hypersetup{
	pdftitle={Construction by bosonization of a fermion-phonon model},
	pdfauthor={Edwin Langmann and Per Moosavi},
	colorlinks=true,
	linkcolor=black,
	citecolor=black,
	filecolor=black,
	urlcolor=blue,
	bookmarksopen,
	bookmarksopenlevel=1,
}

\numberwithin{equation}{section}

\newcommand{\gF}{\lambda}
\newcommand{\gP}{g}
\newcommand{\WW}{W}
\newcommand{\BB}{\tilde{b}}

\renewcommand{\Re}{\operatorname{Re}}

\newcommand{\pdag}{^{\vphantom{\dagger}}}
\newcommand{\pp}{^{\vphantom{\prime}}}
\newcommand{\define}{\equiv}
\newcommand{\tightoverset}[2]{%
	\mathop{#2}\limits^{\vbox to -.5ex{\kern-0.55ex\hbox{$#1$}\vss}}
}
\newcommand{\harpoon}[1]{\tightoverset{\rightharpoonup}{#1}} 

\newcommand{\dop}[2]{\mathop{}\!\mathrm{d}^{#1}{#2}\,}
\newcommand{\abs}[1]{\lvert#1\rvert}
\newcommand{\diag}{\operatorname{diag}}
\newcommand{\sgn}{\operatorname{sgn}}

\newcommand{\noord}[1]{\left. :\! \hspace{-0.5pt} #1 \hspace{-0.5pt} \!: \right.}
\newcommand{\bosonnoord}[1]{\left. \xxa\! \hspace{0.3pt} #1 \hspace{0.3pt} \!\xxe \right.}

\newcommand\xqed[1]{
	\leavevmode\unskip\penalty9999 \hbox{}\nobreak\hfill
	\quad\hbox{#1}
}
\newcommand\remarkend{\xqed{$\Diamond$}}

\theoremstyle{plain}

\theoremstyle{definition}

\theoremstyle{remark}

\newcommand{\thmtemplate}[3]{
	\begin{#1}
		{\bf \ifthenelse{\isempty{#2}}{}{(#2)}
		}#3
	\end{#1}
}

\newcommand{\rmktemplate}[3]{
	\begin{#1}
		{\ifthenelse{\isempty{#2}}{}{[#2]}
		}#3
		\remarkend
	\end{#1}
}

\newcommand{\definition}[2]{\thmtemplate{defin}{#1}{#2}}

\newcommand{\lemma}[2]{\thmtemplate{lem}{#1}{#2}}
\newcommand{\proposition}[2]{\thmtemplate{propos}{#1}{#2}}
\newcommand{\propositionlist}[2]{\thmtemplate{propos}{#1}{\Needspace*{5\baselineskip}#2}}

\newcommand{\corollary}[2]{\thmtemplate{corol}{#1}{#2}}

\newcommand{\remark}[2]{\rmktemplate{rem}{#1}{#2}}

\begin{document}


\title{
	Construction by bosonization of a fermion-phonon model
}

\renewcommand\Authfont{\scshape\large}
\author[]{
	Edwin Langmann\thanks{Electronic address: \texttt{\href{mailto:langmann@kth.se}{langmann@kth.se}}}
}

\author[]{
	Per Moosavi\thanks{Electronic address: \texttt{\href{mailto:pmoosavi@kth.se}{pmoosavi@kth.se}}}
}

\renewcommand\Affilfont{\itshape\small}
\affil[]{Department of Theoretical Physics, KTH Royal Institute of Technology\\ SE-106 91 Stockholm, Sweden}

\date{December 3, 2015}


\maketitle


\begin{abstract}
We discuss an extension of the (massless) Thirring model describing interacting fermions in one dimension which are coupled to phonons and where all interactions are local. 
This fermion-phonon model can be solved exactly by bosonization.
We present a construction and solution of this model which is mathematically rigorous by treating it as a continuum limit of a Luttinger-phonon model. 
A self-contained account of the mathematical results underlying bosonization is included, together with complete proofs.
\end{abstract}


\pagestyle{fancy}
\fancyhf{}
\fancyhead[RE, LO]{\nouppercase\leftmark}
\fancyhead[RO, LE]{\thepage}
\renewcommand{\headrulewidth}{0pt}
\renewcommand{\footrulewidth}{0pt}




\section{Introduction}
\label{Sec:Introduction}
Exactly solvable models have played an important role in the history of condensed matter physics.
A famous example is the Luttinger model \cite{Luttinger:1963, Thirring:1958, Tomonaga:1950} solved in an important paper by Mattis and Lieb \cite{MattisLieb:1965}.
The key to this solution is a collection of mathematical results commonly known as \emph{bosonization} (another common name is \emph{boson-fermion correspondence}).
In the case of the Luttinger model, bosonization is particularly powerful: not only all energy eigenstates and eigenvalues but any quantity of physical interest can be computed exactly by analytical means; see \cite{HeidenreichSeilerUhlenbrock:1980} and references therein.
Moreover, as shown in \cite{Mastropietro:1993}, a non-trivial limit of this solution of the Luttinger model is equivalent to Johnson's solution \cite{Johnson:1961} of the massless Thirring model \cite{Thirring:1958}.

Bosonization is today a standard tool in condensed matter physics, and it has allowed physicists to obtain interesting results even in situations where exact solutions are not available. However, it is not always easy to judge the mathematical status of such results in the physics literature.
One difficulty is that bosonization has a complicated history, and papers developing its mathematical foundations were written with a focus on other applications,
notably the representation theory of infinite dimensional Lie algebras and string theory; see e.g.\ \cite{CareyHurst:1985, CareyRuijsenaars:1987, FrenkelKac:1980, Kac:1998, PressleySegal:1988}.
There exist excellent reviews and textbooks on bosonization in the condensed matter physics literature, including \cite{Emery:1979, Giamarchi:2004, GogolinNersesyanTsvelik:1998, Haldane:1981, Kopietz:1997, Schonhammer:2002, Schulz:LesHouches1994:1995, Senechal:2003, Solyom:1979, Voit:1995, DelftSchoeller:1998}, which provide the reader with practical knowledge on how bosonization is used in applications.
However, to our knowledge, the existing literature lacks a self-contained account of all mathematical details needed to promote formal bosonization results in condensed matter physics to rigorous mathematical statements.
One aim of the present paper is to fill this gap, and for this we combine different approaches found in the mathematics literature \cite{CareyRuijsenaars:1987, Kac:1998, Segal:1981}.

We discuss a model of interacting fermions coupled to acoustic phonons which can be solved exactly by bosonization.
Our program is to develop all mathematical tools which are needed to give this result, known in the condensed matter physics literature for a long time \cite{ChenLeeLuchini:1988, EngelsbergVarga:1964, VoitSchulz:1985}, a rigorous mathematical foundation. 
In particular, we include a self-contained account of bosonization, where we try to remain close to how bosonization is used in condensed matter physics, but, at the same time, also address the relevant functional analytic issues.

The fermion-phonon model we consider is (formally) defined on the real line $\mathbb{R}$ by the Hamiltonian\footnote{The notation used here is common in the physics literature but suppresses mathematical details.}
\begin{multline} 
\label{Eq:Fermion-phonon_formal_Hamiltonian}
	H = \int \dop{}{x} \biggl( v_F\sum_{r = \pm}
			\noord{ \psi^{\dagger}_r(x) r \left( -i\partial_{x} \right) \psi_{r}(x) }
		+ \frac{1}{2} \!
			\noord{ \Big( \Pi_{P}(x)^2 + v_{P}^2 [\partial_x\Phi_{P}(x)]^2 \Big) } \\ 
 	+ \gF \! \noord{ \psi^{\dagger}_{+}(x) \psi\pdag_{+}(x) } \!
			\noord{ \psi^{\dagger}_{-}(x)\psi\pdag_{-}(x) }
		+ \gP{} \sum_{r = \pm}
			\! \noord{ \psi^{\dagger}_{r}(x)\psi_{r}\pdag(x) } \!
			\partial_{x}\Phi_{P}(x) \biggr)
\end{multline}
with fermion fields $\psi^{(\dagger)}_{\pm}(x)$ satisfying canonical anticommutation relations (CAR):
\begin{equation}
\label{Eq:CAR_formal}
	\bigl\{ \psi\pdag_{r\pp}(x), \psi^{\dagger}_{r'}(y) \bigr\} = \delta_{r,r'} \delta(x-y),
	\qquad
	\bigl\{ \psi\pdag_{r\pp}(x), \psi\pdag_{r'}(y) \bigr\} = 0, 
\end{equation}
and phonons described by Hermitian boson fields $\Phi_{P}(x)$ and $\Pi_{P}(x)$ satisfying canonical commutation relations (CCR):
\begin{equation}
\label{Eq:CCR_formal}
	\bigl[ \Phi_{P}(x), \Pi_{P}(y) \bigr] = i \delta(x-y),
	\qquad
	\bigl[ \Phi_{P}(x), \Phi_{P}(y) \bigr] = \bigl[ \Pi_{P}(x), \Pi_{P}(y) \bigr] = 0,
\end{equation}
and which commute with the fermion fields.
The colons in \eqref{Eq:Fermion-phonon_formal_Hamiltonian} indicate (Wick) normal ordering. 
The fermion-phonon model has four parameters: the Fermi velocity $v_F > 0$, the phonon velocity $v_{P} > 0$ with $v_{P} < v_{F}$ (typically $v_{P} \ll v_F$), the fermion-fermion coupling constant $\gF{}$, and the fermion-phonon coupling constant $\gP$.
As will be shown, if the coupling constants are in the range

\begin{equation} 
\label{Eq:Restrictions}
	\gF < 2\pi v_{F},
	\qquad
	2 \left( \gP / v_{P} \right)^2 < 2\pi v_{F} + \gF, 
\end{equation}
then this model can be constructed and solved by bosonization.
(If these conditions are not fulfilled, the model describes an unstable system.) 

We note that, in our approach to the fermion-phonon model, we follow previous work on the construction and exact solution of a two-dimensional interacting fermion model by bosonization in \cite{deWoulLangmann:2012}; see \cite{deWoulLangmann:2012b} for a review. 
However, these papers only stated the mathematical results underlying bosonization, and thus one motivation for the present paper is to provide self-contained proofs of these results. 

The fermion-phonon model in \eqref{Eq:Fermion-phonon_formal_Hamiltonian}--\eqref{Eq:Restrictions} is interesting since it has zero-temperature phase transitions to charge-density-wave and superconducting states, and its phase diagram can be determined from certain fermion four-point correlation functions \cite{ChenLeeLuchini:1988, VoitSchulz:1985}.
From a mathematical point of view, this model is an extension of the massless Thirring model, which (essentially) corresponds to the limiting case $\gP{}=0$.
However, as opposed to this limiting case, the fermion-phonon model is \emph{not} relativistically invariant, which is obvious since its correlation functions depend on two different velocities $\tilde v_F$ and $\tilde v_P$; see \eqref{Eq:Green2} below.
Nevertheless, as will be discussed, this model shares many properties with relativistically invariant quantum field theory models. 
Since many effective low-energy models in condensed matter physics happen to be relativistically invariant, this is an interesting counterexample.

As a further motivation of the bosonization techniques presented in this paper, we mention recent mathematical work on bosonization based on functional integral techniques; see e.g.\ \cite{BenfattoMastropietro:2005} and references therein.
These methods allow for treating even models which need not be exactly solvable, but they have been restricted, up to now, to models with fermions only.
It would be interesting to extend these results to models where fermions are coupled to bosons.
We hope that the results in the present paper can serve as a useful guide in this direction.

As will be shown, the fermion-phonon model can be made mathematically precise by introducing cutoffs $L > 0$ and $a > 0$, which have the physical interpretation as system size and interaction range, respectively.
The regularized model thus obtained is an exactly solvable extension of the Luttinger model.
To obtain the correlation functions for the model in \eqref{Eq:Fermion-phonon_formal_Hamiltonian}--\eqref{Eq:Restrictions}, one has to pass to both the \emph{continuum limit} $a \to 0^{+}$ and the \emph{thermodynamic limit} $L \to \infty$; this is a generalization of the limit in \cite{Mastropietro:1993} where the Thirring model is constructed from the Luttinger model.
As will be shown, the continuum limit is non-trivial in that it requires additive and multiplicative renormalizations.
However, passing to these limits simplifies the correlation functions significantly, and physical arguments suggest that the result is universal in the following sense: the physical properties of the fermion-phonon model at intermediate distances do not depend on the regularization, i.e.\ on short-distance details.
For example, after multiplicative renormalization, the fermion two-point correlation functions we obtain in these limits are
\begin{equation} 
\label{Eq:Green2}
	\langle \psi\pdag_{r\pp}(t, x) \psi_{r'}^{\dagger}(0, 0) \rangle
		= \delta_{r,r'}\frac{1}{2\pi\ell} \prod_{X = F, P}
			\left( \frac{i\ell}{rx - \tilde{v}_{X}t + i0^{+}} \right)^{\rho_{X}^2}
			\left( \frac{i\ell}{-rx - \tilde{v}_{X}t + i0^{+}} \right)^{\sigma_{X}^2}
\end{equation}
with parameters $\tilde{v}_F$, $\rho_F^2$, $\sigma_F^2$ and  $\tilde{v}_P$, $\rho_P^2$, $\sigma_P^2$ determined by the model parameters (see Proposition~\ref{Proposition:Diagonalization_of_fermion-phonon_model}); $\ell>0$ is an arbitrary length parameter introduced by the multiplicative renormalization. 

The rest of the paper is organized as follows.
The collection of mathematical results known as bosonization are presented and proved in Sections~\ref{Sec:Mathematics_of_bosonization_I}~and~\ref{Sec:Mathematics_of_bosonization_II}.
For this we use an infrared cutoff $L > 0$ corresponding to the length of space, i.e.\ the real line $\mathbb{R}$ is replaced by the interval $[-L/2, L/2]$.
These results are used in Section~\ref{Sec:Fermion-phonon_model} to construct and solve the fermion-phonon model described above.
Section~\ref{Sec:Concluding_remarks} contains concluding remarks.
In order to abbreviate our formulas certain variables are reserved to belong to specific sets as given by the following table:
\begin{equation*}
\begin{array}{cc}
	\hline\hline
	\multicolumn{1}{c}{\text{Variables}}	& \multicolumn{1}{c}{\text{Set}}	\\
	\hline
	r, r'		& \{+, -\} 												\\
	x, x'		& [-L/2, L/2]											\\
	k, k'		& (2\pi/L)(\mathbb{Z}+1/2)								\\
	p, p'		& (2\pi/L)\mathbb{Z}			 							\\
	X, X'		& \left\{ F,P \right\}									\\
	\hline\hline
\end{array}
\end{equation*}
We also use the convention that sums over variables in this table range over all allowed values, unless specified otherwise.
These and other conventions are summarized in Appendix~\ref{Appendix:Notation}.
In Appendix~\ref{Appendix:Bosonization_nutshell} we give a summary of the results in the paper 
using a notation common in the physics literature, in order to connect our results with the latter. 
It also contains comments on the history of bosonization. 
Results from functional analysis which we need are collected in Appendix~\ref{Appendix:Quadratic_forms}.


\section{From fermions to bosons}
\label{Sec:Mathematics_of_bosonization_I}
\begin{figure}[!htbp]
	\centering
	\includegraphics[scale=1, clip=true, trim=3cm 3.65cm 3cm 3.65cm]{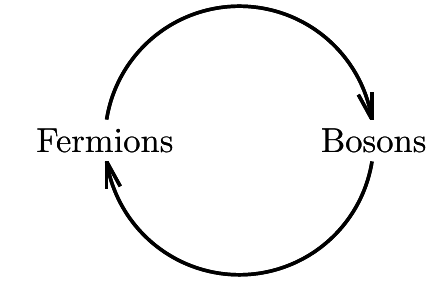}
	\caption{Symbolic illustration of bosonization.}
	\label{Fig:Bosonization_circle}
\end{figure}

The results referred to as bosonization state that two different quantum field theory models, one describing fermions and the other bosons, are actually the same.
There are two approaches to this: one can start with the fermions and construct the bosons, or one can start with the bosons and construct the fermions (see Figure~\ref{Fig:Bosonization_circle}).
Regardless of this choice, the same collection of results is obtained in the end. 
In our opinion, the second approach is simpler from a mathematical point of view.
The first approach is, however, more natural from the point of view of condensed matter physics, and this is why we follow it in this paper; see e.g.\ \cite{CareyLangmann:GeoAnalysis:2002, Senechal:2003} for the second approach.

The present section is concerned with the construction of the bosons from the fermions; the reconstruction of the fermions from the bosons is addressed in Section~\ref{Sec:Mathematics_of_bosonization_II}.
We remind the reader that some notational conventions used throughout this paper are explained in Appendix~\ref{Appendix:Notation}.

\subsection{Fermion Fock space and Hamiltonian}
\label{SubSec:H0}
Our first aim is to construct a Hilbert space $\mathcal{F}$, the so-called \emph{fermion Fock space}, giving a precise mathematical meaning to the quantum field theory model defined by the Hamiltonian\footnote{This Hamiltonian corresponds to an important model in particle physics describing relativistic fermions in 1+1 dimensions; cf.\ Remark~\ref{Remark:Dirac_fermions_KG_bosons} in Appendix~\ref{Appendix:Bosonization_nutshell}.}
\begin{equation}
\label{Eq:Free_fermion_Hamiltonian}
	H_0
	\define
	\sum_{r} \sum_{k} \frac{2\pi}{L}
		rk \noord{ \hat{\psi}^{\dagger}_{r}(k)\hat{\psi}\pdag_{r}(k) }
\end{equation}
for $r = \pm$ and $k \in (2\pi/L) (\mathbb{Z} + 1/2)$, with fermion field operators $\hat{\psi}\pdag_{r}(k)$ and $ \hat{\psi}_{r}^{\dagger}(k) \define \hat{\psi}_{r}(k)^{\dagger}_{\vphantom{r}}$ characterized by the CAR
\begin{equation}
\label{Eq:Fermion_field_anticommutation_relations}
	\left\{ \hat{\psi}\pdag_{r\pp}(k),
		\hat{\psi}^{\dagger}_{r'}(k') \right\}
	= \frac{L}{2\pi} \delta_{r,r'} \delta_{k,k'},
	\qquad
	\left\{ \hat{\psi}\pdag_{r\pp}(k),
		\hat{\psi}\pdag_{r'}(k') \right\}
	= 0.
\end{equation}
The physical requirement underlying this construction is that $H_0$ must have a ground state.
This is ensured by requiring that there is a distinguished element $\Omega$ in $\mathcal{F}$, the so-called \emph{vacuum}, which satisfies 
\begin{equation}
\label{Eq:Fermion_field_action_on_Omega}
	\hat{\psi}\pdag_{r}(rk) \Omega	= \hat{\psi}^{\dagger}_{r}(-rk) \Omega = 0
	\qquad
	\forall k > 0
\end{equation}
and is normalized, i.e.\ $\langle \Omega, \Omega \rangle = 1$, where $\langle\cdot,\cdot\rangle$ is the inner product in $\mathcal{F}$.
The Hilbert space $\mathcal{F}$ is fully determined by these conditions, and it is such that $H_0$ defines a self-adjoint operator on $\mathcal{F}$ bounded from below.
We now recall how this can be substantiated.

To construct $\mathcal{F}$ we use fermion creation operators $c^{\dagger}_{r}(k)$ satisfying the conditions:
\begin{equation}
\label{Eq:c-operator_anticommutation_and_annihilation_relations}
	\left\{ c\pdag_{r\pp}(k), c^{\dagger}_{r'}(k') \right\} = \delta_{r,r'} \delta_{k,k'},
	\qquad
	\left\{ c\pdag_{r\pp}(k), c\pdag_{r'}(k') \right\} = 0,
	\qquad
	c\pdag_{r}(k) \Omega = 0,
\end{equation}
with $c\pdag_{r}(k) = c^{\dagger}_{r}(k)^{\dagger}_{\vphantom{r}}$ and a normalized state $\Omega \in \mathcal{F}$.
Assuming that $c^{(\dagger)}_{r}(k)$ are operators on $\mathcal{F}$, one can construct other states in $\mathcal{F}$ as follows:
\begin{equation}
\label{Eq:Fermion_states1}
	c_{+}^{\dagger}(k_1) \cdots c_{+}^{\dagger}(k_{N_{+}})
		c_{-}^{\dagger}(k'_1) \cdots c_{-}^{\dagger}(k'_{N_-}) \Omega
	\qquad
	\text{for}
	\;\,
	k_1 > \ldots > k_{N_{+}},
	\;\,
	k'_1 > \ldots > k'_{N_{-}},
\end{equation}
where $N_{\pm}$ are non-negative integers.
Using the relations in \eqref{Eq:c-operator_anticommutation_and_annihilation_relations}, it is straightforward to prove by induction over $N_+$ and $N_-$ that these states are orthonormal.
Let $\mathcal{D}_F$ denote the vector space of all finite linear combinations of the states in \eqref{Eq:Fermion_states1}.
This set forms a pre-Hilbert space, and $\mathcal{F}$ is obtained from $\mathcal{D}_F$ by norm-completion; see e.g.\ \cite{ReedSimon:1972}.

We find it convenient to introduce the following notation for the states in \eqref{Eq:Fermion_states1}: 
\begin{equation} 
\label{Eq:Fermion_states}
	\eta^{F}_{\mathbf{n}}
	\define
	\Biggl( \harpoon{\prod}_{r, k} c^{\dagger}_{r}(k)^{n_{r}(k)} \Biggr) \Omega,
	\qquad
	n_r(k) \in \{ 0, 1 \},
\end{equation}
where only finitely many of the quantum numbers $n_r(k)$ are non-zero and $\mathbf{n}$ denotes the infinite vector $\left(n_r(k)\right)_{r,k}$.
The arrow over the product symbol indicates that the fermion operators are ordered as in \eqref{Eq:Fermion_states1}. 

The above definitions allow us to write formulas in a compact way.
For instance, orthonormality and completeness of the states in \eqref{Eq:Fermion_states1} can be expressed as
\begin{equation} 
\label{Eq:F-basis} 
	\langle \eta^F_{\mathbf{n}\pp},\eta^F_{\mathbf{n}'} \rangle
		= \delta_{\mathbf{n},\mathbf{n}'},
	\qquad
	\sum_{\mathbf{n}} \eta^F_{\mathbf{n}} \langle \eta^F_{\mathbf{n}},\cdot \rangle = I. 
\end{equation} 
Note that the states $\eta^F_{\mathbf{n}}$ are common eigenstates of $c_r^{\dagger}(k) c_r\pdag(k)$, i.e.
\begin{equation} 
\label{Eq:neta}
	c_r^{\dagger}(k) c_r\pdag(k)\eta^{F}_{\mathbf{n}} = n_r(k) \eta^{F}_{\mathbf{n}}
\end{equation} 
(this follows from the CAR in \eqref{Eq:c-operator_anticommutation_and_annihilation_relations} and the definition in \eqref{Eq:Fermion_states}).

\remark{}{
It is not difficult to construct a bijection between the set of quantum numbers $\mathbf{n}$ and $\mathbb{N}$ to prove that the Hilbert space $\mathcal{F}$ thus constructed is separable.
}

We are now ready to give a precise meaning to the model defined in \eqref{Eq:Free_fermion_Hamiltonian}--\eqref{Eq:Fermion_field_action_on_Omega}.
The method we employ is \emph{construction by exact solution} in the sense that self-adjointness of the Hamiltonian is shown by constructing all its eigenstates and eigenvalues.

Using the relations in \eqref{Eq:c-operator_anticommutation_and_annihilation_relations} it is easy to check that the operators 
\begin{equation}
\label{Eq:psi_representation}
	\hat{\psi}\pdag_{r}(k)
	\define \begin{cases}
		\sqrt{\frac{L}{2\pi}} c\pdag_{r}(k)		& \text{if} \;\, rk > 0, \\
		\sqrt{\frac{L}{2\pi}} c^{\dagger}_{r}(k)	& \text{if} \;\, rk < 0,
	\end{cases}
	\qquad
	\hat{\psi}^{\dagger}_{r}(k) \define \hat{\psi}\pdag_{r}(k)^{\dagger},
\end{equation}
satisfy the relations in \eqref{Eq:Fermion_field_anticommutation_relations}--\eqref{Eq:Fermion_field_action_on_Omega}.
Moreover, for the operators we consider, normal ordering can be made precise as follows:  
\begin{equation}
\label{Eq:noord1} 
	\noord{A}
	\define
	A - \langle \Omega, A\Omega \rangle. 
\end{equation}
This implies 
\begin{equation}
	\noord{ \hat{\psi}^{\dagger}_{r}(k)\hat{\psi}\pdag_{r}(k) }
		= \sgn(rk) \frac{L}{2\pi} c^{\dagger}_{r}(k) c\pdag_{r}(k), 
\end{equation}
which allows us to cast the Hamiltonian in \eqref{Eq:Free_fermion_Hamiltonian} in a manifestly positive-definite form:
\begin{equation}
\label{Eq:H01} 
	H_0	= \sum_{r} \sum_{k} \abs{k} c^{\dagger}_{r}(k) c\pdag_{r}(k).
\end{equation}

\remark{}{
\label{Remark:Boundedness}
The CAR in \eqref{Eq:c-operator_anticommutation_and_annihilation_relations} and the eigenvalue equation \eqref{Eq:neta} imply that the fermion operators $c^{(\dagger)}_r(k)$ are bounded on $\mathcal{F}$ with operator norm equal to unity,
and thus $c^{\dagger}_r(k)$ is the Hilbert space adjoint of $c\pdag_r(k)$.
It follows that $\hat{\psi}^{(\dagger)}_{r}(k)$ are also bounded operators.
This implies that the algebra generated by the fermion field algebra is a $C^*$-algebra, which is the starting point of an elegant mathematical approach to fermion quantum field theories; see e.g.\ \cite{BratteliRobinson:1979}.
However, to remain self-contained, we follow a more pedestrian approach avoiding $C^*$-algebra techniques. 
}

From the equations above one can deduce several important properties of $H_0$, which we summarize as follows.

\lemma{}{
\label{Lemma:H0}
The Hamiltonian $H_0$ in \eqref{Eq:Free_fermion_Hamiltonian} defines a self-adjoint operator on $\mathcal{F}$, and it satisfies the following commutation relation on $\mathcal{D}_F$: 
\begin{equation}
\label{Eq:H0_psi_dagger_commutation_relation}
	\left[ H_0, \hat{\psi}^{\dagger}_{r}(k) \right]
	= rk \hat{\psi}^{\dagger}_{r}(k).
\end{equation}
Moreover, all states in \eqref{Eq:Fermion_states} are eigenstates of $H_0$ with corresponding eigenvalues 
\begin{equation}
\label{Eq:Fermion_states_eigenvalues}
	E^{F}_{\mathbf{n}} \define \sum_{r} \sum_{k} \abs{k} n_{r}(k) . 
\end{equation}
}

In particular, $\Omega$ is the ground state of $H_0$ with ground state energy zero: $H_0\Omega = 0$. 

\begin{proof}[Proof of Lemma~\ref{Lemma:H0}]
The eigenvalue equation 
\begin{equation} 
\label{Hetan} 
H_0\eta^F_{\mathbf{n}} = E^{F}_{\mathbf{n}} \eta^F_{\mathbf{n}} 
\end{equation} 
follows from \eqref{Eq:neta} and \eqref{Eq:H01}.
This implies that $H_0$ is a self-adjoint operator (this is a consequence of a simple general result stated in Lemma~\ref{Lemma:Self_adjointness}).

We note that the commutation relation in \eqref{Eq:H0_psi_dagger_commutation_relation} is easily motivated by formal computations using the definition of $H_0$ in \eqref{Eq:Free_fermion_Hamiltonian} and the CAR in \eqref{Eq:Fermion_field_anticommutation_relations} if we ignore normal ordering.
To make these computations precise we introduce a cutoff $\Lambda>0$ and study the regularized Hamiltonian
\begin{equation}
\label{Eq:H0reg}
\begin{aligned}
	H_{0}^{\Lambda}
	& \define
	\sum_{r} \sum_{k} \frac{2\pi}{L}
		rk \noord{ \hat{\psi}_{r}^{\dagger}(k)\hat{\psi}\pdag_{r}(k) }
		\Theta( \Lambda - \abs{k} ) \\
	& = \sum_{r} \sum_{\abs{k} \leq \Lambda} \frac{2\pi}{L}
		rk \hat{\psi}_{r}^{\dagger}(k)\hat{\psi}\pdag_{r}(k)
		+ E_{\Lambda},
\end{aligned} 
\end{equation}
where $E_{\Lambda} = (L/(2\pi)) \Lambda^2 + O(\Lambda)$ is a positive constant which diverges with increasing $\Lambda$.
The second equality is obtained using $\noord{ \hat{\psi}_{r}^{\dagger}(k)\hat{\psi}\pdag_{r}(k) } =  \hat{\psi}^{\dagger}_{r}(k)\hat{\psi}\pdag_{r}(k) - (L/(2\pi))\Theta(-rk)$, where $\Theta(\cdot)$ is the Heaviside function (this is a simple consequence of the definitions).

Since the regularized Hamiltonian $H_0^\Lambda$ is a \emph{finite} sum of bounded operators, it is bounded, and thus commutation relations, etc., can be safely derived by formal computations using the CAR in \eqref{Eq:Fermion_field_anticommutation_relations}. 
In particular, it is obvious that $\lim_{\Lambda\to\infty}H_0^\Lambda = H_0\pdag$ on $\mathcal{D}_F$ (in a sense made precise in Definition~\ref{Definition:Operator_convergence}).

We now observe that, for each state $\eta^F_{\mathbf{n}}$ in \eqref{Eq:Fermion_states}, 
\begin{equation} 
\left[ H\pdag_0, \hat{\psi}^{\dagger}_{r}(k) \right]\eta_{\mathbf{n}} = \left[ H^\Lambda_0, \hat{\psi}^{\dagger}_{r}(k) \right] \eta_{\mathbf{n}} 
\end{equation} 
independent of $\Lambda$ for sufficiently large values.
It follows that the result obtained by formal computations is correct if interpreted as an identity on the domain $\mathcal{D}_F$
(this follows from a simple general result stated in Lemma~\ref{Lemma:Commutators}). 
\end{proof} 

\remark{}{
\label{Remark:Commutators}
Throughout this paper we use the same strategy to establish commutation relations for unbounded operators $A$: we introduce a cutoff $\Lambda > 0$ and approximate $A$ by a family of bounded operators $\{ A^{\Lambda} \}_{\Lambda > 0}$ such that $A = \lim_{\Lambda \to \infty} A^{\Lambda}$.  
This strategy is simple in that it avoids topological arguments to make precise mathematical sense of such limits.
To not have to repeat the same technicalities each time we use this strategy, we state the precise definition of this limit, together with a simple abstract lemma, in Appendix~\ref{Appendix:Quadratic_forms}; see Definition~\ref{Definition:Operator_convergence} and Lemma~\ref{Lemma:Commutators}.
}

\subsection{Fermion densities}
\label{SubSec:Fermion_densities} 
We now introduce the \emph{fermion densities}
\begin{equation} 
\label{Eq:Jr}
	\hat{J}_r(p)
	\define
	\sum_{k}\frac{2\pi}{L}\noord{ \hat{\psi}^{\dagger}_r(k)\hat{\psi}\pdag_r(k+p) }
\end{equation}
for $p \in (2\pi/L)\mathbb{Z}$, which we regard as a definition motivated by the Fourier transform of the (formal) fermion densities $\noord{ \psi^{\dagger}_{r}(x)\psi\pdag_{r}(x) }$ in the introduction (cf.\ \eqref{Eq:Jr_nutshell} and \eqref{Eq:Psi_Jr_Fourier_transforms}).

\proposition{}{
\label{Proposition:Bosons_from_fermions}
The fermion densities $\hat{J}_{r}(p)$ in \eqref{Eq:Jr} are well-defined operators on $\mathcal{F}$ mapping $\mathcal{D}_F$ to $\mathcal{D}_F$.
Moreover, they satisfy
\begin{gather}
	\hat{J}\pdag_{r}(p)^{\dagger} = \hat{J}\pdag_{r}(-p),
		\label{Eq:Densities_property_1} \\
	\hat{J}_{r}(rp) \Omega = 0 \qquad \forall p \geq 0,
		\label{Eq:Densities_property_2}
\end{gather}
together with the following commutation relations on $\mathcal{D}_F$:
\begin{subequations}
\begin{align}
	\left[ \hat{J}\pdag_{r\pp}(p), \hat{\psi}^{\dagger}_{r'}(k) \right]
		& = \delta_{r,r'}
			\hat{\psi}^{\dagger}_{r\pp}(k-p), \label{Eq:Jr_psi_dagger_commutation_relation}\\
	\left[ H_0, \hat{J}_{r}(p) \right]
		& = - rp \hat{J}_{r}(p), \label{Eq:H0_and_Jr_commutation_relation}\\
			\left[ \hat{J}_{r\pp}(p), \hat{J}_{r'}(p') \right]
		& = r\delta_{r,r'} \frac{Lp}{2\pi} 
			\delta_{p,-p'}. \label{Eq:Jr_commutation_relation}
\end{align}
\end{subequations}
In particular, the zero modes $\hat{J}_{r}(0)$
are self-adjoint operators, and all states in \eqref{Eq:Fermion_states} are eigenstates of $\hat{J}_r(0)$ with corresponding eigenvalues
\begin{equation} 
\label{Eq:qrn} 
	q_{r,\mathbf{n}}
	\define
	\sum_k \sgn(rk)n_r(k). 
\end{equation}
}

\begin{proof}
Using \eqref{Eq:psi_representation} one can write
\begin{equation} 
\label{Eq:Jr01}
	\hat{J}_{r}(0) = \sum_{k} \sgn(rk)c^{\dagger}_{r}(k) c\pdag_{r}(k), 
\end{equation} 
and thus \eqref{Eq:neta} imply that the states in \eqref{Eq:Fermion_states} are eigenstates of $\hat{J}_r(0)$ with eigenvalues given by \eqref{Eq:qrn}.
This also proves that $\hat{J}_r(0)$ are self-adjoint operators (cf.\ Lemma~\ref{Lemma:Self_adjointness}).

It is easy to motivate \eqref{Eq:Densities_property_1} and the commutation relations in \eqref{Eq:Jr_psi_dagger_commutation_relation}--\eqref{Eq:H0_and_Jr_commutation_relation} by formal computations using $\hat{\psi}_{r}^{\dagger}(k) = \hat{\psi}\pdag_{r}(k)^{\dagger}$ and the CAR in \eqref{Eq:Fermion_field_anticommutation_relations}.
However, the commutation relation in \eqref{Eq:Jr_commutation_relation} is non-trivial (a formal computation suggests that it is zero), and this highlights the need for precise arguments.  

A potential difficulty with the fermion densities is that the sum on the r.h.s.\ in \eqref{Eq:Jr} is infinite. 
As for $H_0$ in Section~\ref{SubSec:H0} (recall Remark~\ref{Remark:Commutators}), we deal with this by introducing \emph{regularized fermion densities}
\begin{equation} 
\label{Eq:Jrreg}
	\hat{J}^\Lambda_r(p)
	\define \sum_{k}\frac{2\pi}{L}\noord{ \hat{\psi}^{\dagger}_r(k) \hat{\psi}\pdag_r(k+p) }
		\Theta( \Lambda - \abs{k+\tfrac{p}{2}} ),
\end{equation}
where $\Lambda > 0$ is a cutoff and $\Theta(\cdot)$ is the Heaviside function.
Note that these operators are bounded for $\Lambda < \infty$ by the same argument as for $H_0^\Lambda$ in the proof of Lemma~\ref{Lemma:H0}.

We use \eqref{Eq:psi_representation} to write, for all $p> 0$,
\begin{equation}
\label{Eq:Jrp}
\begin{aligned}
	\hat{J}_{r}(rp) &= \sum_{rk>0}c_r^{\dagger}(k)c_r\pdag(k+rp)
		+ \sum_{-p<rk<0}c_r\pdag(k)c_r\pdag(k+rp)
		- \sum_{rk<-p}c_r^{\dagger}(k+rp)c_r\pdag(k), \\
	\hat{J}_{r}(-rp) &= \sum_{rk>p}c_r^{\dagger}(k)c_r\pdag(k-rp)
		+ \sum_{0<rk<p}c_r^{\dagger}(k)c_r^{\dagger}(k-rp)
		- \sum_{rk<0}c_r^{\dagger}(k-rp)c_r\pdag(k). 
\end{aligned}
\end{equation}
This and \eqref{Eq:Jr01} make manifest that all $\hat{J}_r(p)$ are well-defined operators mapping $\mathcal{D}_F$ to $\mathcal{D}_F$. 
It also proves \eqref{Eq:Densities_property_2} and implies that, for each state $\eta^F_{\mathbf{n}}$ in \eqref{Eq:Fermion_states}, 
\begin{equation} 
\label{Eq:reg}
\hat{J}\pdag_r(p)\eta^F_{\mathbf{n}} = \hat{J}^\Lambda_r(p)\eta^F_{\mathbf{n}} 
\end{equation} 
independent of $\Lambda$ for sufficiently large values. 
Thus, $\lim_{\Lambda\to\infty} \hat{J}^\Lambda_r(p) = \hat{J}\pdag_r(p)$ on $\mathcal{D}_F$ (in the sense of Definition~\ref{Definition:Operator_convergence}). 
Since it follows from the definitions that $\hat{J}^\Lambda_r(p)^{\dagger}=\hat{J}^\Lambda_r(-p)$, this implies \eqref{Eq:Densities_property_1}.

Straightforward computations using the CAR in \eqref{Eq:Fermion_field_anticommutation_relations} imply
\begin{equation}
\label{Eq:Jr_psi_dagger_commutation_relation1} 
	\left[ \hat{J}^{\Lambda\vphantom{\dagger}}_{r\pp}(p), \hat{\psi}^{\dagger}_{r'}(k) \right]
		 = \delta_{r,r'} \hat{\psi}^{\dagger}_{r\pp}(k-p)
		 	\Theta(\Lambda - \abs{k-\tfrac{p}{2}}). 
\end{equation} 
Acting with this identity on a fixed state $\eta^F_{\mathbf{n}}$, for fixed $k$ and $p$, the r.h.s.\ is independent of $\Lambda$ for sufficiently large values.  
This proves the commutation relation in \eqref{Eq:Jr_psi_dagger_commutation_relation} on $\mathcal{D}_F$ (cf.\ Lemma~\ref{Lemma:Commutators}). 

To prove \eqref{Eq:H0_and_Jr_commutation_relation} we use the CAR in \eqref{Eq:Fermion_field_anticommutation_relations} to  compute 
\begin{equation}
\label{Eq:H0_and_Jr_commutation_relation1}
	\left[ H^\Lambda_0, \hat{J}^{\Lambda'}_{r}(p) \right]
		 = \sum_{k} \frac{2\pi}{L} r \left( k_{\Lambda} - (k+p)_{\Lambda} \right)
		 	\hat{\psi}^{\dagger}_{r}(k) \hat{\psi}\pdag_r(k+p)
		 	\Theta(\Lambda' - \abs{k+\tfrac{p}{2}}),
\end{equation} 
where $k_\Lambda\define k\Theta(\Lambda - \abs{k})$.
As a simple consequence of the definition of normal ordering,
\begin{equation} 
\label{Eq:noord_psipsi}
	\hat{\psi}^{\dagger}_{r}(k)\hat{\psi}\pdag_{r}(k')
		= \noord{ \hat{\psi}_{r}^{\dagger}(k) \hat{\psi}\pdag_{r}(k') } \!
			+ \frac{L}{2\pi}\delta_{k,k'}\Theta(-rk).
\end{equation} 
It follows that $\hat{\psi}^{\dagger}_{r}(k) \hat{\psi}\pdag_{r}(k+p)$ on the r.h.s.\ of \eqref{Eq:H0_and_Jr_commutation_relation1} can be replaced by  $\noord{ \hat{\psi}^{\dagger}_{r}(k) \hat{\psi}\pdag_{r}(k+p) }$ (this is obvious for $p \neq 0$, and it is also true for $p=0$ since then the r.h.s.\ is identically zero anyway). 
Acting with the identity thus obtained on a fixed state $\eta^F_{\mathbf{n}}$, for fixed $p$, the r.h.s.\ becomes independent of $\Lambda$ and $\Lambda'$ when both are sufficiently large, and this implies \eqref{Eq:H0_and_Jr_commutation_relation} (cf.\ Lemma~\ref{Lemma:Commutators}).

We are left to prove \eqref{Eq:Jr_commutation_relation}. 
The case $r\neq r'$ is trivial, and we therefore only consider the case $r = r'$. 
By straightforward computations, 
\begin{multline} 
\label{Eq:JrJr}
	\left[ \hat{J}^{\Lambda}_{r}(p), \hat{J}^{\Lambda'}_{r}(p') \right]
		= \sum_k \frac{2\pi}{L}
		\Bigl( \Theta( \Lambda - \abs{k+\tfrac{p}{2}} )
			\Theta( \Lambda' - \abs{k+p+\tfrac{p'}{2}} ) \\
		- \Theta( \Lambda' - \abs{k+\tfrac{p'}{2}} )
			\Theta( \Lambda - \abs{k+p'+\tfrac{p}{2}} ) \Bigr)
		\hat{\psi}^{\dagger}_{r}(k) \hat{\psi}\pdag_{r}(k+p+p'). 
\end{multline} 
The r.h.s.\ can be cast in a normal-ordered form by inserting \eqref{Eq:noord_psipsi}.
It is thus equal to a sum of two terms: the first is equal to the r.h.s.\ in \eqref{Eq:JrJr} but with $\hat{\psi}^{\dagger}_{r}(k) \hat{\psi}\pdag_{r}(k+p+p')$ replaced by $\noord{ \hat{\psi}^{\dagger}_{r}(k) \hat{\psi}\pdag_{r}(k+p+p')}$, and the second is the $\mathbb{C}$-number term coming from normal ordering.
Acting with the identity thus obtained on a fixed state $\eta^F_{\mathbf{n}}$ we find that, for fixed $p$ and $p'$ and sufficiently large $\Lambda$ and $\Lambda'$, the contribution of the first term is independent of the cutoffs and thus zero.
The only remaining contribution is therefore the second term:
\begin{equation} 
\label{Eq:JrJr1}
	\left[ \hat{J}^{\Lambda}_{r}(p), \hat{J}^{\Lambda'}_{r}(p') \right] \eta^F_{\mathbf{n}}
		= \delta_{p+p',0} \sum_{k} \Bigl( \Theta( \Lambda - \abs{k+\tfrac{p}{2}} ) -
			\Theta( \Lambda - \abs{k-\tfrac{p}{2}} ) \Bigr) \Theta(-rk) \eta^F_{\mathbf{n}} 
\end{equation} 
assuming $\Lambda' > \Lambda$, without loss of generality. 
For sufficiently large $\Lambda$, the r.h.s.\ in \eqref{Eq:JrJr1} is independent of $\Lambda$ and equal to $\delta_{p+p',0}\left(rpL/(2\pi)\right)\eta^F_{\mathbf{n}}$.
This implies the result (cf.\ Lemma~\ref{Lemma:Commutators}).
\end{proof}

\subsection{Bosons, charges, and Klein factors}
\label{SubSec:Bosons_charges_Klain_factors}
In this section, the physical interpretation of the  fermion densities $\hat{J}_r(p)$ as bosons (for $p\neq 0$) and chiral charge operators (for $p=0$) is explained. 
We also discuss how this motivates the introduction of the \emph{Klein factors} $R_{\pm}$ as charge-changing operators.

Consider operators $b^{(\dagger)}(p)$ defined for $p \neq 0$ as follows:
\begin{equation}
\label{Eq:Harmonic_oscillator_operators}
	b(p) \define \begin{cases}
		-i \sqrt{\tfrac{2\pi}{L\abs{p}}} \hat{J}_{+}(p)
			& \text{if} \;\, p > 0, \\
		\phantom{-} i \sqrt{\tfrac{2\pi}{L\abs{p}}} \hat{J}_{-}(p)
			& \text{if} \;\, p < 0,
	\end{cases}
	\qquad
	b^{\dagger}(p) \define b(p)^{\dagger}.
\end{equation}
Proposition~\ref{Proposition:Bosons_from_fermions} implies that these are well-defined and obey standard relations for boson annihilation and creation operators on $\mathcal{D}_F$:
\begin{equation}
\label{Eq:Harmonic_oscillator_operator_properties}
	\left[ b(p), b^{\dagger}(p') \right] = \delta_{p,p'},
	\qquad
	\Big[ b(p), b(p') \Big] = 0,
	\qquad
	b(p)\Omega = 0. 
\end{equation}

The zero modes $\hat{J}_{\pm}(0)$ play a special role as \emph{chiral charge operators}, and the notation
\begin{equation}
\label{Eq:Qr}
	Q_{\pm} \define \hat{J}_{\pm}(0)
\end{equation}
will sometimes be used to emphasize this.
Indeed, they are self-adjoint operators with integer eigenvalues (cf.\ Proposition~\ref{Proposition:Bosons_from_fermions}) which, as discussed below, have a natural physical interpretation as chiral charges.

Our aim is to generate the fermion Fock space $\mathcal{F}$ from the vacuum $\Omega$ using bosons.
However, since the zero modes $Q_{\pm}$ commute with the boson operators $b^{(\dagger)}(p)$ and since $Q_{\pm}\Omega = 0$ (cf.\ \eqref{Eq:Jr_commutation_relation} and \eqref{Eq:Densities_property_2}), the boson operators in \eqref{Eq:Harmonic_oscillator_operators} are not enough: they can only generate states $\eta$ with charge zero ($Q_{\pm}\eta = 0$). 
This problem is solved by the Klein factors $R_{\pm}$, which are unitary operators on $\mathcal{F}$ commuting with all boson operators but not with the zero modes:  
\begin{equation} 
\label{Eq:QrRr}
	Q_{r\pp} R_{r'} = R_{r'}(Q_{r\pp} + r\delta_{r,r'}).  
\end{equation}
This implies that, if $\eta$ is a zero-charge state ($Q_{\pm}\eta = 0$), then $R_{+}^{q_{+}}R_{-}^{-q_{-}}\eta$ is a state with charges $(q_{+},q_{-})$:
\begin{equation} 
	Q_{r}^{\vphantom{q}} R_{+}^{q_+}R_{-}^{-q_-}\eta
		= q_{r}^{\vphantom{q}} R_{+}^{q_{+}}R_{-}^{-q_{-}}\eta
\end{equation} 
for all integer pairs $(q_+,q_-)$, where $R_{r}^{-q\vphantom{\dagger}} \define (R_r^{\dagger})^{q}$ for all integers $q$.

\remark{}{
The picture above can be summarized as follows: 
\begin{equation} 
	\mathcal{F} = \bigoplus_{(q_+,q_-) \in \mathbb{Z}^2} \mathcal{F}^{(q_+,q_-)},
\end{equation} 
where $ \mathcal{F}^{(q_+,q_-)}$ is the subspace of $\mathcal{F}$ containing all states $\eta$ such that $Q_{r}\eta = q_{r}\eta$, and the unitary operator $R_+^{q_+}R_-^{-q_-}$ maps the zero-charge subspace $\mathcal{F}^{(0,0)}$ bijectively to $\mathcal{F}^{(q_+,q_-)}$. 
The subspaces $\mathcal{F}^{(q_+,q_-)}$ are known as \emph{charge sectors} or \emph{superselection sectors}. 
}

A construction of the Klein factors, together with proofs of properties needed in practical computations, is given in Section~\ref{SubSec:Klein_factors}; see Proposition~\ref{Proposition:Klein_factors}.
One important result underlying bosonization is that the boson operators $b^{\dagger}(p)$, together with the Klein factors $R_{\pm}$, generate $\mathcal{F}$; see Proposition~\ref{Proposition:BF}.

In the remainder of this section we elaborate on the physical interpretation of $Q_{\pm}$ as chiral charge operators.
This interpretation is based on Dirac's hole theory and can be skipped without loss of continuity.

The model in \eqref{Eq:Free_fermion_Hamiltonian}--\eqref{Eq:Fermion_field_action_on_Omega} describes a system of fermions with one-particle states labeled by pairs of quantum numbers $(r,k)$ and with one-particle energies $rk$.
The many-body ground state $\Omega$ is a \emph{Dirac sea} where all one-particle states with negative energy are filled and all one-particle states with positive energy are empty; this is expressed by \eqref{Eq:Fermion_field_action_on_Omega}, where the second relation is a consequence of the Pauli principle.
The physical interpretation of \eqref{Eq:psi_representation} is that $\hat{\psi}^{\dagger}_{r}(k)$ is a \emph{creation operator for a particle} if $rk > 0$ and an \emph{annihilation operator for a hole} if $rk < 0$.
Similarly, $\hat{\psi}\pdag_{r}(k)$ is an annihilation operator for a particle if $rk > 0$ and a creation operator for a hole if $rk < 0$.  
Moreover, the quantum numbers $n_r(k)$ used to label the states in \eqref{Eq:Fermion_states} correspond to the excitation numbers of the one-particle states $(r,k)$; the excitations are particles for $rk > 0$ and holes for $rk < 0$.
Finally, $c^{\dagger}_r(k)c\pdag_r(k)$ are excitation-number operators with eigenvalues equal to either 0 or 1 (this corresponds to the Pauli principle).
It is natural, from a physical point of view, to assign charge numbers $+1$ to particles and $-1$ to holes (Dirac proposed this interpretation in a particle physics context where the holes correspond to antiparticles). 
This suggests to interpret $Q_{r}$ as an operator which measures the charge of particles and holes with chirality $r$ (cf.\ \eqref{Eq:Jr01} and \eqref{Eq:Qr}).

It is also worth to mention a beautiful mathematical interpretation of the fermion densities and the Klein factors as implementers of gauge transformations.
This interpretation clarifies that the Klein factors exist due to topological reasons; see the final paragraph in Section~\ref{SubSec:Vertex_operators1}.

\subsection{Construction of the Klein factors}
\label{SubSec:Klein_factors}
This section contains a construction of the Klein factors $R_{\pm}$.
We start with a summary of the main results:

\proposition{Klein factors}{
\label{Proposition:Klein_factors}
There exist unitary operators $R_{\pm}$ on $\mathcal{F}$ mapping $\mathcal{D}_F$ to $\mathcal{D}_F$ and satisfying the commutation relations
\begin{align}
	\Bigl[ \hat{J}_{r}(p), R_{r'} \Bigr]
	& = r \delta_{r,r'}\delta_{p,0} R_{r}, \label{Eq:JrRr} \\
	\Bigl[ H_0, R_r \Bigr]
	& = r\frac{\pi}{L} \Bigl\{ \hat{J}_r(0), R_{r} \Bigr\} \label{Eq:H0Rr}
\end{align}
on $\mathcal{D}_F$, together with
\begin{equation} 
\label{Eq:RR}
	R_{+}R_{-} = - R_{-}R_{+}.
\end{equation}
}
\noindent (A constructive proof is given further below.)

Note that \eqref{Eq:JrRr} summarizes properties of the Klein factors mentioned in Section~\ref{SubSec:Bosons_charges_Klain_factors}: choosing $p\neq 0$, it implies that $R_{\pm}$ commute with the boson operators in \eqref{Eq:Harmonic_oscillator_operators}, and choosing $p=0$, it implies \eqref{Eq:QrRr} (recall that $Q_{r} \define \hat{J}_{r}(0)$).  

The following corollary of Proposition~\ref{Proposition:Klein_factors} contains a list of identities which we state for later reference:

\corollary{}{
\label{Corollary:Rr}
For all $q_{\pm} \in \mathbb{Z}$ and $c \in \mathbb{C}$, the following identities hold true:
\begin{subequations} 
\begin{align} 
	R_{+}^{q_{+}}R_{-}^{-q_{-}}
		& = (-1)^{-q_{+}q_{-}}R_{-}^{-q_{-}}R_{+}^{q_+}, \label{Eq:RqRq} \\
	\Bigl[ Q_r , R_{+}^{q_{+}}R_{-}^{-q_{-}} \Bigr] 
		& = q_{r}R_{+}^{q_{+}}R_{-}^{-q_{-}}, \label{Eq:QRqRq} \\
	e^{icQ_{r}} R^{\pm q_{\pm}}_{\pm}
		& = e^{icq_{r}\delta_{r,\pm}} R^{\pm q_{\pm}}_{\pm} e^{icQ_{r}} \label{Eq:ExpQRq}
\end{align}
\end{subequations}
on $\mathcal{D}_F$, together with:
\begin{subequations} 
\begin{align} 
	Q_r R_+^{q_+}R_-^{-q_-}\Omega
		& = q_rR_+^{q_+}R_-^{-q_-}\Omega, \label{Eq:QRqRqOmega} \\
	H_0 R_+^{q_+}R_-^{-q_-}\Omega
		& = \frac{\pi}{L}( q_{+}^2 + q_{-}^2 )
			R_{+}^{q_{+}}R_{-}^{-q_{-}}\Omega, \label{Eq:H0RqRq}
\end{align}
and
\begin{equation}
\label{Eq:Klein_factor_VEV}
	\langle \Omega, R^{q_{+}}_{+}R^{-q_{-}}_{-} \Omega \rangle
		= \delta_{q_{+}, 0} \delta_{q_{-}, 0}. 
\end{equation}
\end{subequations}
}
\noindent (The proofs of these identities are straightforward and thus omitted.)
 
It is interesting to note that the Klein factors can be fully characterized by the following conditions: 
\begin{gather}
	R\pdag_{\pm} \hat{\psi}^{(\dagger)}_{\pm}(k)
	\define
	\hat{\psi}^{(\dagger)}_{\pm}(k + \tfrac{2\pi}{L}) R\pdag_{\pm} ,
	\qquad
	R\pdag_{\pm} \hat{\psi}^{(\dagger)}_{\mp}(k)
	\define
	-\hat{\psi}^{(\dagger)}_{\mp}(k) R\pdag_{\pm},
		\label{Eq:Klein_factor_defining_condition_1} \\
	R\pdag_{+} \Omega
	\define
	\sqrt{\tfrac{2\pi}{L}} \hat{\psi}^{\dagger}_{+}(\tfrac{\pi}{L}) \Omega,
	\qquad
	R\pdag_{-} \Omega
	\define
	\sqrt{\tfrac{2\pi}{L}} \hat{\psi}\pdag_{-}(\tfrac{\pi}{L}) \Omega.
		\label{Eq:Klein_factor_defining_condition_2}
\end{gather}
In our proof of Proposition~\ref{Proposition:Klein_factors} below we show that these conditions define unique operators with the stated properties.  

\remark{}{
Equation~\eqref{Eq:Klein_factor_defining_condition_1} is a simple example of a defining condition for an implementer of a Bogoliubov transformation; see \cite{Ruijsenaars:1977} for a unified construction of such implementers.
}

\remark{}{
The conditions in \eqref{Eq:Klein_factor_defining_condition_1}--\eqref{Eq:Klein_factor_defining_condition_2} are not minimal \cite{Ruijsenaars:1977} and are chosen so as to allow for a simple proof.
} 

In our proof of Proposition~\ref{Proposition:Klein_factors} we make repeated use of the following technical result:

\lemma{}{
\label{Lemma:Unitarity_lemma}
Let $C$ be a linear operator from $\mathcal{D}_F$ to $\mathcal{D}_F$ satisfying the following two conditions: $C$ commutes with all fermion field operators $\hat{\psi}^{(\dagger)}_{r}(k)$, and $C\Omega = c\Omega$ for some constant $c \in \mathbb{C}$.
Then $C = cI$.
}

\begin{proof}
The first condition and \eqref{Eq:psi_representation} imply that $C$ commutes with all operators $c^{(\dagger)}_r(k)$. 
The action of $C$ on a fermion state $\eta^{F}_{\mathbf{n}}$ in \eqref{Eq:Fermion_states} is then fully determined by its action on $\Omega$, and the second condition then implies that $C \eta^{F}_{\mathbf{n}} = c\eta^{F}_{\mathbf{n}}$.
The assertion follows from linearity.
\end{proof}

\begin{proof}[Proof of Proposition~\ref{Proposition:Klein_factors}]
We rewrite the defining conditions of $R_{\pm}$ in terms of the operators $c^{(\dagger)}_{r}(k)$ using \eqref{Eq:psi_representation}.
The identities in \eqref{Eq:Klein_factor_defining_condition_1} take the following form:
\begin{equation}
\label{Eq:Rrcr}
	\begin{aligned}
	R\pdag_{r}c\pdag_{r}(k) & = \begin{cases}
		c^{\dagger}_{r}(\tfrac{\pi}{L}) R\pdag_{r} & \text{if} \;\, k = -\tfrac{\pi}{L}, \\
		c\pdag_{r}(k + \tfrac{2\pi}{L}) R\pdag_{r} & \text{otherwise},
	\end{cases}\\
	R\pdag_{r}c^{\dagger}_{r}(k) & = \begin{cases}
		c\pdag_{r}(\tfrac{\pi}{L}) R\pdag_{r} & \text{if} \;\, k = -\tfrac{\pi}{L}, \\
		c^{\dagger}_{r}(k + \tfrac{2\pi}{L}) R\pdag_{r} & \text{otherwise},
	\end{cases}
	\end{aligned}
	\qquad
	R\pdag_{r} c^{(\dagger)}_{-r}(k) = -c^{(\dagger)}_{-r}(k) R\pdag_{r},
\end{equation}
and \eqref{Eq:Klein_factor_defining_condition_2} becomes
\begin{equation}
\label{Eq:RrOmega1}
	R\pdag_{r} \Omega = c^{\dagger}_{r}(\tfrac{\pi}{L}) \Omega.
\end{equation}
Using this, the action of $R_{\pm}$ on the fermion states $\eta^F_{\mathbf{n}}$ in \eqref{Eq:Fermion_states} can be computed, and the result is obviously in $\mathcal{D}_F$. 
It then follows that \eqref{Eq:Klein_factor_defining_condition_1}--\eqref{Eq:Klein_factor_defining_condition_2} determine two unique linear operators $R_{\pm}$ mapping $\mathcal{D}_F$ to $\mathcal{D}_F$. 

We now show that both $R_r\pdag R_r^{\dagger}$ and $R_r^{\dagger} R_r\pdag$ satisfy the conditions of the operator $C$ in Lemma~\ref{Lemma:Unitarity_lemma} with the constant $c=1$, which then implies unitarity: $R_r^{\dagger}=R_r^{-1}$ (cf.\ Lemma~\ref{Lemma:Unitarity}).
By taking the adjoint of \eqref{Eq:Klein_factor_defining_condition_1} and replacing $k$ with $k-\tfrac{2\pi}{L}$ we obtain
\begin{equation} 
\label{Eq:Klein_factor_defining_condition_3}
	R_{\pm}^{\dagger} \hat{\psi}^{(\dagger)}_{\pm}(k)
		= \hat{\psi}^{(\dagger)}_{\pm}(k - \tfrac{2\pi}{L})R_{\pm}^{\dagger} ,
	\qquad
	R_{\pm}^{\dagger} \hat{\psi}^{(\dagger)}_{\mp}(k)
		= -\hat{\psi}^{(\dagger)}_{\mp}(k)R_{\pm}^{\dagger} ,  
\end{equation} 
which together with \eqref{Eq:Klein_factor_defining_condition_1} implies that $R_r\pdag R_r^{\dagger}$ and $R_r^{\dagger} R_r\pdag$ commute with all fermion field operators $\hat{\psi}^{(\dagger)}_r(k)$.
To check the second condition, we compute the inner product of $R_{r}^{\dagger}\Omega$ with the fermion states in \eqref{Eq:Fermion_states}.
Using \eqref{Eq:Rrcr}, \eqref{Eq:RrOmega1}, and the CAR in \eqref{Eq:c-operator_anticommutation_and_annihilation_relations} we find
\begin{equation} 
	\langle \eta^F_{\mathbf{n}},R_{r}^{\dagger}\Omega\rangle
	= \langle R_{r}\eta^F_{\mathbf{n}},\Omega\rangle
	= \begin{cases}
		1 & \text{if} \;\, \eta^F_{\mathbf{n}} = c_r^{\dagger}(-\tfrac{\pi}{L})\Omega, \\
		0 & \text{otherwise}
	\end{cases} 
\end{equation} 
(to see this, we observe that $R_{r}c^{\dagger}_{r}(-\tfrac{\pi}{L})\Omega = c\pdag_{r}(\tfrac{\pi}{L}) R_{r}\Omega = c\pdag_{r}(\tfrac{\pi}{L}) c^{\dagger}_{r}(\tfrac{\pi}{L}) \Omega = \Omega$ and that no other inner product can yield a non-zero result),
which is only possible if
\begin{equation}
\label{Eq:RrOmega2}
	R_{r}^{\dagger}\Omega = c^{\dagger}_{r}(-\tfrac{\pi}{L}) \Omega.
\end{equation}
It follows that $R_{r}^{\dagger} R\pdag_{r} \Omega = R^{\dagger}_{r} c^{\dagger}_{r}(\tfrac{\pi}{L}) \Omega = c\pdag_{r}(-\tfrac{\pi}{L}) R^{\dagger}_{r} \Omega = \Omega$ and similarly $R_r\pdag R_r^{\dagger}\Omega=\Omega$.
This concludes the proof of the unitarity of the Klein factors.

We now turn to the commutation relation in \eqref{Eq:JrRr}. 
Using the defining properties of the Klein factors in \eqref{Eq:Klein_factor_defining_condition_1} and the regularized fermion densities in \eqref{Eq:Jrreg} to compute
\begin{equation}
\label{Eq:JrRr_compute} 
	\left( R\pdag_{r'} \hat{J}_{r\pp}^{\Lambda}(p) R^{\dagger}_{r'}
		- \hat{J}_{r}^{\Lambda}(p) \right) \eta
\end{equation} 
for an arbitrary $\eta \in \mathcal{D}_F$, one finds that this is equal to 
\begin{equation}
\label{Eq:JrRr_compute2}
	\delta_{r,r'} \sum_{k} \frac{2\pi}{L}
		\Bigl( \Theta( \Lambda - \abs{k - \tfrac{2\pi}{L} + \tfrac{p}{2}} ) -
			\Theta( \Lambda - \abs{k + \tfrac{p}{2}} ) \Bigr)
		\hat{\psi}^{\dagger}_{r}(k) \hat{\psi}\pdag_{r}(k+p)\eta. 
\end{equation} 
The fermion bilinear above can be cast in a normal-ordered form by inserting \eqref{Eq:noord_psipsi}, and, by a similar argument as the one following \eqref{Eq:JrJr1}, we conclude that \eqref{Eq:JrRr_compute2} is equal to   
\begin{equation} 
	\delta_{r,r'}\sum_{k} 
		\Bigl( \Theta( \Lambda - \abs{k - \tfrac{2\pi}{L} + \tfrac{p}{2}} ) -
			\Theta( \Lambda - \abs{k + \tfrac{p}{2}} ) \Bigr)\Theta(-rk)\eta
	= -r\delta_{r,r'} \eta
\end{equation} 
for sufficiently large $\Lambda$.
This proves $R\pdag_{r'} \hat{J}\pdag_{r\pp}(p) R^{\dagger}_{r'} - \hat{J}\pdag_{r\pp}(p) = -r\delta_{r,r'}$ on $\mathcal{D}_F$ (cf.\ Lemma~\ref{Lemma:Commutators}), which is equivalent to \eqref{Eq:JrRr}.

The commutation relation in \eqref{Eq:H0Rr} can be proved in a similar manner by showing that $H_{0} - R\pdag_r H_{0} R^{\dagger}_r = r (\pi/L) ( \hat{J}\pdag_{r}(0) + R\pdag_{r} \hat{J}\pdag_r(0) R^{\dagger}_{r} )$ on $\mathcal{D}_F$.
For this we use \eqref{Eq:H0reg}, \eqref{Eq:Jrreg}, and \eqref{Eq:noord_psipsi} to compute
\begin{multline} 
\label{Eq:H0Rr_compute}
	\left( H^\Lambda_{0}  - R\pdag_r H^\Lambda_{0} R^{\dagger}_r - r \frac{\pi}{L}
		\left[ \hat{J}_{r}^{\Lambda}(0)
			+ R^{\vphantom{\Lambda}}_{r} \hat{J}_{r}^{\Lambda}(0) R^{\dagger}_{r}
		\right] \right) \eta \\
	= \sum_{k} \Biggl( \frac{2\pi}{L} r \left(k -\tfrac{\pi}{L} \right)
		 \Bigl(
		 	\Theta( \Lambda - \abs{k}) - \Theta( \Lambda - \abs{k - \tfrac{2\pi}{L}} ) 
		\Bigr) \hat{\psi}^{\dagger}_{r}(k) \hat{\psi}\pdag_{r}(k) \\
	+ r \frac{\pi}{L} \Bigl( \Theta( \Lambda - \abs{k - \tfrac{2\pi}{L}} ) +
			\Theta( \Lambda - \abs{k}) \Bigr) \Theta(-rk) \Biggr)  \eta 
\end{multline} 
for an arbitrary $\eta \in \mathcal{D}_F$.
As before, the fermion bilinear can be cast in a normal-ordered form by inserting \eqref{Eq:noord_psipsi}, and, by a similar argument as the one following \eqref{Eq:JrJr1}, we find that the r.h.s.\ in \eqref{Eq:H0Rr_compute} is equal to
\begin{equation} 
	r \sum_{k} 
	\Bigl( k \Theta( \Lambda - \abs{k} )
		- \left(k - \tfrac{2\pi}{L} \right) \Theta( \Lambda - \abs{k - \tfrac{2\pi}{L}} )
	\Bigr) \Theta(-rk)\eta = 0 
\end{equation} 
for sufficiently large $\Lambda$, which proves the result (cf.\ Lemma~\ref{Lemma:Commutators}).

Finally, equation \eqref{Eq:RR} can be proved by showing that $R_{+}\pdag R_{-}\pdag R_{+}^{\dagger} R_{-}^{\dagger}$ satisfies the conditions of the operator $C$ in Lemma~\ref{Lemma:Unitarity_lemma} with the constant $c=-1$.
This implies that $R\pdag_{+} R\pdag_{-} R^{\dagger}_{+} R^{\dagger}_{-} = -I$, and the result then follows by multiplication with $R_{-}R_{+}$ from the right.
Indeed, it is easily verified using \eqref{Eq:Klein_factor_defining_condition_1} and  \eqref{Eq:Klein_factor_defining_condition_3} that $R_{+}\pdag R_{-}\pdag R_{+}^{\dagger} R_{-}^{\dagger}$ commutes with all fermion field operators $\hat{\psi}^{(\dagger)}_r(k)$.
Moreover, \eqref{Eq:Rrcr}--\eqref{Eq:RrOmega1} and \eqref{Eq:RrOmega2} imply that $R_{+}\pdag R_{-}\pdag R_{+}^{\dagger} R_{-}^{\dagger} \Omega = c\pdag_{-}(\tfrac{\pi}{L}) c\pdag_{+}(\tfrac{\pi}{L}) c^{\dagger}_{-}(\tfrac{\pi}{L}) c^{\dagger}_{+}(\tfrac{\pi}{L})\Omega = -\Omega$, where the CAR in \eqref{Eq:c-operator_anticommutation_and_annihilation_relations} was used in the last step.
\end{proof}

\subsection{Boson-fermion correspondence} 
\label{SubSec:BF_correspondence}
We are now ready to explain in which sense the fermion Fock space $\mathcal{F}$ is identical with a boson Fock space.
For this we recall the boson creation and annihilation operators $b^{\dagger}(p)$ and $b(p)$ in \eqref{Eq:Harmonic_oscillator_operators}.
Since the Klein factors commute with the boson operators (this is a simple consequence of \eqref{Eq:JrRr}), all states $R_{+}^{q_+}R_{-}^{-q_-}\Omega$, where $q_{\pm} \in \mathbb{Z}$, constitute a possible boson vacuum:
\begin{equation} 
\label{Eq:Boson_vacua}
	b(p)R_{+}^{q_+}R_{-}^{-q_-}\Omega = 0
	\qquad
	\forall p\neq 0. 
\end{equation} 
This suggests to introduce boson states 
\begin{equation}
\label{Eq:Boson_states}
	\eta_{\mathbf{m}}^{B}
	\define
	\left( \prod_{p \neq 0} \frac{b^{\dagger}(p)^{m(p)}}{\sqrt{m(p)!}} \right)
	R_{+}^{q_{+}} R_{-}^{-q_{-}} \Omega ,
	\qquad 
	q_{\pm} \in \mathbb{Z},
	\;\,
	m(p) \in \mathbb{N}_{0},
\end{equation}
where $\mathbf{m}$ is short for the vector $( \left( m(p)\right)_{p \neq 0}, q_{+}, q_{-} )$ and where only a finite number of the quantum numbers $m(p)$ are non-zero.

Our main result in this section is that the boson states in \eqref{Eq:Boson_states} comprise a complete orthonormal basis in the fermion Fock space $\mathcal{F}$: 
\begin{equation} 
\label{Eq:B-basis} 
	\langle \eta^B_{\mathbf{m}\pp}, \eta^B_{\mathbf{m}'} \rangle
		= \delta_{\mathbf{m},\mathbf{m}'},
	\qquad
	\sum_{\mathbf{m}} \eta^B_{\mathbf{m}}\langle \eta^B_{\mathbf{m}},\cdot \rangle = I. 
\end{equation} 
Orthonormality is easy to check (it follows from the commutation relations in \eqref{Eq:Harmonic_oscillator_operator_properties}, the vacuum condition in \eqref{Eq:Boson_vacua}, and the unitarity of the Klein factors).
The non-trivial part is therefore completeness. 

Recall that $\mathcal{D}_F$ is the vector space of all finite linear combinations of the fermion states in \eqref{Eq:Fermion_states}. 
Since the boson creation operators $b^{\dagger}(p)$ and the Klein factors map $\mathcal{D}_F$ to $\mathcal{D}_F$, it is clear that all boson states in \eqref{Eq:Boson_states} are in $\mathcal{D}_F$. 
The converse is also true: 

\proposition{}{
\label{Proposition:BF}
The vector space of all finite linear combinations of the boson states in \eqref{Eq:Boson_states} is identical with $\mathcal{D}_F$.
}

Our proof of Proposition~\ref{Proposition:BF} is based on the following:

\lemma{}{
\label{Lemma:Boson_eigenvalues}
All boson states $\eta^B_{\mathbf{m}}$ in \eqref{Eq:Boson_states} are eigenstates of the fermion Hamiltonian $H_0$ in \eqref{Eq:Free_fermion_Hamiltonian} with corresponding eigenvalues
\begin{equation} 
\label{Eq:Boson_states_eigenvalues} 
	E^{B}_{\mathbf{m}}
	\define
	\frac{\pi}{L} \left( q_{+}^2 + q_{-}^2 \right) + \sum_{p \neq 0} \abs{p} m(p). 
\end{equation} 
} 

\begin{proof}
The definition of the boson operators $b^{(\dagger)}(p)$ in \eqref{Eq:Harmonic_oscillator_operators} and the commutation relations in \eqref{Eq:H0_and_Jr_commutation_relation} imply
\begin{equation} 
	\left[ H_0 , b^{\dagger}(p) \right] = \abs{p} b^{\dagger}(p)
	\qquad
	\forall p\neq 0
\end{equation} 
on $\mathcal{D}_F$. This, together with \eqref{Eq:H0RqRq}, yields the eigenvalue equation $H_0 \eta^B_{\mathbf{m}} = E^{B}_{\mathbf{m}} \eta^B_{\mathbf{m}}$.  
\end{proof} 

\begin{proof}[Proof of Proposition~\ref{Proposition:BF}]
Let $E$ be a fixed eigenvalue of $H_0$.
Denote by $\dim_{F}(E)$ the number of distinct fermion eigenstates $\eta_{\mathbf{n}}^{F}$ such that $E^{F}_{\mathbf{n}} = E$ (cf.\ Lemma~\ref{Lemma:H0}).
Correspondingly, denote by $\dim_{B}(E)$ the number of distinct boson eigenstates $\eta^{B}_{\mathbf{m}}$ such that $E^{B}_{\mathbf{m}} = E$ (cf.\ Lemma~\ref{Lemma:Boson_eigenvalues}).
It follows from \eqref{Eq:Fermion_states_eigenvalues} and \eqref{Eq:Boson_states_eigenvalues}  that both $\dim_{F}(E)$ and $\dim_{B}(E)$ are finite.

We now prove that
\begin{equation}
\label{Eq:Equality_of_number_of_eigenstates}
	\dim_{F}(E) = \dim_{B}(E)
\end{equation}
for all eigenvalues $E$ of $H_0$. 
To this end, we introduce the generating functions
\begin{equation}
\label{Eq:Generating_functions_special_case}
	Z_{X}^{\beta} \define \sum_{E} \dim_{X}(E) e^{-\beta E}
	\qquad
	(X = B,F) 
\end{equation}
for some parameter $\beta > 0$, where the sum is over all distinct eigenvalues $E$ of $H_0$. 
(As the symbols suggest, $\beta$ and $Z_{X}^{\beta}$ have physical interpretations as inverse temperature and partition functions, respectively.)

The generating functions can be computed as follows:
\begin{equation}
	Z_{F}^{\beta}
		= \sum_{\mathbf{n}} e^{-\beta E_{\mathbf{n}}^{F}},
	\qquad
	Z_{B}^{\beta}
		= \sum_{\mathbf{m}}	e^{-\beta E_{\mathbf{m}}^{B}}. 
\end{equation}
From the explicit expression for the eigenvalues in \eqref{Eq:Fermion_states_eigenvalues} we obtain
\begin{equation}
	\begin {aligned}
	Z_{F}^{\beta}
		& = \prod_{r} \prod_{k} \sum_{n_{r}(k) \in \left\{ 0,1 \right\}}
			e^{-\beta\abs{k} n_{r}(k)}  \\
		& = \Biggl( \prod_{n \in \mathbb{Z}}
			\left( 1 + e^{-\beta \frac{2\pi}{L}\abs{n+\frac{1}{2}}}
			\right) \Biggr)^{2}
		= \Biggl( \prod_{n=1}^{\infty} \left( 1 + e^{-\beta \frac{\pi}{L}(2n-1)}
			\right) \Biggr)^{4},
	\end{aligned}
\end{equation}
and similarly, using \eqref{Eq:Boson_states_eigenvalues},
\begin{equation}
	\begin{aligned}
	Z_{B}^{\beta}
		& = \sum_{q_{+},q_{-} \in \mathbb{Z}}
			\prod_{p \neq 0} \sum_{m(p) = 0}^{\infty}
			e^{-\beta \frac{\pi}{L}(q_{+}^{2} + q_{-}^{2})}
			e^{-\beta \abs{p} m(p)} \\
		& = \Biggl( \sum_{q \in \mathbb{Z}} e^{-\beta \frac{\pi}{L}q^2}
			\prod_{n=1}^{\infty} \sum_{m = 0}^{\infty}
			e^{-\beta \tfrac{2\pi}{L}n m} \Biggr)^{2}
		= \Biggl( \sum_{q \in \mathbb{Z}} e^{-\beta \frac{\pi}{L}q^2}
			\prod_{n=1}^{\infty} \frac{1}{1 - e^{-\beta \tfrac{2\pi}{L}n}} \Biggr)^{2}.
	\end{aligned}
\end{equation}
Hence, $Z_{F}^{\beta} = Z_{B}^{\beta}$ is equivalent to
\begin{equation}
\label{Eq:Jacobi_triple_product_special_case}
	\Biggl( \prod_{n=1}^{\infty} \left( 1 + z^{2n-1} \right) \Biggr)^{2}
		= \Biggl( \sum_{q \in \mathbb{Z}} z^{q^2} \Biggr)
			\Biggl( \prod_{n=1}^{\infty} \frac{1}{1 - z^{2n}} \Biggr)
\end{equation}
with $z = e^{-\beta \pi/L}$.
The latter is a special case of a mathematical identity known as Jacobi's triple product; see e.g.\ Chapter 16 in \cite{AbramowitzStegun:1972}.
It follows that $Z_{F}^{\beta} = Z_{B}^{\beta}$ holds true, which proves \eqref{Eq:Equality_of_number_of_eigenstates}.

In conclusion, we have proved that every fermion state $\eta^F_{\mathbf{n}}$ can be written as a linear combination of boson states $\eta^B_{\mathbf{m}}$ such that $E^B_{\mathbf{m}}=E^F_{\mathbf{n}}$, and there is only a finite number of such boson states.
The converse is obvious (see the above). 
\end{proof}

\subsection{Boson Hamiltonian}
From the results in Section~\ref{SubSec:BF_correspondence} it is easy to deduce that the fermion Hamiltonian $H_0$ in \eqref{Eq:Free_fermion_Hamiltonian} can be written as a boson Hamiltonian: the commutation relations in \eqref{Eq:Harmonic_oscillator_operator_properties} and \eqref{Eq:H0RqRq} imply that the boson states $\eta^B_{\mathbf{m}}$ in \eqref{Eq:Boson_states} are eigenstates of the operator 
\begin{equation}
\label{Eq:Free_bosonized_Hamiltonian}
	H^B_{0} = \frac{\pi}{L} \left( Q_{+}^2 + Q_{-}^2 \right)
		+ \sum_{p \neq 0}
		\abs{p} b^{\dagger}(p) b(p)
\end{equation}
with corresponding eigenvalues $E^B_{\mathbf{m}}$ in \eqref{Eq:Boson_states_eigenvalues}. 
Thus, due to Lemma~\ref{Lemma:Boson_eigenvalues}, it follows that $(H_0 - H_0^B)\eta^{B}_{\mathbf{m}}=0$ for all boson states.
Since these states form a complete orthonormal basis in $\mathcal{F}$, we conclude that $H_0 = H_0^B$ on $\mathcal{F}$. 
Expressing the operators $b^{(\dagger)}(p)$ and $Q_{\pm}$ in terms of the fermion densities (cf.\ \eqref{Eq:Harmonic_oscillator_operators} and \eqref{Eq:Qr}) we can state this result as follows. 

\proposition{Kronig's identity}{
The following holds true as an identity of self-adjoint operators on $\mathcal{F}$: 
\begin{equation}
\label{Eq:Kronigs_identity}
	H_0 = \sum_{r} \Biggl( \frac{\pi}{L}\hat{J}_r(0)^2
		+ \sum_{p>0} \frac{2\pi}{L} \hat{J}_{r}(-rp)\hat{J}_r(rp) \Biggr)
	= \sum_{r} \sum_{p} \frac{\pi}{L} \noord{\hat{J}_{r}(-p)\hat{J}_{r}(p)}.  
\end{equation} 
}

\remark{}{
\label{Remark:Normal_ordering}
Mathematical physicists often write Kronig's identity in \eqref{Eq:Kronigs_identity} with boson normal ordering, i.e.\ with $\bosonnoord{ \hat{J}_{r}(-p) \hat{J}_{r}(p) }$ instead of $\noord{ \hat{J}_{r}(-p) \hat{J}_{r}(p) }$.
We note that for expressions containing only factors with at most two field operators, as in this case, both normal orderings give the same result.
However, for other expressions it is important to distinguish between boson and fermion normal ordering.
}

We conclude this section by making precise how this is related to a conventional boson Hamiltonian.
It is natural to introduce the following operators:
\begin{equation}
\label{Eq:Phi_Pi_operators}
\begin{aligned}
	\hat{\Pi}(p)
		& = -\frac{1}{\sqrt{2}} \left( \hat{J}_{+}(p) - \hat{J}_{-}(p) \right), \\
	\hat{\Phi}(p)
		& = \frac{1}{ip\sqrt{2}} \left( \hat{J}_{+}(p) + \hat{J}_{-}(p) \right)
		\qquad (p \neq 0)
\end{aligned}
\end{equation} 
which obey CCR of neutral bosons:
\begin{equation} 
\label{Eq:CCR_in_Fourier_space} 
	\left[ \hat{\Phi}(p), \hat{\Pi}^{\dagger}(p') \right]
		= i\frac{L}{2\pi} \delta_{p,p'},
	\qquad 
	\left[ \hat{\Phi}(p), \hat{\Phi}^{\dagger}(p') \right]
		= \left[ \hat{\Pi}(p), \hat{\Pi}^{\dagger}(p') \right]
		= 0 , 
\end{equation}
where $\hat{\Pi}^{\dagger}(p) \define \hat{\Pi}(-p)$ and $\hat{\Phi}^{\dagger}(p) \define \hat{\Phi}(-p)$.
Furthermore, they allow Kronig's identity to be written in the form\footnote{This Hamiltonian corresponds to an important model in particle physics describing relativistic bosons in 1+1 dimensions; cf.\ Remark~\ref{Remark:Dirac_fermions_KG_bosons} in Appendix~\ref{Appendix:Bosonization_nutshell}.}
\begin{equation}
\label{Eq:Kronig2}
	H_0 = \frac{\pi}{L} \left( Q_{+}^2 + Q_{-}^2 \right)
		+ \sum_{p} \frac{\pi}{L} \! \noord{ \left( \hat{\Pi}^{\dagger}(p) \hat{\Pi}(p)
		+ p^2 \hat{\Phi}^{\dagger}(p) \hat{\Phi}(p) \right) }.
\end{equation}  
A difficulty with this interpretation is that the operator $\hat{\Phi}(0)$ is not defined.
This does not matter for Kronig's identity, but it is a problem for the CCR. 
One natural solution to this is as follows.
The non-trivial commutation relation in \eqref{Eq:QrRr} implies
\begin{subequations}
\label{Eqs:RRPiQ_comm_rel}
\begin{align}
	(R_{+}R_{-}) (Q_{+} + Q_{-}) (R_{+}R_{-})^{\dagger}
		& = Q_{+} + Q_{-}, \\
	(R_{+}R_{-}) \hat{\Pi}(0) (R_{+}R_{-})^{\dagger}
		& = \hat{\Pi}(0) + \sqrt{2}.
\end{align}
\end{subequations} 
This suggests to formally identify $R_{+} R_{-}$ with
\begin{equation} 
	\exp \left( -i \frac{2\sqrt{2}\pi}{L} \hat{\Phi}(0) \right) 
\end{equation}  
since this is compatible with \eqref{Eqs:RRPiQ_comm_rel} and the CCR in \eqref{Eq:CCR_in_Fourier_space} for $p=p'=0$.
The natural interpretation of $\hat{\Phi}(0)$ is thus in the form of a variable on a circle with radius $L/(2\sqrt{2}\pi)$, i.e.\ $\hat{\Phi}(0)$ should be identified with $\hat{\Phi}(0) + {L}/{\sqrt{2}}$.
We note that this interpretation is also compatible with the discrete spectrum $(1/\sqrt{2})\mathbb{Z}$ of $\hat{\Pi}(0)$.


\section{From bosons to fermions}
\label{Sec:Mathematics_of_bosonization_II}
This section is dedicated to the reconstruction of the fermion field operators $\hat{\psi}^{(\dagger)}_{r}(k)$ from the fermion densities $\hat{J}_r(p)$ and the Klein factors $R_r$.
The main objects of study will be so-called \emph{vertex operators}
\begin{equation} 
\label{Eq:Gamma0}
	\bosonnoord{ R_{r}^{w} \exp\left( i \sum_{p}\frac{2\pi}{L}\alpha(p)\hat{J}_r(-p) \right) } 
\end{equation}
with integers $w$ and complex parameters $\alpha(p)$.
The crosses in \eqref{Eq:Gamma0} indicate boson normal ordering which is defined in \eqref{Eq:Bosonnoord} below.
Our goal is to prove that the vertex operators
\begin{equation} 
\label{Eq:Regularized_psi_explicit0}
	\psi_{r}(x;\epsilon)
	= \frac{1}{\sqrt{L}} \! \bosonnoord{ R_{r}^{-r}
		\exp \left( r\frac{2\pi i}{L}\left( \hat{J}_{r}(0)x  - \sum_{p \neq 0} 
			\frac{\hat{J}_{r}(-p)}{ip} e^{-ipx - \epsilon \abs{p}/2} \right) \right)}
\end{equation}
converge in the limit $\epsilon \to 0^{+}$ to the formal fields $\psi_r(x)$ in the introduction, in a sense made precise in Proposition~\ref{Proposition:Fermions_from_bosons}. 
Using this we will also develop tools needed to compute fermion correlation functions.
We note that the formula in \eqref{Eq:Regularized_psi_explicit0} appeared in the condensed matter physics literature in \cite{LutherPeschel:1974, Mattis:1974} and in the mathematics literature in \cite{Segal:1981}.

\remark{}{
The name of the objects in \eqref{Eq:Gamma0} is motivated by the study of vertex operator algebras in the mathematics literature; see e.g.\ \cite{Kac:1998}.
}

\subsection{Vertex operators}
\label{SubSec:Vertex_operators1}
We consider periodic functions $e^{if}$ on the space $[-L/2, L/2]$ given by 
\begin{equation} 
\label{Eq:f}
	f(y) = w\frac{2\pi}{L} y + \sum_{p} \frac{2\pi}{L} \alpha(p) e^{ipy}
	\qquad
	( y \in [-L/2,L/2] ),
\end{equation}
where $w \in \mathbb{Z}$ and $\alpha(p) \in \mathbb{C}$ satisfying the following conditions:
\begin{equation}
\label{Eq:G}  
	\overline{\alpha(p)} = \alpha(-p),
	\qquad
	\sum_{p} \frac{2\pi}{L} \abs{p} \abs{\alpha(p)}^2 < \infty.
\end{equation}
The first condition is equivalent to requiring that $f$ must be real-valued, i.e.\ $f(y) = \overline{f(y)}$ for all $y \in [-L/2,L/2]$, and the second is a weak smoothness condition on $f$.\footnote{Functions satisfying the second condition in \eqref{Eq:G} are sometimes said to be $C^{1/2}$; this condition is weaker than differentiability, but stronger than continuity.}
We denote by $\mathcal{G}$ the set of all functions $e^{if}$ satisfying these conditions.

For each $e^{if} \in \mathcal{G}$, we define
\begin{equation}
\label{Eq:Gamma}
	\Gamma_r(e^{if})
	\define
	e^{i\pi\alpha(0) \hat{J}_r(0)/L} R_r^{w} e^{i\pi\alpha(0) \hat{J}_r(0)/L}
	\exp \left( i \sum_{p\neq 0}\frac{2\pi}{L}\alpha(p)\hat{J}_r(-p) \right).
\end{equation} 
If $f$ is real-valued, then the properties of the fermion densities $\hat{J}_{r}(p)$ and the Klein factors $R_r$ derived in Section~\ref{Sec:Mathematics_of_bosonization_I} suggest that $\Gamma(e^{if})$ defines a unitary operator.
As we will show, this is indeed the case.

\remark{}{
We note that, in order for $e^{if}$ to be periodic with period $L$, it is enough to require that $f$ is periodic on $[-L/2, L/2]$ up to an integer multiple of $2\pi$. 
It follows that, since $w = [f(L/2) - f(-L/2)]/(2\pi)$ and since $\alpha(p)$ are the Fourier coefficients of the periodic function $f(y) - w 2\pi x/L$, the parameters $\left(w, ( \alpha(p)\right)_{p} )$ are in a one-to-one correspondence with such functions $e^{if}$.
This integer $w$ is often referred to as \emph{winding number}. 
}

\remark{}{
It is easy to verify that $\mathcal{G}$ forms a group under multiplication (with $e^{if_{1}}e^{if_{2}} = e^{i[f_{1} + f_{2}]}$, $(e^{if})^{-1} = e^{-if}$, etc.).
In fact, this group is an important example of a so-called \emph{loop group}, and $\Gamma_r(e^{if})$ in \eqref{Eq:Gamma} corresponds to a projective representation of $\mathcal{G}$ \cite{CareyLangmann:GeoAnalysis:2002}.
}

The vertex operators in \eqref{Eq:Gamma0} are related to $\Gamma_{r}(e^{if})$ in the sense that the former can be written as $\bosonnoord{\Gamma_{r}(e^{if})}$.
The precise definition of $\bosonnoord{\cdots}$ for such operators is
\begin{equation} 
\label{Eq:Bosonnoord}
	\bosonnoord{\Gamma_r(e^{if})}
	\define
	e^{iJ_{r}^{+}(\alpha)} e^{i J_{r}^{0}(\alpha)/2}
		R_{r}^{w} e^{i J_{r}^{0}(\alpha)/2} e^{iJ_{r}^{-}(\alpha)}
\end{equation}
with $\alpha$ short for $(\alpha(p))_{p}$ and 
\begin{equation}
\label{Eq:J_operators}
	J_{r}^{\pm}(\alpha)
	\define
	\sum_{p>0} \frac{2\pi}{L} \alpha(\pm rp) \hat{J}_{r}(\mp rp),
	\qquad
	J_{r}^{0}(\alpha)
	\define
	\frac{2\pi}{L} \alpha(0) \hat{J}_{r}(0).
\end{equation}
As a motivation, we note that Proposition~\ref{Proposition:Bosons_from_fermions} suggests that\footnote{As will be shown, the $J_r^\pm(\alpha)$ are well-defined as quadratic forms on $\mathcal{D}_F$, and this gives a precise meaning to the relations in \eqref{Eq:J_operators_annihilation_creation}; see Section~\ref{SubSec:Proof_of_vertex_operator_proposition} and also Appendix~\ref{Appendix:Quadratic_forms}.}
\begin{equation}
\label{Eq:J_operators_annihilation_creation}
	J_{r}^{-}(\alpha)\Omega = J_{r}^{+}(\alpha)^{\dagger}\Omega = 0,
\end{equation}
and it is therefore natural to regard $J_{r}^{+}(\alpha)$ and $J_{r}^{-}(\alpha)$ as creation and annihilation operators, respectively.
We conclude that \eqref{Eq:Bosonnoord} indeed is a normal ordering prescription: all creation operators are to the left of the annihilation operators, and the zero-mode parts, namely $J^{0}_{r}(\alpha)$ and the Klein factors $R_r$, are placed symmetrically in the middle.

We find it convenient to introduce the shorthand notation
\begin{align} 
	\label{Eq:cr} 
	c_{r}(\alpha_1,\alpha_2)
	& \define \sum_{p>0}
	\frac{2\pi}{L} p \alpha_1(-rp)\alpha_2(rp), \\
	\label{Eq:Cr} 
	C_{r}(e^{if_{1}},e^{if_{2}})
	& \define
	e^{i\pi r \left( w_{2}\alpha_{1}(0) - w_{1}\alpha_{2}(0) \right)/L} e^{-c_{r}(\alpha_{1},\alpha_{2})}.
\end{align}
Note that the second condition in \eqref{Eq:G} is equivalent to requiring that $c_r(\alpha,\alpha)$ is finite.
Moreover, if both $\alpha_{1}(p)$ and $\alpha_{2}(p)$ satisfy this condition, then $c_r(\alpha_{1},\alpha_{2})$ is finite (by the Cauchy-Schwarz inequality).

The following proposition summarizes important properties of the vertex operators which we will need:

\proposition{}{
\label{Proposition:Vertex_operators}
For each $e^{if} \in \mathcal{G}$, $\Gamma_r(e^{if})$ in \eqref{Eq:Gamma} defines a unitary operator on $\mathcal{F}$ satisfying
\begin{subequations} 
\begin{align} 
	\label{Eq:Bosonnoord1}
	\Gamma_r(e^{if})
		& = e^{-c_r(\alpha,\alpha)/2} \! \bosonnoord{ \Gamma_r(e^{if}) }, \\
	\label{Eq:Gamma_dag}
	\bosonnoord{\Gamma_r(e^{if})}^{\dagger}
		& = \bosonnoord{ \Gamma_r(e^{-if}) }, 
\end{align} 
\end{subequations}
and
\begin{subequations} 
\begin{align} 
	\label{Eq:RrGamma1}
	R_{\pm} \! \bosonnoord{\Gamma_{\pm}(e^{if})}
		& = e^{\mp 2\pi i \alpha_{\pm}(0)/L} \! \bosonnoord{\Gamma_{\pm}(e^{if})} \! R_{\pm}, \\
	\label{Eq:RrGamma2}
	R_{\pm} \! \bosonnoord{\Gamma_{\mp}(e^{if})}
		& = (-1)^{w} \! \bosonnoord{\Gamma_{\mp}(e^{if})} \! R_{\pm},
\end{align}
\end{subequations}
together with the following commutation relation on $\mathcal{D}_F$: 
\begin{equation}
\label{Eq:JrGamma}
	\left[ \hat{J}_{r'}(p), \! \bosonnoord{\Gamma_{r}(e^{if})} \right]
		= r\delta_{r,r'} \left( w\delta_{p,0} + i p \alpha(p) \right)
			\! \bosonnoord{\Gamma_{r}(e^{if})}.
\end{equation}
Moreover, the following identities hold true: 
\begin{subequations}
\begin{align} 
	\label{Eq:Gamma_pm_Gamma_pm}
	\bosonnoord{\Gamma_{\pm}(e^{if})} \!\! \bosonnoord{\Gamma_{\pm}(e^{if'})}
		& = C_{\pm}(e^{if},e^{if'})
			\! \bosonnoord{\Gamma_{\pm}(e^{i[f+f']})}, \\
	\label{Eq:Gamma_pm_Gamma_mp}
	\bosonnoord{\Gamma_{\pm}(e^{if})} \!\! \bosonnoord{\Gamma_{\mp}(e^{if'})}
		& = (-1)^{ww'}
			\! \bosonnoord{\Gamma_{\mp}(e^{if'})} \!\! \bosonnoord{\Gamma_{\pm}(e^{if})},
\end{align}
and
\begin{equation}
\label{Eq:OmGammaOm}
	\langle\Omega, \!  \bosonnoord{\Gamma_+(e^{if})} \! 
		\! \bosonnoord{\Gamma_-(e^{if'})} \! \Omega\rangle
	= \delta_{w\pp,0} \delta_{w',0}
\end{equation}
\end{subequations} 
for all $e^{if},e^{if'} \in \mathcal{G}$.
}
\noindent (The proof is given in Section~\ref{SubSec:Proof_of_vertex_operator_proposition}.)

Equation \eqref{Eq:Bosonnoord1} implies that the operator $\bosonnoord{\Gamma_{r}(e^{if})}$ is bounded for all $e^{if} \in \mathcal{G}$ (since it is equal to a unitary operator multiplied by a finite constant).
Therefore, as long as the conditions in \eqref{Eq:G} are satisfied, formal computations with vertex operators are safe. 
However, as we will see, in order to reconstruct the fermion field operators using vertex operators, it is necessary to pass to the non-trivial limit $\abs{\alpha(p)} \to 1/\abs{p}$. 

We now state a simple implication of \eqref{Eq:Gamma_pm_Gamma_pm}, which, together with \eqref{Eq:Gamma_pm_Gamma_mp}--\eqref{Eq:OmGammaOm}, functions as a working horse for computing fermion correlation functions.
 
\corollary{}{
\label{Corollary:Workhorse}
Let $N > 1$ be an integer and consider $e^{if_{j}} \in \mathcal{G}$ for $j = 1, 2, \ldots, N$.
Then the following holds true:
\begin{equation} 
\label{Eq:Gamma^n}
	\bosonnoord{\Gamma_{r}(e^{if_{1}})} \! \ldots \! \bosonnoord{\Gamma_{r}(e^{if_{N}})}
		= \left( \prod_{1 \leq j < j' \leq N} C_{r}( e^{if_{j}}, e^{if_{j'}} ) \right)
			\! \bosonnoord{\Gamma_{r}(e^{i[f_{1} + \ldots + f_{N}]})}
\end{equation} 
with $C_{r}(\cdot,\cdot)$ defined in \eqref{Eq:Cr}.
}
\noindent (The proof follows by a simple induction argument.) 

As a closing remark we mention that, for the formal fermion field operators $\psi_{r}(x)$ in the introduction, the following identity holds true:
\begin{equation} 
	\Gamma_{r}(e^{if})^{\dagger}\psi_{r'}(x)\Gamma_{r}(e^{if})
	= \begin{cases}
		e^{if(x)} \psi_{r}(x)	& \text{if} \;\, r' = r, \\
		\psi_{r'}(x)				& \text{if} \;\, r' \neq r
	\end{cases} 
\end{equation} 
(this follows from \eqref{Eq:Regularized_psi_explicit0} using identities given in Proposition~\ref{Proposition:Vertex_operators}).  
The r.h.s.\ of this has a natural interpretation as a chiral gauge transformation, and, for this reason, $\Gamma_{r}(e^{if})$ are sometimes referred to as \emph{implementers of gauge transformations} in the literature.
In the case of the space $[-L/2, L/2]$, there exist so-called \emph{large gauge transformations} where $f(L/2) - f(-L/2)$ is a non-zero integer multiple of $2\pi$.
It follows that the Klein factors $R^{w}_{r}$ correspond to the implementers of large gauge transformations given by the functions $f(y) = w 2\pi y/L$.
It is also interesting to note that \eqref{Eq:JrGamma} implies 
\begin{equation} 
\Gamma_{r}(e^{if})^{\dagger} J_{r'}(x)\Gamma_{r}(e^{if})
	= J_{r'}(x) + r\delta_{r,r'}\frac1{2\pi} \partial_x f(x)
\end{equation}
for the operator-valued distributions $J_r(x)$ related to $\hat J_r(p)$ by the Fourier transform in \eqref{Eq:Psi_Jr_Fourier_transforms} (cf.\ also \eqref{Eq:Jr_nutshell}).

\remark{}{
The observation that the Klein factors $R_{r}$ and the fermion densities $\hat{J}_{r}(p)$ function as implementers of gauge transformations makes clear that they should be treated on an equal footing.
This is also the reason why we sometimes use the word \emph{bosons} to refer to both of these operators together.
}

\subsection{Reconstruction of the fermion field operators}
\label{SubSec:Vertex_operators2}
Let $\epsilon > 0$ be a regularization parameter and consider functions
\begin{equation} 
\label{Eq:feps}
	f_{x, \epsilon}(y)
		= \frac{2\pi}{L}(y-x)
		+ \sum_{p \neq 0} \frac{2\pi}{L} \frac{1}{ip} e^{ip(y-x) - \epsilon \abs{p}/2}
	\qquad
	(y \in [-L/2, L/2])
\end{equation}
parametrized by a variable $x \in [-L/2, L/2]$.
Since these functions satisfy the conditions in \eqref{Eq:G}, it follows that they can be used to define vertex operators $\bosonnoord{\Gamma_r(e^{-irf_{x,\epsilon}})}$.
Our main result in this section is that these allow us to reconstruct the fermion field operators.

\proposition{}{
\label{Proposition:Fermions_from_bosons}
The following holds true as an identity on $\mathcal{D}_F$\textnormal{:}\footnote{The precise meaning of the limit in \eqref{Eq:Fermions_from_bosons} is given in Remark~\ref{Remark:Strong_topology_limit}.}
\begin{equation}
\label{Eq:Fermions_from_bosons}
	\hat{\psi}_{r}(k)
		 = \lim_{\epsilon \to 0} \frac{1}{\sqrt{2\pi L}}\int_{-L/2}^{L/2} \dop{}{x}
			\! \bosonnoord{\Gamma_r(e^{-irf_{x,\epsilon}})} \! e^{-ikx}.
\end{equation}
}
\noindent (The proof is given in Section~\ref{SubSec:Proof_of_reconstruction_proposition}.)

One important observation is that the r.h.s.\ in \eqref{Eq:Fermions_from_bosons} is a well-defined operator mapping $\mathcal{D}_{F}$ to $\mathcal{D}_{F}$ (since this is true for the l.h.s.).
Combining this with the Fourier transform in \eqref{Eq:Psi_Jr_Fourier_transforms} justifies \eqref{Eq:Regularized_psi_explicit0} since the r.h.s.\ in the latter is equal to $(1/\sqrt{L}) \! \bosonnoord{ \Gamma_{r}(e^{-irf_{x,\epsilon}}) }$, and it is natural to denote this operator by $\psi_{r}(x;\epsilon)$ since it converges to $\psi_r(x)$ in the limit $\epsilon \to 0^+$ in a sense made precise by \eqref{Eq:Fermions_from_bosons} and Remark~\ref{Remark:Strong_topology_limit}.

For the sake of computations, it is useful to write \eqref{Eq:Regularized_psi_explicit0} in the following form:
\begin{equation}
\label{Eq:Regularized_psi_explicit}
	\psi_{r}(x; \epsilon)
		= \frac{\mathcal{N}_{\epsilon}}{\sqrt{2\pi\epsilon}}
			e^{i\pi rx Q_{r}/L} R_{r}^{-r} e^{i\pi rx Q_{r}/L}
		\exp \left( -r \sum_{p \neq 0} \frac{2\pi}{Lp}
			\hat{J}_{r}(-p) e^{-ipx - \epsilon \abs{p}/2} \right)
\end{equation}
with
\begin{equation}
\label{Eq:cNe}
	\mathcal{N}_{\epsilon}
	\define
	\sqrt{ \frac{2\pi\epsilon/L} {1 - e^{-2\pi\epsilon/L}}}
		= 1 + O(\epsilon/L)
\end{equation}
(this follows from \eqref{Eq:Bosonnoord1}; cf.\ also \eqref{Eq:Log_Taylor_expansion}).
Since one is usually only interested in the limit $\epsilon \to 0$, the constant $\mathcal{N}_{\epsilon}$ is often ignored.
For later reference we also state the Hermitian conjugate of \eqref{Eq:Fermions_from_bosons}:
\begin{equation}
\label{Eq:Fermions_from_bosons_conjugate}
	\hat{\psi}^{\dagger}_{r}(k)
		 = \lim_{\epsilon \to 0} \frac{1}{\sqrt{2\pi L}}
		 	\int_{-L/2}^{L/2} \dop{}{x}
			\! \bosonnoord{\Gamma_{r}(e^{irf_{x,\epsilon}})} \! e^{ikx} 
\end{equation}
(this follows from \eqref{Eq:Gamma_dag}).

\remark{}{
Note that $f_{x,\epsilon}(y)$ are $C^{\infty}$-functions which converge to the distribution $\pi\sgn(y-x)$ in the limit $\epsilon \to 0^{+}$. 
This is the key observation which allow the operator-valued distribution $\psi_{r}(x)$ to be obtained as a limit of a bounded operator $\psi_{r}(x;\epsilon)$, which, as we will see later, is very useful to define and compute fermion correlation functions. 
}

\remark{}{
\label{Remark:Strong_topology_limit}
The limit in \eqref{Eq:Fermions_from_bosons} is in the strong operator topology \cite{ReedSimon:1972}.
We emphasize that one can use other regularizations than the damping factor $\exp(-\epsilon\abs{p}/2)$ in \eqref{Eq:feps}.
For instance, this factor can be replaced with a sharp cutoff $\Theta(\Lambda - \abs{p})$.
This would have the advantage that the corresponding limit $\Lambda \to \infty$ in \eqref{Eq:Fermions_from_bosons} would be in the sense of Definition~\ref{Definition:Operator_convergence}, which is somewhat stronger.
However, the $\epsilon$-regularization is easier from a computational point of view.
}

\subsection{Correlation functions}
\label{SubSec:Mathematics_of_bosonization_II:Correlation_functions}
To introduce some notation and techniques needed in Section~\ref{Sec:Fermion-phonon_model}, we demonstrate how the results in Sections~\ref{SubSec:Vertex_operators1}~and~\ref{SubSec:Vertex_operators2} can be used to compute fermion correlation functions. 
For simplicity, we restrict ourselves to equal-time correlation functions. 

Proposition~\ref{Proposition:Fermions_from_bosons} and the discussion directly thereafter suggest to use the regularized fermion operators $\psi^{(\dagger)}_{r}(x; \epsilon)$ in \eqref{Eq:Regularized_psi_explicit0} to define the correlation functions
\begin{equation}
\label{Eq:Fermion_correlation_functions0} 
	\langle \Omega, \psi_{r_1}^{q_1}(x_1;0^+) \ldots
		\psi_{r_N}^{q_N}(x_N;0^+) \Omega\rangle
	\define
	\lim_{\epsilon_1,\ldots,\epsilon_N\to 0^+}
		\langle \Omega,  \psi_{r_1}^{q_1}(x_1;\epsilon_1) \ldots
		\psi_{r_N}^{q_N}(x_N;\epsilon_N)\Omega \rangle
\end{equation} 
for $r_j, q_j = \pm$ and $x_j \in [-L/2, L/2]$, where we use the following convenient notation: 
\begin{equation}
\label{Eq:Psi_pm_operators}
	\psi^{+}_{r}(x;\epsilon) \define \psi^{\dagger}_{r}(x; \epsilon),
	\qquad
	\psi^{-}_{r}(x;\epsilon) \define \psi\pdag_{r}(x; \epsilon).
\end{equation}
As will be seen, with this definition, one automatically obtains prescriptions of how to interpret certain singular functions as distributions; see Proposition~\ref{Proposition:CF0}.
Moreover, it follows from \eqref{Eq:Fermions_from_bosons} and \eqref{Eq:Fermions_from_bosons_conjugate} that $\psi^q_r(x;\epsilon)=(1/\sqrt{L}) \! \bosonnoord{\Gamma(e^{irqf_{x,\epsilon}})}$ with $f_{x,\epsilon}(y)$ in \eqref{Eq:feps}.

To compute the fermion correlation functions we use the following two special cases of \eqref{Eq:Gamma_pm_Gamma_pm} and \eqref{Eq:OmGammaOm}, respectively:
\begin{equation} 
	\psi^q_r(x;\epsilon)\psi^{q'}_r(x';\epsilon')
	= \left( \frac{ i\exp \left(\frac{\pi}{2L}(\epsilon+\epsilon') \right) }
		{ 2\sin \left( \frac{\pi}{L}[r(x-x') + \tfrac{i}{2}(\epsilon+\epsilon')] \right) }
		\right)^{-qq'}
		\! \bosonnoord{ \psi^q_r(x;\epsilon) \psi^{q'}_r(x';\epsilon') }
\end{equation} 
(this follows using \eqref{Eq:Log_Taylor_expansion}) and 
\begin{equation} 
\label{Eq:OmRRROm}
	\langle \Omega, \! \bosonnoord{ \psi_{r_1}^{q_1}(x_1;0^+) \ldots
		\psi_{r_N}^{q_N}(x_N;0^+) } \! \Omega\rangle
	= L^{-N/2}\langle\Omega, R_{r_1}^{q_1r_1} \ldots R_{r_N}^{q_Nr_N} \Omega\rangle.
\end{equation}
The r.h.s.\ in \eqref{Eq:OmRRROm} is a sign factor which is either $0$, $1$, or $-1$, and we note that it is zero unless both $\sum_{n=1}^N q_n \delta_{r_n,+}$ and $\sum_{n=1}^N q_n \delta_{r_n,-}$ are zero (this follows from the identities in Corollary~\ref{Corollary:Rr}). 
In particular, 
\begin{align}  
	\langle\Omega, R_{r_1}^{q_1r_1}R_{r_2}^{q_2r_2} \Omega\rangle
		& = \delta_{r_1,r_2}\delta_{q_1,-q_2}, \\
	\langle\Omega,R_{r_1}^{q_1r_1}R_{r_2}^{q_2r_2}R_{r_3}^{q_3r_3}R_{r_4}^{q_4r_4}\Omega\rangle
		& = \begin{aligned}[t]
			\delta_{r_1,r_2} \delta_{r_3,r_4}\delta_{q_1,-q_2}\delta_{q_3,-q_4}
			& - \delta_{r_1,r_3} \delta_{r_2,r_4}\delta_{q_1,-q_3}\delta_{q_2,-q_4} \\
			& + \delta_{r_1,r_4} \delta_{r_2,r_3}\delta_{q_1,-q_4}\delta_{q_2,-q_3}.
		\end{aligned}
\end{align}
Using Corollary~\ref{Corollary:Workhorse} we get the following result:

\proposition{}{
\label{Proposition:CF0}
The fermion correlation functions in \eqref{Eq:Fermion_correlation_functions0}\textnormal{--}\eqref{Eq:Psi_pm_operators} are equal to
\begin{equation} 
\label{Eq:free_fermion_CF} 
	\langle\Omega, R_{r_1}^{q_1r_1} \ldots R_{r_N}^{q_Nr_N} \Omega\rangle
	\prod_{1\leq n<m\leq N} \left( \frac{ i }{ 2L\sin \left( \tfrac{\pi}{L}[r(x_n-x_m) + i 0^+] \right) } \right)^{-q_n q_m \delta_{r_n,r_m}}. 
\end{equation}
}

In the proof of this, and also in Section~\ref{Sec:Fermion-phonon_model}, we need the following: 

\lemma{}{
\label{Lemma:qqrr}
Let $r_n, q_n = \pm$, $n = 1, \ldots, N$, be such that $\langle\Omega,R_{r_1}^{q_1r_1} \ldots R_{r_N}^{q_Nr_N}\Omega\rangle$ is non-zero for some $N \in \mathbb{N}$.
Then
\begin{equation} 
	\sum_{1\leq n<m\leq N} q_n q_m \delta_{r_n,r_m} = - N/2,
	\qquad
	\sum_{1\leq n<m\leq N} q_n q_m \delta_{r_n,-r_m} = 0. 
\end{equation} 
} 

\begin{proof}
As mentioned above, the condition implies that $\sum_m q_m \delta_{r_m,r} = 0$ for $r=\pm$, and thus $\sum_{n<m} q_n q_m \delta_{r_n,r_m}$ is equal to
\begin{equation} 
	\frac{1}{2} \sum_r \sum_{n} q_n \delta_{r_n,r}
		\left( \sum_{m\neq n} q_m\delta_{r_m,r} \right)
	= \frac{1}{2} \sum_{n,r} q_n \delta_{r_n,r} \left( - q_n \delta_{r_n,r} \right)
	= -\frac{N}2.
\end{equation} 
Similarly, $\sum_{n<m} q_n q_m \delta_{r_n,-r_m} = (1/2)  \sum_{n,r} q_n \delta_{r_n,r} \left(-q_n \delta_{r_n,-r} \right)=0$. 
\end{proof} 

\begin{proof}[Proof of Proposition~\ref{Proposition:CF0}]
We are left to show that
\begin{equation} 
\label{Eq:ABC}
	\left( 1/\sqrt{L} \right)^{N}
	= \prod_{1\leq n<m\leq N} L^{q_n q_m \delta_{r_n,r_m}}
	\qquad
	\text{if} \;\, \langle\Omega,R_{r_1}^{q_1r_1} \ldots R_{r_N}^{q_Nr_N} \Omega\rangle \neq 0, 
\end{equation} 
but this is true according to Lemma~\ref{Lemma:qqrr}. 
\end{proof} 

Note that, due to \eqref{Eq:ABC}, the thermodynamic limit $L \to \infty$ of the correlation functions in \eqref{Eq:free_fermion_CF} exists and is equal to
\begin{equation} 
\label{Eq:free_fermion_CF_Linfty} 
	\langle\Omega, R_{r_1}^{q_1r_1} \ldots R_{r_N}^{q_Nr_N} \Omega\rangle
	\left( \frac{1}{2\pi} \right)^{N/2}
	\prod_{1\leq n<m\leq N}
	\left( \frac{i}{r(x_n-x_m) + i 0^+} \right)^{-q_n q_m \delta_{r_n,r_m}}. 
\end{equation}

\remark{}{
A careful reader might wonder how the result in Proposition~\ref{Proposition:CF0} can be consistent with Wick's theorem, which applies to the present model and states that any $N$-point correlation function can be expressed as a linear combination of products of two-point correlation functions.
It can be shown that one only needs to check the cases $N=2M$ with $(r_n,q_n) = (r,+)$ and $(r_{M+n},q_{M+n}) = (r,-)$ for $1 \leq n\leq M$.
The result in Proposition~\ref{Proposition:CF0} is then equivalent to the one obtained by Wick's theorem if and only if 
\begin{equation} 
\label{Eq:CauchyIdentity}
	\frac{\prod_{1\leq n<m\leq M}\sin(U_n-U_m)\sin(V_m-V_n) }{\prod_{n,m=1}^M \sin(U_n-V_m)}
	= \det_{1\leq n,m\leq M} \left( \frac1{\sin(U_n-V_m)} \right)
\end{equation} 
for $U_n=\pi x_n/L$ and $V_n=\pi x_{2M+1-n}/L$.
The latter holds true due to a well-known variant of \emph{Cauchy's determinant identity}.
}

\subsection{Proof of Proposition \ref{Proposition:Vertex_operators}}
\label{SubSec:Proof_of_vertex_operator_proposition}

The identities in Proposition \ref{Proposition:Vertex_operators} can easily be derived using the following special cases of the Baker-Campbell-Hausdorff formula:\footnote{See e.g.\ Appendix~C in \cite{DelftSchoeller:1998} for proofs of these identities valid for \emph{bounded} operators $A$ and $B$.}
\begin{equation} 
\label{Eq:BCH1} 
	\left. \begin{aligned} 
		e^A e^B & = e^c e^B e^A, \\
		e^A e^B & = e^{c/2} e^{A+B},  \\
		[A,e^B] & = c e^B
	\end{aligned}
	\;\, \right\}
	\text{ if } \;\, [A,B] = c I, \;\, c \in \mathbb{C},
\end{equation} 
together with properties of the fermion densities and the Klein factors summarized in Proposition~\ref{Proposition:Bosons_from_fermions} and Corollary~\ref{Corollary:Rr}; see e.g.\ \cite{DelftSchoeller:1998}.
However, this would not constitute a proof: the $J_{r}^{\pm}(\alpha)$ are unbounded operators, but the identities in \eqref{Eq:BCH1} can only be used in such a way for bounded operators.
Moreover, it is not obvious how the exponentials of the unbounded operators $J_{r}^{\pm}(\alpha)$ should be defined (or that they can be defined at all).
To prove these identities we give a precise meaning to the vertex operators in \eqref{Eq:Bosonnoord} as quadratic forms; see Appendix~\ref{Appendix:Quadratic_forms}.
We first concentrate on vertex operators with trivial zero-mode parts, i.e.\ $\alpha(0) = w = 0$.

Our strategy of proof is as follows.
We start by showing that the vertex operators  
\begin{equation} 
\label{Eq:SVdef} 
	\bosonnoord{ e^{i J_{r}^{+}(\alpha) + iJ_{r}^{-}(\alpha)} }
		= e^{i J_{r}^{+}(\alpha)}e^{i J_{r}^{-}(\alpha)}
\end{equation} 
are well-defined as quadratic forms on $\mathcal{D}_F$ and that the identities
\begin{subequations}
\label{Eqs:SV_equations}
\begin{align}
	\bigl( \bosonnoord{ e^{i J_{r}^{+}(\alpha) + i J_{r}^{-}(\alpha)} } \bigr)^{\dagger}
	& = \bosonnoord{ e^{-i J_{r}^{+}(\alpha) - i J_{r}^{-}(\alpha)} }, \label{Eq:SV0} \\
		\bosonnoord{ e^{i J^{+}(\alpha_1)+iJ^{-}(\alpha_1)} } \!\! 
		\bosonnoord{ e^{i J_{r}^{+}(\alpha_2)+iJ_{r}^{-}(\alpha_2)} }
	& = e^{ - c_r(\alpha_1,\alpha_2)}
		\! \bosonnoord{ e^{i J_{r}^{+}(\alpha_1+\alpha_2)
			+ iJ_{r}^{-}(\alpha_1+\alpha_2)} }, \label{Eq:SV1} \\
	\Bigl[ \hat{J}_{r}(p), \! \bosonnoord{ e^{i J_{r}^{+}(\alpha) + iJ_{r}^{-}(\alpha)} } \Bigr]
	& = i rp \alpha(p)
		\! \bosonnoord{ e^{i J_{r}^{+}(\alpha) + iJ_{r}^{-}(\alpha)} } \label{Eq:SV2}
\end{align}
\end{subequations}
hold true for such quadratic forms; see Lemma~\ref{Lemma:SV1}.
Note that, formally, the definition of $J^{\pm}_{r}(\alpha)$ in \eqref{Eq:J_operators} and the commutation relation in \eqref{Eq:Jr_commutation_relation} imply
\begin{subequations}
\begin{align}
	\Bigl[ J_{r}^{-}(\alpha_1), J_{r}^{+}(\alpha_2) \Bigr]
	& = c_{r}(\alpha_1, \alpha_2), \label{Eq:JmJp} \\
	\Bigl[ \hat{J}_{r}(p), J_{r}^{+}(\alpha) + J_{r}^{-}(\alpha) \Bigr]
	& = irp \alpha(p),
\end{align}
\end{subequations}
with $c_{r}(\alpha_1, \alpha_2)$ in \eqref{Eq:cr}.
The identities in \eqref{Eq:SV1} and \eqref{Eq:SV2} thus follow from the first and third identities in \eqref{Eq:BCH1}. 
However, we need to justify the use of these identities for quadratic forms. 

In the next step, we show that, for $\alpha = ( \alpha(p) )_{p}$ satisfying the conditions in \eqref{Eq:G}, the quadratic form 
\begin{equation} 
\label{Eq:USVdef} 
	e^{i J_{r}^{+}(\alpha)+ iJ_{r}^{-}(\alpha)}
	\define
	e^{-c_{r}(\alpha,\alpha)/2} \! \bosonnoord{ e^{i J_{r}^{+}(\alpha) + iJ_{r}^{-}(\alpha)} }
\end{equation} 
can be identified with a unitary operator on $\mathcal{F}$; see Lemma~\ref{Lemma:V2}.
(We emphasize that we regard \eqref{Eq:USVdef} as a definition motivated by the second identity in \eqref{Eq:BCH1}.)
Since there is no risk of confusion, we can thereafter use the same symbol $e^{i J_{r}^{+}(\alpha)+ iJ_{r}^{-}(\alpha)}$ for this unitary operator.
With these results in place, the proof of Proposition \ref{Proposition:Vertex_operators} is straightforward. 

\remark{}{
It is interesting to note that our proof of \eqref{Eq:SV0}--\eqref{Eq:SV2} does not require the conditions in \eqref{Eq:G}: the quadratic form in \eqref{Eq:SVdef} is well-defined for \emph{any} $\alpha \define ( \alpha(p) )_{p}$, i.e.\ no reality or decay condition is needed, and \eqref{Eq:SV0} (with $\alpha$ on the r.h.s.\ replaced by $\overline{\alpha}$) and \eqref{Eq:SV2} are always true.
Moreover, \eqref{Eq:SV1} holds true provided that 
\begin{equation} 
	\sum_{p > 0} p \abs{\alpha_1(-rp)} \abs{\alpha_2(rp)} < \infty. 
\end{equation}   
While we do not make use of these more general results, they are important in conformal field theory; see e.g.\ \cite{Kac:1998}.
}

A simple way to make the vertex operators in \eqref{Eq:SVdef} precise is as quadratic forms, with the exponentials defined as Taylor series:  
 
\lemma{}{
\label{Lemma:SV1} 
Let $\alpha \define ( \alpha(p) )_{p}$ satisfy the conditions in \eqref{Eq:G}.
Then the vertex operator
\begin{equation} 
\label{Eq:SV3} 
	\bosonnoord{ e^{i J_{r}^{+}(\alpha)+iJ_{r}^{-}(\alpha)} }
	\define
	\sum_{n,m = 0}^{\infty} \frac{[i J_r^+(\alpha)]^{n}}{n!} \frac{[i J_r^-(\alpha)]^{m}}{m!}
\end{equation} 
is well-defined as a quadratic form on $\mathcal{D}_F$.
Moreover, for $\alpha_1$ and $\alpha_2$ similarly defined, \eqref{Eq:SV0}\textnormal{--}\eqref{Eq:SV2} hold true as identities of quadratic forms on $\mathcal{D}_F$.
} 
  
\begin{proof}
Note that, for boson states $\eta^B_{\mathbf{m}}$ and  $\eta^B_{\mathbf{m}'}$ as in \eqref{Eq:Boson_states}, 
\begin{equation} 
	\langle \eta^B_{\mathbf{m}\pp},
		(J_r^+(\alpha))^n (J_r^-(\alpha))^m \eta^B_{\mathbf{m}'}\rangle
	= \langle (J_r^-(\bar{\alpha}))^n \eta^B_{\mathbf{m}\pp},
		(J_r^-(\alpha))^m \eta^B_{\mathbf{m}'}\rangle
\end{equation} 
is well-defined for all $(n,m) \in \mathbb{N}_0^2$ and that it is non-zero only for finitely many such pairs (this follows from \eqref{Eq:Jr_commutation_relation} and \eqref{Eq:J_operators_annihilation_creation}).
This together with Proposition~\ref{Proposition:BF} implies that \eqref{Eq:SV3} is well-defined as a quadratic form on $\mathcal{D}_F$.

As a consequence of this and $(J_{r}^{\pm}(\alpha))^{\dagger} = J_{r}^{\mp}(\overline{\alpha})$ (this follows from \eqref{Eq:Densities_property_1} and \eqref{Eq:J_operators}) we conclude that \eqref{Eq:SV0} holds true. 
 
To prove \eqref{Eq:SV1}, we note that, by a similar argument as above, 
\begin{equation}
\label{Eq:SV4}  
	e^{i J_{r}^{-}(\alpha)}\eta \in  \mathcal{D}_F
	\qquad
	\forall \eta \in \mathcal{D}_F. 
\end{equation} 
Indeed, for all boson states $\eta^B_{\mathbf{m}}$ as in \eqref{Eq:Boson_states} and $n \in \mathbb{N}$, $(J_r^-(\alpha))^n\eta^B_{\mathbf{m}}$ is a linear combination of such boson states, and it is non-zero only for finitely many $n$ (this follows again from \eqref{Eq:Jr_commutation_relation} and \eqref{Eq:J_operators_annihilation_creation}).
Thus, by Proposition~\ref{Proposition:BF}, $e^{i J_{r}^{-}(\alpha)}\eta  = \sum_{n=0}^\infty (1/n!) (i J_r^-(\alpha))^n \eta$ is well-defined and in $\mathcal{D}_F$.
Therefore, to prove the validity of \eqref{Eq:SV1} in the sense of quadratic forms, we only need to show that
\begin{equation} 
\label{Eq:SV10} 
	\langle \eta, e^A e^B \eta' \rangle = e^c \langle \eta, e^B e^A \eta' \rangle
	\qquad
	\forall \eta,\eta' \in \mathcal{D}_F
\end{equation} 
for $A \define iJ_{r}^{-}(\alpha_1)$, $B \define iJ_{r}^{+}(\alpha_2)$, and $c \define -c_{r}(\alpha_1, \alpha_2)$.
We note that \eqref{Eq:JmJp} is equivalent to $[A,B] = c$, which is obviously well-defined as an identity of quadratic forms on $\mathcal{D}_F$.
Moreover, this implies that\footnote{The identity in \eqref{Eq:AnBm} can be proved by induction over $m \in \mathbb{N}_{0}$ for fixed $n \in \mathbb{N}_{0}$ using standard relations for the binomial coefficients.} 
\begin{equation} 
\label{Eq:AnBm} 
	A^n B^m = \sum_{l=0}^{\min(n,m)} l! \binom{m}{l} \binom{n}{l} c^l B^{m-l}A^{n-l}
\end{equation} 
holds true as an identity of quadratic forms on $\mathcal{D}_F$.
Recalling that $\langle\eta, B^m A^n\eta'\rangle$ is non-zero only for finitely many $(n,m) \in \mathbb{N}_n^2$, one can expand the exponentials on both sides in \eqref{Eq:SV10} as Taylor series and use \eqref{Eq:AnBm} to compute the l.h.s. 
One thus finds that both sides are well-defined and identical.\footnote{The key identity is 
\begin{equation*} 
	\sum_{n=0}^N \sum_{m=0}^M \frac{1}{n!m!} \langle\eta, A^n B^m \eta'\rangle
	= \sum_{n=0}^N \sum_{m=0}^M \sum_{l=0}^{\min(N,M)}
		\frac{1}{n!m!l!} c^l \langle\eta, B^m A^n \eta'\rangle
	\qquad
	(N,M \in \mathbb{N}).
\end{equation*} 
} 

The proof of \eqref{Eq:SV2} is similar, but simpler, and thus omitted. 
\end{proof} 

\lemma{}{
\label{Lemma:SV2}  
Given $\alpha \define ( \alpha(p) )_p$ satisfying the conditions in \eqref{Eq:G}, the quadratic form $e^{i J_{r}^{+}(\alpha)+ iJ_{r}^{-}(\alpha)}$ defined in \eqref{Eq:USVdef} can be identified with a unitary operator on $\mathcal{F}$.
} 
  
\begin{proof} 
Using the definition and \eqref{Eq:SV0}--\eqref{Eq:SV1} we compute
\begin{equation} 
	e^{ iJ_{r}^{+}(\alpha) + iJ_{r}^{-}(\alpha) }
		\left( e^{ iJ_{r}^{+}(\alpha) + iJ_{r}^{-}(\alpha)} \right)^{\dagger}
	= e^{-c_r(\alpha,\alpha)} \! 
		\bosonnoord{ e^{ iJ_{r}^{+}(\alpha) + iJ_{r}^{-}(\alpha)} } \!\! 
		\bosonnoord{ e^{-iJ_{r}^{+}(\alpha) - iJ_{r}^{-}(\alpha)} }
	= I, 
\end{equation}  
and similarly
\begin{equation} 
	\left( e^{i J_r^+(\alpha)+ iJ_r^-(\alpha)} \right)^{\dagger}
		e^{i J_r^+(\alpha)+ iJ_r^-(\alpha)} =I. 
\end{equation} 
This and a general result for quadratic forms (see Lemma~\ref{Lemma:Unitarity}) conclude the proof. 
\end{proof} 
  
\begin{proof}[Proof of Proposition~\ref{Proposition:Vertex_operators}]
It follows from the self-adjointness of $Q_{r}$ and the unitarity of $R_{r}$ that $e^{iJ^{0}_{r}(\alpha)/2} R_{r}^{w} e^{iJ^{0}_{r}(\alpha)/2}$ is a unitary operator.
Equations \eqref{Eq:Jr_commutation_relation} and \eqref{Eq:JrRr} imply that this operator commutes with $e^{iJ_{r}^{+}(\alpha) + iJ_{r}^{-}(\alpha)}$.
From this and Lemma~\ref{Lemma:SV2} it follows that $\Gamma_{r}(e^{if})$ in \eqref{Eq:Gamma} defines a unitary operator on $\mathcal{F}$, and \eqref{Eq:USVdef} then implies that \eqref{Eq:Bosonnoord1} holds true.
The identity in \eqref{Eq:Gamma_dag} thus follows directly from \eqref{Eq:G}--\eqref{Eq:Gamma} and \eqref{Eq:SV0}.

The remaining identities can be proven using Proposition~\ref{Proposition:Bosons_from_fermions}, Proposition~\ref{Proposition:Klein_factors}, and Corollary~\ref{Corollary:Rr}.
As an example we prove the last three identities.
Equation \eqref{Eq:ExpQRq} implies that
\begin{multline}
\label{Eq:Product_rule_for_unitary_R_tilde-operators}
	e^{iJ^{0}_{r}(\alpha_1)/2} R_{r}^{w_1} e^{iJ^{0}_{r}(\alpha_1)/2}
		e^{iJ^{0}_{r}(\alpha_2)/2} R_{r}^{w_2} e^{iJ^{0}_{r}(\alpha_2)/2} \\
	= e^{i\pi r \left( w_2\alpha_1(0) - w_1\alpha_2(0) \right)/L}
		e^{iJ^{0}_{r}(\alpha_1+\alpha_2)/2} R_{r}^{w_1+w_2} e^{iJ^{0}_{r}(\alpha_1+\alpha_2)/2},
\end{multline}
which together with \eqref{Eq:SV1} yields the identity in \eqref{Eq:Gamma_pm_Gamma_pm}.
The identity in \eqref{Eq:Gamma_pm_Gamma_mp} follows from \eqref{Eq:RqRq}.
Lastly, we note that the l.h.s.\ of \eqref{Eq:OmGammaOm} is equal to
\begin{multline}
	\langle e^{iJ_{-}^{+}(\alpha')^{\dagger}} e^{iJ_{+}^{+}(\alpha)^{\dagger}} \Omega,
		\left( e^{iJ^{0}_{+}(\alpha)/2} R_{+}^{w} e^{iJ^{0}_{+}(\alpha)/2}
		e^{iJ^{0}_{-}(\alpha')/2} R_{-}^{w'} e^{iJ^{0}_{-}(\alpha')/2} \right)
		e^{iJ_{+}^{-}(\alpha)} e^{iJ_{-}^{-}(\alpha')} \Omega \rangle \\
	= \langle \Omega, e^{iJ^{0}_{+}(\alpha)/2} e^{iJ^{0}_{-}(\alpha')/2} R_{+}^{w}
		R_{-}^{w'} e^{iJ^{0}_{+}(\alpha)/2} e^{iJ^{0}_{-}(\alpha')/2} \Omega \rangle
	= \langle \Omega, R_{+}^{w} R_{-}^{w'} \Omega \rangle
\end{multline}
(where we have used \eqref{Eq:J_operators_annihilation_creation} and $Q_{r}\Omega = 0$), and the r.h.s.\ then follows from \eqref{Eq:Klein_factor_VEV}.
\end{proof}

\subsection{Proof of Proposition \ref{Proposition:Fermions_from_bosons}}
\label{SubSec:Proof_of_reconstruction_proposition}
We divide the proof into three steps.
In the first step we show that the r.h.s.\ in \eqref{Eq:Fermions_from_bosons} is a well-defined operator, which we denote by $\hat{V}_{r}(k)$, satisfying a list of four identities (see Lemma~\ref{Lemma:V1}) which are identical with identities satisfied by the fermion field operators $\hat{\psi}_{r}(k)$.
In the second step we show that all matrix elements of $\hat{V}_{r}(k)$ with respect to fermion states in \eqref{Eq:Fermion_states} can be computed using only the identities in this list (see Lemma~\ref{Lemma:V2}).
We thus conclude in the third step that both sides in \eqref{Eq:Fermions_from_bosons} agree as quadratic forms on $\mathcal{D}_{F}$, and we argue that this is only possible if they agree as operators on $\mathcal{D}_{F}$.

\lemma{}{
\label{Lemma:V1}
Denote  the r.h.s.\ in \eqref{Eq:Fermions_from_bosons} by $\hat{V}_{r}(k)$.
Then $\hat{V}_{r}(k)$ is a well-defined operator mapping $\mathcal{D}_F$ to $\mathcal{D}_F$ and satisfying the following identities:
\begin{equation}
\label{Eq:V1}
	\hat{V}\pdag_{r}(rk) \Omega = \hat{V}^{\dagger}_{r}(-rk) \Omega = 0
	\qquad
	\forall k  > 0,
\end{equation}
with $\hat{V}^{\dagger}_{r}(k) \define \hat{V}\pdag_{r}(k)^{\dagger}$, together with
\begin{gather} 
	\label{Eq:V2}
	\left[\hat{J}\pdag_{r\pp}(p), \hat{V}^{\dagger}_{r'}(k) \right]
		= \delta_{r,r'} \hat{V}^{\dagger}_{r\pp}(k-p), \\
	\label{Eq:V3}
	R_{\pm} \hat{V}_{\pm}(k) = \hat{V}_{\pm}(k + \tfrac{2\pi}{L}) R_{\pm},
	\qquad
	R_{\pm} \hat{V}_{\mp}(k) = -\hat{V}_{\mp}(k) R_{\pm}
\end{gather}
on $\mathcal{D}_F$, and
\begin{equation} 
\label{Eq:V4}
	\langle R^{\dagger}_{+} \Omega, \hat{V}\pdag_{+}(-\tfrac{\pi}{L}) \Omega \rangle
		= \sqrt{\frac{L}{2\pi}},
	\qquad
	\langle R\pdag_{-} \Omega, \hat{V}\pdag_{-}(\tfrac{\pi}{L}) \Omega \rangle
		= \sqrt{\frac{L}{2\pi}}.
\end{equation} 
}
\noindent (The proof is given below.)

It is important to note that the identities in Lemma~\ref{Lemma:V1} are also satisfied by the fermion field operators $\hat{\psi}_r(k)$: \eqref{Eq:V1}, \eqref{Eq:V2}, and \eqref{Eq:V3} correspond to \eqref{Eq:Fermion_field_action_on_Omega}, \eqref{Eq:Jr_psi_dagger_commutation_relation}, and \eqref{Eq:Klein_factor_defining_condition_1}, respectively, and the identities corresponding to \eqref{Eq:V4} are
\begin{equation} 
	\langle R^{\dagger}_{+} \Omega, \hat{\psi}_{+}(-\tfrac{\pi}{L}) \Omega \rangle
		= \sqrt{\frac{L}{2\pi}},
	\qquad
	\langle R\pdag_{-} \Omega, \hat{\psi}_{-}(\tfrac{\pi}{L}) \Omega \rangle
		=  \sqrt{\frac{L}{2\pi}}
\end{equation}  
(the latter follow from \eqref{Eq:Fermion_field_anticommutation_relations}--\eqref{Eq:Fermion_field_action_on_Omega} and \eqref{Eq:Klein_factor_defining_condition_1}--\eqref{Eq:Klein_factor_defining_condition_2}).

\begin{proof}[Proof of Lemma~\ref{Lemma:V1}]
The vertex operators $\bosonnoord{\Gamma_r(e^{-irf_{x,\epsilon}})}$ can be written in the form
\begin{align}
	\label{Eq:V_operator_factorized}
	\bosonnoord{\Gamma_r(e^{-irf_{x,\epsilon}})}
	& = V_{r}^{+}(x; \epsilon) 
		e^{i\pi rx Q_{r}/L} R_{r}^{-r} e^{i\pi rx Q_{r}/L}
		V_{r}^{-}(x; \epsilon), \\
	\label{Eq:V_operators}
	V_{r}^{\pm}(x; \epsilon)
	& \define
	\prod_{p > 0} \exp \left( \mp \frac{2\pi}{Lp} \hat{J}_{r}(\mp rp)
		e^{\mp irpx - \epsilon p/2} \right).
\end{align}
We note that $V_{r}^{+}(x; \epsilon)$ and $V_{r}^{-}(x; \epsilon)$ correspond to creation and annihilation operators, respectively, since
\begin{equation}
\label{Eq:V_operators_annihilation_creation}
	 V_{r}^{-}(x; \epsilon) \Omega = V_{r}^{+}(x; \epsilon)^{\dagger} \Omega = \Omega,
\end{equation}
as suggested by \eqref{Eq:J_operators_annihilation_creation} (this is proved below).
The exponentials in \eqref{Eq:V_operators} are well-defined as Taylor series: we can represent $V_{r}^{\pm}(x; \epsilon)$ as sums over integer vectors $\boldsymbol{\nu}_{\pm} \define ( \nu_{\pm}(p) )_{p > 0}$ with elements $\nu_{\pm}(p) \in \mathbb{N}_{0}$,
\begin{equation}
\label{Eq:V_operators_as_series}
	V_{r}^{\pm}(x; \epsilon)
		= \sum_{\boldsymbol{\nu}_{\pm}}
			U_{r}(\boldsymbol{\nu}_{\pm})
			e^{\mp irx P(\boldsymbol{\nu}_{\pm}) - \epsilon P(\boldsymbol{\nu}_{\pm})/2},
\end{equation}
using the shorthand notation
\begin{equation}
\label{Eq:PU_shorthand_notation}
	U_{r}(\boldsymbol{\nu}_{\pm})
	\define
	\prod_{p>0} \frac{1}{\nu_{\pm}(p)!}
			\left( \mp \frac{2\pi}{Lp} \hat{J}_{r}(\mp rp) \right)^{\nu_{\pm}(p)},
	\qquad
	P({\boldsymbol{\nu}_{\pm}})
	\define
	\sum_{p>0} p\nu_{\pm}(p).
\end{equation}
We note that \eqref{Eq:V_operators_as_series} is well-defined as a quadratic form on $\mathcal{D}_{F}$: for fixed finite linear combinations $\eta$ and $\eta'$ of boson states in \eqref{Eq:Boson_states}, $\langle \eta, \! \bosonnoord{\Gamma_r(e^{-irf_{x,\epsilon}})}\eta' \rangle$ can be computed using \eqref{Eq:V_operator_factorized} and \eqref{Eq:V_operators_as_series}--\eqref{Eq:PU_shorthand_notation}, and one finds that only a finite number of the terms in the resulting $\boldsymbol{\nu}_{\pm}$-sums are non-zero, which implies that the same holds true for all $\eta, \eta' \in \mathcal{D}_{F}$ (cf.\ Proposition \ref{Proposition:BF}).
Moreover, using \eqref{Eq:V_operators_as_series}--\eqref{Eq:PU_shorthand_notation}, the identity in \eqref{Eq:V_operators_annihilation_creation} follows from Proposition~\ref{Proposition:Bosons_from_fermions}.

We now fix a boson state
\begin{equation}
\label{Eq:Fixed_boson_state}
	\eta_{\mathbf{m}}^{B}
		= B_{\mathbf{m}} R_{+}^{q_{+}} R_{-}^{-q_{-}} \Omega,
	\qquad
	B_{\mathbf{m}} \define \prod_{p \neq 0} \frac{b^{\dagger}(p)^{m(p)}}{\sqrt{m(p)!}},
\end{equation}
and compute $\bosonnoord{\Gamma_r(e^{-irf_{x,\epsilon}})}\! \eta_{\mathbf{m}}^{B}$ with the help of \eqref{Eq:V_operator_factorized}--\eqref{Eq:PU_shorthand_notation}.
It follows from \eqref{Eq:ExpQRq} that
\begin{equation}
	e^{ia Q_{r}} R_{+}^{q_{+}} R_{-}^{-q_{-}} \Omega
		= e^{iaq_{r}} R_{+}^{q_{+}} R_{-}^{-q_{-}} \Omega
	\qquad
	\forall a \in \mathbb{R},
\end{equation}
which together with the above identities yields
\begin{equation}
\label{Eq:V_operator_action_on_eta}
	\bosonnoord{\Gamma_r(e^{-irf_{x,\epsilon}})} \! \eta_{\mathbf{m}}^{B}
	= \sum_{\boldsymbol{\nu}_{+}, \boldsymbol{\nu}_{-}}
		U_{r}(\boldsymbol{\nu}_{+}) U_{r}(\boldsymbol{\nu}_{-})
		e^{ix W_{r}(q_{r}, \left\{ \boldsymbol{\nu}_{\pm} \right\})}
			e^{-\epsilon \left[ P(\boldsymbol{\nu}_{+}) + P(\boldsymbol{\nu}_{-}) \right]/2}
			R_{r}^{-r} \eta_{\mathbf{m}}^{B}
\end{equation}
with the shorthand notation
\begin{equation}
	W_{r}(q_{r}, \left\{ \boldsymbol{\nu}_{\pm} \right\})
		\define \frac{2\pi}{L} \left( q_{r} - \frac{r}{2} \right)
			- r \left[ P(\boldsymbol{\nu}_{+}) - P(\boldsymbol{\nu}_{-}) \right].
\end{equation}
The integral over $[-L/2, L/2]$ in \eqref{Eq:Fermions_from_bosons} and the $\boldsymbol{\nu}_{\pm}$-sums can now be exchanged (this is allowed due to the $\epsilon$-regularization).
Hence, for fixed $k$, the integral
\begin{equation}
	\frac{1}{\sqrt{2\pi L}} \int_{-L/2}^{L/2} e^{-ikx}
		e^{ix W_{r}(q_{r}, \left\{ \boldsymbol{\nu}_{\pm} \right\})}
	= \sqrt{\frac{L}{2\pi}}
		\delta_{k, W_{r}(q_{r}, \left\{ \boldsymbol{\nu}_{\pm} \right\})}
\end{equation}
imposes the following constraint on the $\boldsymbol{\nu}_{\pm}$-sums in \eqref{Eq:V_operator_action_on_eta}:
\begin{equation}
\label{Eq:nu_sum_constraint_1}
	k = \frac{2\pi}{L} \left( q_{r} - \frac{r}{2} \right)
		- r \left[ P(\boldsymbol{\nu}_{+}) - P(\boldsymbol{\nu}_{-}) \right].
\end{equation}
For any boson state, there is at most a finite number of $\boldsymbol{\nu}_{-}$-terms which give a non-zero contribution to \eqref{Eq:V_operator_action_on_eta}
(this follows from \eqref{Eq:Densities_property_2} and \eqref{Eq:Boson_states}).
Hence, there exists some $P < \infty$ such that $P(\boldsymbol{\nu}_{-}) \leq P$ for all $\boldsymbol{\nu}_{-}$-terms.
It follows that \eqref{Eq:nu_sum_constraint_1} restricts the $\boldsymbol{\nu}_{+}$-sum in \eqref{Eq:V_operator_action_on_eta} to a finite number, which means that all non-zero contributions are given by pairs $(\boldsymbol{\nu}_{+}, \boldsymbol{\nu}_{-})$ belonging to some finite set determined by the constraint in \eqref{Eq:nu_sum_constraint_1}.

In conclusion, both $\boldsymbol{\nu}_{\pm}$-sums in \eqref{Eq:V_operator_action_on_eta} are finite, and the limit $\epsilon \to 0^{+}$ in \eqref{Eq:Fermions_from_bosons} can therefore be taken without problem.
The result is
\begin{equation}
\label{Eq:V_hat_operator_explicit}
	\hat{V}_{r}(k) \eta_{\mathbf{m}}^{B}
		= \sqrt{\frac{L}{2\pi}}
			\underset{\boldsymbol{\nu}_{+}, \boldsymbol{\nu}_{-}}{\widetilde{\sum}}
			U_{r}(\boldsymbol{\nu}_{+}) U_{r}(\boldsymbol{\nu}_{-})
			R_{r}^{-r} \eta_{\mathbf{m}}^{B},
\end{equation}
where the tilde above the summation symbol indicates that the $\boldsymbol{\nu}_{\pm}$-sums are restricted by \eqref{Eq:nu_sum_constraint_1}; the corresponding formula for $\hat{V}^{\dagger}_{r}(k) \eta_{\mathbf{m}}^{B}$ can be derived in a similar manner (we omit the details).
It follows that $\hat{V}_{r}(k)$ is a well-defined operator mapping $\mathcal{D}_{F}$ to $\mathcal{D}_{F}$, and we are only left to show that this operator satisfies the stated identities.

Let $\eta^{B}_{\mathbf{m}} = \Omega$ in \eqref{Eq:Fixed_boson_state} and use \eqref{Eq:V_hat_operator_explicit} to compute $\hat{V}_{r}(k)\Omega$.
All non-zero $\boldsymbol{\nu}_{-}$-vectors will give a zero contribution to the $\boldsymbol{\nu}_{-}$-sum in \eqref{Eq:V_hat_operator_explicit}, and the constraint in \eqref{Eq:nu_sum_constraint_1} therefore simplifies to
\begin{equation}
\label{Eq:nu_sum_constraint_1_special_case}
	k = -r\frac{\pi}{L} - r P({\boldsymbol{\nu}_{+}}).
\end{equation}
If $rk > 0$, then no $\boldsymbol{\nu}_{+}$-vector can satisfy \eqref{Eq:nu_sum_constraint_1_special_case}, which implies that
\begin{equation}
\label{Eq:V_hat_property_1}
	\hat{V}_{r}(k) \Omega = 0
	\qquad
	\forall k : rk > 0.
\end{equation}
This proves the first identity in \eqref{Eq:V1}; the second follows from a similar argument applied to $\hat{V}^{\dagger}_{r}(k) \Omega$.
If $k = -r\pi/L$, only the solution $P({\boldsymbol{\nu}_{+}}) = 0$ is allowed, in which case \eqref{Eq:V_hat_operator_explicit} yields
\begin{equation}
\label{Eq:V_hat_property_2}
	R_{+}^{-1} \Omega = \sqrt{\frac{2\pi}{L}} \hat{V}_{+}(-\tfrac{\pi}{L}) \Omega,
	\qquad
	R_{-} \Omega = \sqrt{\frac{2\pi}{L}} \hat{V}_{-}(\tfrac{\pi}{L}) \Omega.
\end{equation}
This and the unitarity of $R_{+}$ imply \eqref{Eq:V4}.

Equation \eqref{Eq:JrGamma} implies that
\begin{equation}
	\left[ \hat{J}_{r\pp}(p), \! \bosonnoord{\Gamma_{r'}(e^{-ir'f_{x,\epsilon}})} \right]
		= - \delta_{r,r'} e^{-ipx - \epsilon \abs{p}/2}
		\! \bosonnoord{\Gamma_{r'}(e^{-ir'f_{x,\epsilon}})},
\end{equation}
which together with \eqref{Eq:Fermions_from_bosons} yields
\begin{equation}
\label{Eq:V_hat_property_5}
	\left[ \hat{J}_{r\pp}(p), \hat{V}_{r'}(k) \right]
		= - \delta_{r,r'} \hat{V}_{r'}(k+p).
\end{equation}
By taking the adjoint of this identity and using \eqref{Eq:Densities_property_1} we obtain \eqref{Eq:V2}.

From \eqref{Eq:RrGamma1} and \eqref{Eq:RrGamma2} we obtain
\begin{subequations}
\begin{align}
	R_{\pm} \! \bosonnoord{\Gamma_{\pm}(e^{\mp if_{x,\epsilon}})}
		& = e^{-2\pi ix/L} \! \bosonnoord{\Gamma_{\pm}(e^{\mp if_{x,\epsilon}})} \! R_{\pm}, \\
	R_{\pm} \! \bosonnoord{\Gamma_{\mp}(e^{\pm if_{x,\epsilon}})}
		& = - \! \bosonnoord{\Gamma_{\mp}(e^{\pm if_{x,\epsilon}})} \! R_{\pm},
\end{align}
\end{subequations}
and this yields \eqref{Eq:V3} by recalling that $\hat{V}_{r}(k)$ is defined as the r.h.s.\ of \eqref{Eq:Fermions_from_bosons}.
\end{proof}

\lemma{}{
\label{Lemma:V2}
For all $\eta, \eta' \in \mathcal{D}_{F}$, the matrix element $\langle \eta, \hat{V}_{r}(k) \eta' \rangle$ is uniquely determined by the identities in \eqref{Eq:V1}\textnormal{--}\eqref{Eq:V4}.
}

\begin{proof}
Due to Proposition~\ref{Proposition:BF} we can restrict ourselves to the boson states in \eqref{Eq:Boson_states}.
We thus set $\eta = \eta_{\mathbf{m}}^{B}$ and $\eta' = \eta_{\mathbf{m}'}^{B}$ which we write as (cf.\ \eqref{Eq:Fixed_boson_state})
\begin{equation}
	\eta_{\mathbf{m}}^{B} = B_{(\mathbf{m},{+})} B_{(\mathbf{m},{-})}
		R_{+}^{q_{+}} R_{-}^{q_{-}} \Omega,
	\qquad
	B_{(\mathbf{m},{r})} \define \prod_{p > 0} \frac{b^{\dagger}(rp)^{m(rp)}}{\sqrt{m(rp)!}},
\end{equation}
and similarly for $\eta_{\mathbf{m}'}^{B}$.
The matrix elements $\langle \eta^{B}_{\mathbf{m}}, \hat{V}_{r}(k) \eta^{B}_{\mathbf{m'}} \rangle$ can thus be written
\begin{equation}
\label{Eq:V_operator_matrix_elements_step_1}
	\langle B_{(\mathbf{m},{+})} B_{(\mathbf{m},{-})}
		R_{+}^{q\pp_{+}} R_{-}^{q\pp_{-}} \Omega,
		\hat{V}_{r}(k) B_{(\mathbf{m}',{+})} B_{(\mathbf{m}',{-})}
			R_{+}^{q'_{+}} R_{-}^{q'_{-}} \Omega \rangle,
\end{equation}
which is non-zero only if $q\pp_{-r} = q'_{-r}$ and $q\pp_{r} = q'_{r} - r$ (this follows from \eqref{Eq:RqRq}, \eqref{Eq:Klein_factor_VEV}, and \eqref{Eq:V_hat_operator_explicit} since $R_{\pm}$ and $b^{\dagger}(p)$ commute).
Furthermore, from \eqref{Eq:Harmonic_oscillator_operators} and \eqref{Eq:V_hat_property_5} it follows that
\begin{equation}
\label{Eq:V_hat_commutation_rel_with_b_dagger}
	\hat{V}_{r}(k) b^{\dagger}(r'p) = b^{\dagger}(r'p) \hat{V}_{r}(k)
		+ ir \delta_{r,r'} \sqrt{\frac{2\pi}{L \abs{p}}} \hat{V}_{r}(k-rp),
\end{equation}
which implies that the matrix elements in \eqref{Eq:V_operator_matrix_elements_step_1} are non-zero only if $B_{(\mathbf{m},{-r})} = B_{(\mathbf{m}',{-r})}$.

It follows that we can restrict ourselves to states $\eta_{\mathbf{m}}^{B}$ and $\eta_{\mathbf{m}'}^{B}$ with $(q\pp_{r}, q\pp_{-r}) = (q-r,0)$ and $(q'_{r}, q'_{-r}) = (q,0)$ for some $q \in \mathbb{Z}$ and $B_{(\mathbf{m},{-r})} = B_{(\mathbf{m}',{-r})} = I$: all non-zero matrix elements in \eqref{Eq:V_operator_matrix_elements_step_1} can be written in the form
\begin{equation}
\label{Eq:V_operator_matrix_elements_step_2}
	\langle B_{(\mathbf{m},{r})} R_{r}^{q-r} \Omega,
		\hat{V}_{r}(k)
		B_{(\mathbf{m}',{r})} R_{r}^{q} \Omega \rangle.
\end{equation}
By repeated use of \eqref{Eq:V_hat_commutation_rel_with_b_dagger} we can write \eqref{Eq:V_operator_matrix_elements_step_2} as a linear combination of matrix elements
\begin{equation}
\label{Eq:V_operator_matrix_elements_step_3}
	\langle B_{(\mathbf{m},{r})} R_{r}^{q-r} \Omega,
		B_{(\mathbf{m}'-\mathbf{m}''_{r},{r})}
		\hat{V}_{r} \left( k - r P(\mathbf{m}''_{r}) \right) R_{r}^{q} \Omega \rangle, 
\end{equation}
where $P(\mathbf{m}''_{r})$ is defined as in \eqref{Eq:PU_shorthand_notation} and $\mathbf{m}''_{r}$ is any integer vector satisfying $0 \leq m''_{r}(p) \leq m'(rp)$ for $p>0$ and $m''_{r}(p) = 0$ for $p\leq 0$.
Using \eqref{Eq:Harmonic_oscillator_operator_properties}, we see that \eqref{Eq:V_operator_matrix_elements_step_3} is non-zero only if $B_{(\mathbf{m},{r})} = B_{(\mathbf{m}' - \mathbf{m}''_{r},r)}$.
Hence, only the matrix elements
\begin{equation}
\label{Eq:V_operator_matrix_elements_step_4}
	\langle R_{r}^{q-r} \Omega,
		\hat{V}_{r} \left( k - r P(\mathbf{m}''_{r}) \right) R_{r}^{q} \Omega \rangle
\end{equation}
need to be considered.
These can be computed using \eqref{Eq:V3} as follows:
\begin{equation}
\label{Eq:V_operator_matrix_elements_step_5}
	\begin{aligned}
		\langle R_{r}^{q-r} \Omega,
			\hat{V}_{r} \left( k - r P(\mathbf{m}''_{r}) \right) R_{r}^{q} \Omega \rangle
		& = \langle R_{r}^{-r} \Omega, R_{r}^{-q} \hat{V}_{r}
			\left( k - r P(\mathbf{m}''_{r}) \right) R_{r}^{q} \Omega \rangle \\
		& = \langle R_{r}^{-r} \Omega, \hat{V}_{r}(k') \Omega \rangle,
	\end{aligned}
\end{equation}
where $k' \define k - q \tfrac{2\pi}{L} - r P(\mathbf{m}''_{r})$.
We are left to compute the matrix element on the r.h.s.\ of \eqref{Eq:V_operator_matrix_elements_step_5}.
For $rk' > 0$, this matrix element is zero due to \eqref{Eq:V_hat_property_1}.
For $rk' = -\pi/L$, it is equal to $\sqrt{L/(2\pi)}$ due to \eqref{Eq:V4} and the unitarity of $R_{r}$.
For $rk' < - \pi/L$, we can rearrange the r.h.s.\ of \eqref{Eq:V_operator_matrix_elements_step_5} as
\begin{equation}
	\begin{aligned}
		\langle R_{r}^{-r} \Omega, \hat{V}_{r}(k') \Omega \rangle
			& = \langle \Omega, R_{r}^{r} \hat{V}_{r}(k') R_{r}^{-r} R_{r}^{r} \Omega \rangle\\
			& = \langle \Omega, \hat{V}_{r}(k' + r \tfrac{2\pi}{L}) R_{r}^{r} \Omega \rangle
			= \langle \hat{V}^{\dagger}_{r}(k' + r \tfrac{2\pi}{L}) \Omega,
				R_{r}^{r} \Omega \rangle
	\end{aligned}
\end{equation}
(the second identity follows from \eqref{Eq:V3}), and it thus follows from \eqref{Eq:V1} that this is zero since $r(k' + r \tfrac{2\pi}{L}) < 0$.
\end{proof}

\begin{proof}[Proof of Proposition~\ref{Proposition:Fermions_from_bosons}]
Lemmas \ref{Lemma:V1} and \ref{Lemma:V2} imply that
\begin{equation}
	\langle \eta, \hat{V}_{r}(k) \eta' \rangle
		= \langle \eta, \hat{\psi}_{r}(k) \eta' \rangle
	\qquad
	\forall \eta, \eta' \in \mathcal{D}_{F},
\end{equation}
and by general results for quadratic forms it follows that \eqref{Eq:Fermions_from_bosons} holds true as an operator identity on $\mathcal{D}_{F}$ (see Appendix~\ref{Appendix:Quadratic_forms}).
This concludes the proof of Proposition \ref{Proposition:Fermions_from_bosons}.
\end{proof}


\section{Fermion-phonon model}
\label{Sec:Fermion-phonon_model}
In this section we use the mathematical results developed in Sections~\ref{Sec:Mathematics_of_bosonization_I}~and~\ref{Sec:Mathematics_of_bosonization_II} to study the fermion-phonon model described in the introduction.
As will be shown, this model is exactly solvable in the sense that not only the eigenstates and eigenvalues of the Hamiltonian, but also all correlation functions, can be computed by analytical means; 
we will, however, be content with discussing fermion correlation functions.

\subsection{Definition and summary of results} 
We give a precise definition of the fermion-phonon model in Fourier space using fermion field operators $\hat{\psi}_r^{(\dagger)}(k)$ ($r = \pm$) and phonon field operators $\hat{\Phi}_P(p)$ and $\hat{\Pi}_P(p)$.
We recall that $L>0$ is an infrared cutoff which constrain the fermion and boson momenta such that $k \in (2\pi/L)(\mathbb{Z}+1/2)$ and $p \in (2\pi/L)\mathbb{Z}$, respectively.
The interactions are given by the fermion-fermion and the fermion-phonon coupling constants $\gF$ and $\gP$, respectively, and they are regularized using an ultraviolet cutoff $a > 0$; see \eqref{Eq:gFgP} below.
The other two model parameters $v_F > 0$ and $v_P > 0$ with $v_P < v_F$ correspond to the Fermi velocity and the phonon velocity, respectively.
As will be shown, the four model parameters must satisfy the conditions in \eqref{Eq:Restrictions}.
For reasons explained in Remark~\ref{Remark:Continuous_spectrum}, we also introduce a less important infrared cutoff $\omega_{0} > 0$. 
We remind the reader that some of our notation is summarized in Appendix~\ref{Appendix:Notation}.

\definition{}{
The fermion-phonon model is defined by the Hamiltonian 
\begin{multline}
\label{Eq:Fermion-phonon_Hamiltonian}
	H \define
	v_F\sum_{r} \sum_{k} \frac{2\pi}{L}
			rk \noord{ \hat{\psi}^{\dagger}_{r}(k)\hat{\psi}\pdag_{r}(k) }
		+ \sum_{p} \frac{\pi}{L}
			\! \noord{ \Big( \hat{\Pi}^{\dagger}_{P}(p) \hat{\Pi}\pdag_{P}(p)
			+ \omega^{0}_{P}(p)^2 \hat{\Phi}^{\dagger}_{P}(p) \hat{\Phi}\pdag_{P}(p) \Big) }\\
	+ \sum_{p} \frac{2\pi}{L} \left( \frac{\hat{\gF}_a(p)}{2\pi}
		\hat{J}^{\dagger}_{+}(p) \hat{J}\pdag_{-}(p)
	+ \sum_{r} 
	\frac{\hat{\gP}_a(p)}{\sqrt{2\pi}} ip
		\hat{J}^{\dagger}_{r}(p) \hat{\Phi}_{P}(p) \right)
\end{multline}
with the fermion densities $\hat{J}_r(p) \define \sum_{k}(2\pi/L) \! \noord{ \hat{\psi}^{\dagger}_r(k) \hat{\psi}\pdag_r(k+p) } $, where the colons indicate normal ordering defined in \eqref{Eq:noord1}, the free-phonon dispersion relations
\begin{equation}
\label{Eq:Free-phonon_disp_relations}
	\omega^{0}_{P}(p)
	\define
	\begin{cases}
		\omega_{0}		& \text{if} \;\, p = 0, \\
		v_{P} \abs{p}	& \text{if} \;\, p \neq 0,
	\end{cases}
\end{equation}
and the regularized interaction potentials
\begin{equation}
\label{Eq:gFgP} 
	\hat{\gF}_a(p)
	\define
	\gF \Theta( {\pi}/{a} - \abs{p} ),
	\qquad
	\hat{\gP}_a(p)
	\define
	\gP \Theta( {\pi}/{a} - \abs{p} ) . 
\end{equation}
The fermion and phonon field operators satisfy the following CAR and CCR:
\begin{subequations}
\label{Eqs:Fermion_phonon_CAR_CCR}
\begin{alignat}{3}
	\Big\{ \hat{\psi}\pdag_{r\pp}(k), \hat{\psi}^{\dagger}_{r'}(k') \Big\}
			& = \frac{L}{2\pi} \delta_{r,r'} \delta_{k,k'},
		\qquad
		& \Big\{ \hat{\psi}\pdag_{r\pp}(k), \hat{\psi}\pdag_{r'}(k') \Big\}
			& = &\; \Big\{ \hat{\psi}^{\dagger}_{r\pp}(k), \hat{\psi}^{\dagger}_{r'}(k') \Big\}
			& = 0, \\
	\Big[ \hat{\Phi}\pdag_{P}(p), \hat{\Pi}^{\dagger}_{P}(p') \Big]
			& = i \frac{L}{2\pi} \delta_{p,p'},
		\qquad
		& \Big[ \hat{\Phi}\pdag_{P}(p), \hat{\Phi}\pdag_{P}(p') \Big]
			& = &\; \Big[ \hat{\Pi}\pdag_{P}(p), \hat{\Pi}\pdag_{P}(p') \Big]
			& = 0, \\
	&& \Big[ \hat{\Phi}\pdag_{P}(p), \hat{\psi}^{(\dagger)}_{r}(k) \Big]
			& = &\; \Big[ \hat{\Pi}\pdag_{P}(p), \hat{\psi}^{(\dagger)}_{r}(k) \Big]
			& = 0,
			\vphantom{\frac{L}{2\pi}}
\end{alignat}
\end{subequations}
where $\hat{\psi}_{r}^{\dagger}(k) \define \hat{\psi}_{r}(k)^{\dagger}_{\vphantom{r}}$, 
\begin{equation}
\label{Eq:Fermion_phonon_CAR_CCR_help_eq}
	\hat{\Phi}^{\dagger}_{P}(p)
	\define \hat{\Phi}\pdag_{P}(p)^{\dagger}_{\vphantom{P}}
		= \hat{\Phi}\pdag_{P}(-p),
	\qquad
	\hat{\Pi}^{\dagger}_{P}(p)
	\define \hat{\Pi}\pdag_{P}(p)^{\dagger}_{\vphantom{P}}
		= \hat{\Pi}\pdag_{P}(-p), 
\end{equation}
and these are represented on a Hilbert space $\mathcal{F}$ which contains a normalized state $\Omega$, i.e.\ $\langle\Omega, \Omega\rangle = 1$ with the inner product $\langle \cdot,\cdot \rangle$ in $\mathcal{F}$, such that 
\begin{equation} 
\label{Eq:Vacuum_conditions}
\begin{aligned} 
	\hat{\psi}_{r}^{\phantom{\dagger}}(rk) \Omega	
		= \hat{\psi}_{r}^{\dagger}(-rk) \Omega
	& = 0
	&& \forall k > 0, \\ 
	\left( i \hat{\Pi}\pdag_{P}(p) + \omega^{0}_{P}(p)\hat{\Phi}\pdag_{P}(p) \right) \Omega
	& = 0
	&& \forall p. 
\end{aligned} 
\end{equation}
We refer to $\mathcal{F}$ as the \emph{fermion-phonon Fock space}.
}

It is important to note that we use the regularized interactions in \eqref{Eq:gFgP} only as a specific example to simplify our notation.
As will become clear, all computations in the regularized model can be done without specifying the explicit form of $\hat{\gF}_a(p)$ and $\hat{\gP}_a(p)$.
In fact, our results for the regularized model hold true for a large class of such potentials: the only restriction is that these potentials vanish sufficiently fast as $\abs{p} \to \infty$ and that $\abs{ \hat{\gF}_{a}(p) } < \abs{\gF}$ for all $p$ and similarly for $\hat{\gP}_{a}(p)$. 
Moreover, we expect that our results are universal in the sense that they are (essentially) independent of the regularization in the limit $a \to 0^{+}$ as long as $\lim_{a \to 0^+}\hat{\gF}_a(p) = \gF$ (independent of $p$) and similarly for $\hat{\gP}_a(p)$. 
However, a proof of this is beyond the scope of the present paper. 

We begin with a summary of the results obtained in this section.
The first set of results concerns the regularized model with $\omega_0 > 0$, $a > 0$, and $L < \infty$:
\begin{itemize}

\item We show that the fermion-phonon Hilbert space $\mathcal{F}$ is fully specified by the conditions in \eqref{Eqs:Fermion_phonon_CAR_CCR}--\eqref{Eq:Vacuum_conditions}; see Section~\ref{SubSec:Fermion-phonon-Fock-space}.

\item We prove that the Hamiltonian $H$ in \eqref{Eq:Fermion-phonon_Hamiltonian}--\eqref{Eq:gFgP} defines a self-adjoint operator on $\mathcal{F}$ with pure point spectrum and non-degenerate ground state, and we construct all its eigenstates and eigenvalues; see Corollary~\ref{Corollary:Hdiag}.

\item We develop tools by which any fermion correlation function can be computed; see \eqref{Eq:Fermion_correlation_functions1}--\eqref{Eq:Zaepsilon}.

\end{itemize} 
The second set of results concerns the limit where $\omega_0 \to 0^+$, $a \to 0^+$, and $L \to\infty$, which leads to simpler formulas.
The physical motivation is that this limiting description is adequate for describing properties at intermediate length scales which are much larger than the interaction range $a$ and much smaller than the system size $L$.
The price to pay for this simplification are divergencies which appear in formal treatments.
However, these divergencies can be eliminated by additive and multiplicative renormalizations: 
\begin{itemize}

\item We construct a renormalized fermion-phonon model in the continuum limit $a \to 0^+$; see Proposition~\ref{Proposition:Fermion_phonon_QFT}. 
Our construction makes manifest that the Hilbert spaces for $a>0$ and $a=0$ are \emph{not} unitarily equivalent.

\item We obtain explicit formulas for all fermion correlation functions in the limit $a\to 0^+$ and $L \to \infty$; see Corollary~\ref{Corollary:GreenN1}.

\end{itemize}

\subsection{Fermion-phonon Fock space}
\label{SubSec:Fermion-phonon-Fock-space}
To construct the Hilbert space $\mathcal{F}$ from the vacuum $\Omega$ we follow the method in Section~\ref{Sec:Mathematics_of_bosonization_I}.
In the present case, we need both fermion and boson creation operators $c^{\dagger}_{r}(k)$ and $b^{\dagger}_{P}(p)$, respectively; these are operators on $\mathcal{F}$ satisfying the conditions
\begin{equation}
\begin{aligned}
	\left\{ c\pdag_{r\pp}(k), c^{\dagger}_{r'}(k') \right\} & = \delta_{r,r'} \delta_{k,k'},
		& \left\{ c\pdag_{r\pp}(k), c\pdag_{r'}(k') \right\} & = 0,
		& c\pdag_{r}(k) \Omega & = 0, \\
	\left[ b\pdag_{P}(p), b^{\dagger}_{P}(p') \right] & = \delta_{p,p'},
		& \Big[ b\pdag_{P}(p), b\pdag_{P}(p') \Big] & = 0,
		& b\pdag_{P}(p)\Omega & = 0, \\
	\left[ c\pdag_{r}(k), b^{(\dagger)}_{P}(p) \right] & = 0,
		& \left[ c^{\dagger}_{r}(k), b^{(\dagger)}_{P}(p) \right] & = 0,
\end{aligned}
\end{equation}
with $c\pdag_{r}(k) \define c^{\dagger}_{r}(k)^{\dagger}_{\vphantom{r}}$ and $b\pdag_{P}(p) \define b^{\dagger}_{P}(p)^{\dagger}_{\vphantom{P}}$. 
Using these we can construct the following states in $\mathcal{F}$:  
\begin{equation}
\label{Eq:Fermion_Phonon_states}
	\eta_{\mathbf{n}}^{FP}
	\define 
	\Biggl( \harpoon{\prod}_{r, k} c^{\dagger}_{r}(k)^{n_{r,F}(k)} \Biggr)\left( \prod_{p} \frac{b^{\dagger}_{P}(p)^{m_P(p)}}{\sqrt{m_P(p)!}} \right) \Omega,
	\quad 
	n_{r,F}(k) \in \{ 0, 1 \}, \;\, m_P(p) \in \mathbb{N}_{0}, 
\end{equation}
where only finitely many of the quantum numbers $n_{r,F}(k)$ and $m_P(p)$ are non-zero, and where $\mathbf{n}$ is short for the infinite vector $((n_{r,F}(k))_{r,k}, (m_P(p))_p)$.\footnote{Some of the notation used here is explained after \eqref{Eq:Fermion_states}.} 
The relations above imply that the states in \eqref{Eq:Fermion_Phonon_states} are orthonormal, and they also determine the action of the creation and annihilation operators on these states. 
Let $\mathcal{D}$ denote the vector space of all finite linear combinations of the states in \eqref{Eq:Fermion_Phonon_states}.
This set forms a pre-Hilbert space, and $\mathcal{F}$ is obtained from $\mathcal{D}$ by norm-completion \cite{ReedSimon:1972}.

\remark{}{The Hilbert space $\mathcal{F}$ thus constructed can (of course) be identified with the tensor product $\mathcal{F}_{F} \otimes \mathcal{F}_{P}$ of a fermion Fock space and a phonon Fock space.
The Fock space $\mathcal{F}_F$ is identical with the fermion Fock space constructed in Section~\ref{SubSec:H0}, and the states in \eqref{Eq:Fermion_Phonon_states} correspond to a tensor product of the fermion states in \eqref{Eq:Fermion_states} with phonon states. 
It is thus clear that all bosonization results developed in Sections~\ref{Sec:Mathematics_of_bosonization_I} and \ref{Sec:Mathematics_of_bosonization_II} can be used for the fermion-phonon model. 
} 

The representation of the field operators used in our definition of the fermion-phonon model is such that the free fermion-phonon Hamiltonian $H_0$, which is obtained from the Hamiltonian in \eqref{Eq:Fermion-phonon_Hamiltonian} by setting  $\hat{\gF}_a(p) = \hat{\gP}_a(p)=0$, is a (positive-definite) self-adjoint operator with eigenstates $\eta_{\mathbf{n}}^{FP}$ in \eqref{Eq:Fermion_Phonon_states}. 
To see this we note that the fermion field operators can be constructed as in \eqref{Eq:psi_representation}, and for the phonon operators the corresponding formulas are
\begin{equation}
\label{Eqs:Phonon_Pi_Phi_representation}
\begin{aligned}
	\hat{\Pi}\pdag_{P}(p)
		& \define -i \sqrt{ \frac{L}{2\pi} } \sqrt{ \frac{\omega^{0}_{P}(p)}{2} }
			\left( b\pdag_{P}(p) - b^{\dagger}_{P}(-p) \right),\\
	\hat{\Phi}\pdag_{P}(p)
		& \define \sqrt{ \frac{L}{2\pi} } \frac{1}{\sqrt{2\omega^{0}_{P}(p)}}
			\left( b\pdag_{P}(p) + b^{\dagger}_{P}(-p) \right).
\end{aligned}
\end{equation}
Indeed, one can verify that this implies that all the conditions in \eqref{Eqs:Fermion_phonon_CAR_CCR}--\eqref{Eq:Fermion_phonon_CAR_CCR_help_eq} are fulfilled and that the free fermion-phonon Hamiltonian can be written as
\begin{equation}
\label{Eq:Free_fermion-phonon_Hamiltonian}
	H_0
	= \sum_{r,k} v_F \abs{k} c_r^{\dagger}(k) c_r\pdag(k)
		+ \sum_p \omega^{0}_{P}(p) b_P^{\dagger}(p) b_P\pdag(p). 
\end{equation} 
Moreover, as the latter formula makes manifest, the states in \eqref{Eq:Fermion_Phonon_states} are exact eigenstates of $H_0$ with corresponding  eigenvalues $\sum_{r,k}v_F \abs{k} n_{r,F}(k) + \sum_p \omega^{0}_{P}(p) m_P(p) \geq 0$. 
It follows that $H_0$ can be identified with a self-adjoint operator on $\mathcal{F}$ (cf.\ Lemma~\ref{Lemma:Self_adjointness}) and that it is positive definite. 

\remark{}{
\label{Remark:Continuous_spectrum}
The significance of the cutoff $\omega_0$ can be explained as follows.
The contribution from the zero mode $\hat{\Pi}_P(0)^2 + \omega_0^2\hat{\Phi}_P(0)^2$ to the free fermion-phonon Hamiltonian is a harmonic oscillator Hamiltonian with a discrete spectrum for $\omega_0 > 0$ but a free-particle Hamiltonian with a continuous spectrum for $\omega_0 = 0$. 
Although the latter is easy to treat, with $\omega_0>0$ we avoid a separate discussion of the phonon zero modes which would be of no consequence for our results.
}

\subsection{Bosonized Hamiltonian}
\label{Sec:exact_solution_I}
To bosonize the fermion-phonon Hamiltonian in \eqref{Eq:Fermion-phonon_Hamiltonian}--\eqref{Eq:gFgP}, we use the boson operators arising from the fermion densities and Kronig's identity; see \eqref{Eq:Harmonic_oscillator_operators}, \eqref{Eq:Phi_Pi_operators}, and \eqref{Eq:Kronig2} (note that the latter refers to the free-fermion Hamiltonian $H_0$ in \eqref{Eq:Free_fermion_Hamiltonian}). 
If we use boson flavor indices $X, X' = F, P$ and define
\begin{equation}
\label{Eq:Fermion_Phi_Pi_operators}
\begin{aligned}
	\hat{\Pi}_{F}(p)
		& \define - \sqrt{\frac{v_{F}}{2}} \left( \hat{J}_{+}(p) - \hat{J}_{-}(p) \right)
	= -i \sqrt{ \frac{L}{2\pi} } \sqrt{\frac{\omega^{0}_{F}(p)}{2}}
			\left( b\pdag_{F}(p) - b^{\dagger}_{F}(-p) \right), \\
	\hat{\Phi}_{F}(p)
		& \define \frac{1}{ip\sqrt{2v_{F}}} \left( \hat{J}_{+}(p) + \hat{J}_{-}(p) \right)
	= \sqrt{ \frac{L}{2\pi} } \frac{1}{\sqrt{2\omega^{0}_{F}(p)}}
			\left( b\pdag_{F}(p) + b^{\dagger}_{F}(-p) \right)
\end{aligned} 
\end{equation}
for $p \neq 0$, then the bosonized Hamiltonian can be written  
\begin{multline}
\label{Eq:Bosonized_Hamiltonian_precise}
	H = \frac{\pi}{L} \left( v_{F} \left( Q_{+}^2 + Q_{-}^2 \right)  
			+ \frac{\hat{\gF}_a(0)}{\pi} Q_{+}Q_{-}
			+ \noord{ \left(
				\hat{\Pi}_{P}(0)^2 + \omega_{0}^{2} \hat{\Phi}_{P}(0)^2
			\right) } \right) \\
	\begin{aligned}
		& + \sum_{X} \sum_{p \neq 0} \frac{\pi}{L}
			\! \noord{ \Bigl( \hat{\Pi}^{\dagger}_{X}(p) \hat{\Pi}\pdag_{X}(p)
			+ \omega^{0}_{X}(p)^2 \hat{\Phi}^{\dagger}_{X}(p) \hat{\Phi}\pdag_{X}(p) \Bigr) } \\
		& + \sum_{p \neq 0} \frac{\pi}{L}
			\frac{\hat{\gF}_a(p)}{2\pi v_{F}}
				\Bigl( - \hat{\Pi}^{\dagger}_{F}(p) \hat{\Pi}\pdag_{F}(p)
			+ \omega^{0}_{F}(p)^2 \hat{\Phi}^{\dagger}_{F}(p) \hat{\Phi}\pdag_{F}(p) \Bigr) \\
		& + \sum_{p \neq 0} \frac{\pi}{L}
			\frac{\hat{\gP}_a(p)}{v_{P}\sqrt{\pi v_{F}}}
				\omega^{0}_{F}(p) \omega^{0}_{P}(p)
				\Bigl( \hat{\Phi}^{\dagger}_{F}(p) \hat{\Phi}\pdag_{P}(p)
					+ \hat{\Phi}^{\dagger}_{P}(p) \hat{\Phi}\pdag_{F}(p) \Bigr), 
	\end{aligned}
\end{multline}
where we have collected all zero modes in the first line (recall that $Q_{\pm} \define \hat{J}_{\pm}(0)$) and introduced
\begin{equation} 
	\omega^{0}_F(p)
	\define
	v_F \abs{p}. 
\end{equation}
We note that, for both fermions ($X=F$) and phonons ($X=P$), the operators $\hat{\Phi}_{X}(p)$ and $\hat{\Pi}_{X}(p)$ satisfy the same CCR:
\begin{equation}
\label{Eq:Fermion-phonon_CCR}
	\Big[ \hat{\Phi}\pdag_{X\pp}(p), \hat{\Pi}^{\dagger}_{X'}(p') \Big]
		= i \frac{L}{2\pi} \delta_{X,X'} \delta_{p,p'},
	\quad
	\Big[ \hat{\Phi}\pdag_{X\pp}(p), \hat{\Phi}\pdag_{X'}(p') \Big]
		= \Big[ \hat{\Pi}\pdag_{X\pp}(p), \hat{\Pi}\pdag_{X'}(p') \Big] = 0
\end{equation}
(this follows from \eqref{Eq:CCR_in_Fourier_space} and \eqref{Eqs:Fermion_phonon_CAR_CCR}),
and similarly for the boson creation and annihilation operators: 
\begin{equation}
\label{Eq:CCR_b}
	\left[ b\pdag_{X\pp}(p), b^{\dagger}_{X'}(p') \right] = \delta_{X,X'}\delta_{p,p'},
	\qquad
	\Big[ b\pdag_{X\pp}(p), b\pdag_{X'}(p') \Big] = 0,
	\qquad
	b\pdag_{X\pp}(p)\Omega = 0. 
\end{equation}
As a consequence of the results in Sections~\ref{SubSec:BF_correspondence} and \ref{SubSec:Fermion-phonon-Fock-space}, the states
\begin{equation} 
\label{Eq:etaBP}
	\eta_{\mathbf{m}}^{BP}
	\define
	\left( \prod_{X} \prod_{p\neq 0} \frac{b^{\dagger}_{X}(p)^{m_X(p)}}{\sqrt{m_F(p)!}} \right)
	\frac{b^{\dagger}_{P}(0)^{m_P(0)}}{\sqrt{m_P(0)!}} R_+^{q_+}R_-^{-q_-} \Omega,
	\qquad
	m_{X}(p) \in \mathbb{N}_0, \;\, q_{\pm} \in \mathbb{Z}, 
\end{equation}
where only finitely many of the $m_F(p)$ and $m_P(p)$ are non-zero, and where $\mathbf{m}$ is short for the infinite vector $((m_F(p))_{p\neq 0}, (m_P(p))_{p}, q_+, q_-)$, provide a complete orthonormal basis for the fermion-phonon Fock space.
These states are also exact eigenstates of the free fermion-phonon Hamiltonian $H_0$ (cf.\ \eqref{Eq:Free_fermion-phonon_Hamiltonian}) in its bosonized form.

\subsection{Exact solution}
\label{SubSec:Exact_solution}
We now turn to the exact solution of the regularized fermion-phonon model.
The Hamiltonian in \eqref{Eq:Bosonized_Hamiltonian_precise} describes two boson fields with \emph{linear} coupling terms. 
Such a Hamiltonian can always be diagonalized by a Bogoliubov transformation; see e.g.\ Appendix~C.1 in \cite{deWoulLangmann:2012}. 
The present case is particularly simple since the computations can be reduced to diagonalizing $2\times 2$ matrices (the details can be found in Section~\ref{Sec:Proof_of_diagonalization_proposition}). 

To state our results we find it convenient to introduce the following parameters:
\begin{equation}
\label{Eq:Rescaled_interaction_potentials_in_Fourier_space}
	\gamma_{1}
	\define
	\frac{\gF}{2\pi v_{F}},
	\qquad
	\gamma_{2}
	\define
	\frac{\gP}{v_{P}\sqrt{\pi v_{F}}},
\end{equation}
which are dimensionless, and
\begin{equation}
	W
	\define
	\sqrt{ \left( v_{F}^2 \left( 1 - \gamma_{1}^2 \right) - v_{P}^2 \right)^2
		+ 4 v_{F}^2 v_{P}^2 \gamma_{2}^2 \left( 1 - \gamma_{1} \right) },
\end{equation}
which has the dimension of velocity squared.
We note that the conditions in \eqref{Eq:Restrictions} are equivalent to $\gamma_1 < 1$ and $\gamma_{2}^2 < 1 + \gamma_{1}$, which guarantees that $W$, and all other parameters given below, are well-defined.
We also introduce the operators
\begin{equation} 
\label{Eq:JrX} 
	\hat{J}\pdag_{r,X}(p)
	\define
	\begin{cases}
		i \sqrt{\tfrac{L\abs{p}}{2\pi}} b_X(p) 
			& \text{if} \;\, r = +, \\
		i \sqrt{\tfrac{L\abs{p}}{2\pi}}b_X^{\dagger}(-p)
			& \text{if} \;\, r = -,
	\end{cases}
	\qquad
	\hat{J}^{\dagger}_{r,X}(p)
	\define
	\hat{J}\pdag_{r,X}(-p)
	\qquad
	(p > 0).
\end{equation} 
This notation is motivated by the following: $J_{r,F}(p) = J_{r}(p)$ (cf.\  \eqref{Eq:Harmonic_oscillator_operators}), and
\begin{gather} 
	\hat{J}_{r,X}(rp) \Omega=0
	\qquad
	\forall p>0, \\
	\left[ \hat{J}_{r,X}(p),\hat{J}_{r',X'}(p') \right]
		= r \delta_{r,r'} \delta_{X,X'} \frac{Lp}{2\pi}\delta_{p,-p'}.
		\label{Eq:JrXJrX}
\end{gather} 
The mathematical properties of the phonon operators $\hat{J}_{r,P}(p)$ are thus identical with those of the fermion densities $J_{r,F}(p)$. 
For this reason, they can also be used to construct phonon vertex operators $\bosonnoord{ \exp(i\sum_{p\neq 0} (2\pi/L)\alpha(p)J_{r,P}(-p)) }$ (cf.\ \eqref{Eq:Gamma} for $\alpha(0) = w = 0$), and the mathematical results obtained for the fermion vertex operators in Section~\ref{SubSec:Vertex_operators1} are equally true for such phonon vertex operators. 

We are now ready to formulate a key result towards the exact solution:

\propositionlist{}{
\label{Proposition:Diagonalization_of_fermion-phonon_model}
\
\begin{enumerate}[label=\textnormal{(\alph*)}, leftmargin=0pt, itemindent = 10pt, labelwidth=0pt, align=left, nosep]

\item
\label{Item:Diagonalization_of_fermion-phonon_model_A}
There exists a unitary operator $\mathcal{U}$ on $\mathcal{F}$ diagonalizing the Hamiltonian in \eqref{Eq:Bosonized_Hamiltonian_precise}:
\begin{multline}
\label{Eq:Diagonalized_Hamiltonian}
	\mathcal{U}^{\dagger} H \mathcal{U}
	= \mathcal{E}_{0}
		+ \frac{\pi v_F}{L} \left( Q_{+}^2 + Q_{-}^2 + 2\gamma_1 Q_{+}Q_{-} \right)
		+ \omega_0 b_P^{\dagger}(0)b_P\pdag(0) \\
	+ \sum_{X} \sum_{p \neq 0} \tilde{v}\pdag_{X}(p) \abs{p} b^{\dagger}_{X}(p) b\pdag_{X}(p)
\end{multline}
with
\begin{equation}
\label{Eq:Fermion_phonon_model_ground_state_energy}
	\mathcal{E}_{0}
	\define
	\frac{1}{2} \sum_{X} \sum_{0 < \abs{p} \leq \pi/a}
		\left( \tilde{v}_{X} - v_{X} \right) \abs{p},
\end{equation}	
\begin{equation}
\label{Eq:Fermion-phonon_renormalized_velocities_general}
	\tilde{v}_X(p)
	\define
	\begin{cases}
		\tilde{v}_{X}	& \text{if} \;\, \abs{p} \leq \pi/a, \\
		v_{X}			& \text{otherwise},
	\end{cases}
\end{equation}
where\footnote{Note that $\tilde{v}_{X} > 0$ if and only if $\gamma_1 < 1$ and $\gamma_{2}^2 < 1 + \gamma_{1}$.}
\begin{equation}
\label{Eq:Fermion-phonon_renormalized_velocities}
	\tilde{v}_{F}
	= \sqrt{ \frac{v_{F}^2 \left( 1 - \gamma_{1}^2 \right) + v_{P}^2 + W}{2} },
	\qquad
	\tilde{v}_{P}
	= \sqrt{ \frac{v_{F}^2 \left( 1 - \gamma_{1}^2 \right) + v_{P}^2 - W}{2} }. 
\end{equation}

\item
\label{Item:Diagonalization_of_fermion-phonon_model_B}
The unitary operator $\mathcal{U}$ commutes with the Klein factors $R_r$ and the fermion charge operators $Q_r$.
Moreover,
\begin{gather} 
	\mathcal{U}^{\dagger} \hat{J}_{r}(p)\mathcal{U}
		= \sum_{X} \rho_X (p) \hat{J}_{r,X}(p) - \sigma_X (p) \hat{J}_{-r,X}(p),
		\label{Eq:UJrU} \\
	\rho_{X}(p)
	\define
	\begin{cases}
		\rho_X 			& \text{if} \;\, 0 < \abs{p} \leq \pi/a, \\ 
		\delta_{X,F}		& \text{otherwise},
	\end{cases}
	\qquad
	\sigma_X (p)
	\define
	\begin{cases}
		\sigma_X			& \text{if} \;\, 0 < \abs{p} \leq \pi/a, \\ 
		0				& \text{otherwise}, 
	\end{cases} \label{Eq:rhoXp_sigmaXp}
\end{gather} 	
with the parameters 
\begin{equation}
\label{Eq:CS_matrices_explicit}
\begin{aligned}
	\rho_{F}
	& = \sqrt{\frac{v_{F}}{\tilde{v}_{F}}}
		\frac{ \gamma_{2} v_{P} \left( \tilde{v}_{F} + v_{F} (1 - \gamma_{1}) \right) }
		{ 2\sqrt{W} \sqrt{\tilde{v}_{F}^2 - v_{F}^2 \left( 1 -\gamma_{1}^2 \right)} },
	& \rho_{P}
	& = 	- \sqrt{\frac{v_{F}}{\tilde{v}_{P}}}
		\frac{ \gamma_{2} v_{P} \left( \tilde{v}_{P} + v_{F} (1 - \gamma_{1}) \right) }
		{ 2\sqrt{W} \sqrt{v_{F}^2 \left( 1 -\gamma_{1}^2 \right) - \tilde{v}_{P}^2} },\\
	\sigma_{F}
	& = \sqrt{\frac{v_{F}}{\tilde{v}_{F}}}
		\frac{ \gamma_{2} v_{P} \left( \tilde{v}_{F} - v_{F} (1 - \gamma_{1}) \right) }
		{ 2\sqrt{W} \sqrt{\tilde{v}_{F}^2 - v_{F}^2 \left( 1 -\gamma_{1}^2 \right)} },
	& \sigma_{P}
	& = 	- \sqrt{\frac{v_{F}}{\tilde{v}_{P}}}
		\frac{\gamma_{2} v_{P} \left( \tilde{v}_{P} - v_{F} (1 - \gamma_{1}) \right) }
		{ 2\sqrt{W} \sqrt{v_{F}^2 \left( 1 -\gamma_{1}^2 \right) - \tilde{v}_{P}^2} } ,
\end{aligned}
\end{equation}
which satisfy the identity
\begin{equation}
\label{Eq:Rho_sigma_identities}
	\rho_{F}^2 - \sigma_{F}^2
		+ \rho_{P}^2 - \sigma_{P}^2
	= 1.
\end{equation}

\end{enumerate}
}
\noindent (The proof is given in Section~\ref{Sec:Proof_of_diagonalization_proposition}.)

The formulas for $\tilde{v}_X$, $\rho_X$, and $\sigma_X$ in terms of the model parameters constitute the main computational result.
We note that the notation we use for the parameters in \eqref{Eq:CS_matrices_explicit} is inspired by \cite{CareyRuijsenaarsWright:1985}.
Moreover, the dispersion relations for the fermion-phonon model are
\begin{equation}
\label{Eq:Fermion-phonon_dispersion_relations}
	\omega_{X}(p)
	\define
	\tilde{v}_X(p) \abs{p},
\end{equation}
from which it follows that $\tilde{v}_F$ and $\tilde{v}_P$ have the physical interpretation of renormalized fermion and phonon velocities, respectively.
One can show, although it is not manifest, that our formulas simplify to
\begin{equation}
\label{Eq:Gamma2_limit}
	\begin{aligned}
		\tilde{v}_{F} & = v_{F} \sqrt{1 - \gamma_{1}^{2}},
		\qquad
		& \tilde{v}_{P} & = v_{P}, \\
		\rho_{F} & = \sqrt{ \frac{v_{F} + \tilde{v}_{F}}{2\tilde{v}_{F}} },
		\qquad
		\sigma_{F} = \pm\sqrt{ \frac{v_{F} - \tilde{v}_{F}}{2\tilde{v}_{F}} },
		\qquad
		& \rho_{P} & = \sigma_{P} = 0
	\end{aligned}
\end{equation}
in the limiting case $\gamma_2 = 0$ where the phonons decouple from the fermions; the upper sign for $\sigma_{F}$ in \eqref{Eq:Gamma2_limit} corresponds to $\lambda > 0$ and the lower to $\lambda < 0$.
These can be used to check that we recover Johnson's solution of the massless Thirring model \cite{Johnson:1961, Mastropietro:1993} (cf.\ the special case $\lambda_{\text{Klaiber}} = 0$ of Klaiber's solution as in \cite{CareyRuijsenaarsWright:1985}, which corresponds to the solution by Johnson \cite{Klaiber:LectTheoPhys:1968}).

\remark{}{
\label{Remark:Gamma2_limit}
The limit $\gamma_2 \to 0$ is tricky due to the possibility of incorrectly identifying fermions as phonons and vice versa (cf.\ \eqref{Eq:Fermion-phonon_renormalized_velocities}).
One way to avoid this is to restrict the coupling constants further, as compared to \eqref{Eq:Restrictions}, and require that $\left( \gF/(2\pi) \right)^2 < v_{F}^2 - v_{P}^2$,
which corresponds to $\gamma_1^2 < 1 - v_{P}^2/v_{F}^2$.
It then follows that $W \to v_{F}^2 \left( 1 - \gamma_1^2 \right) - v_{P}^2$ in the limit $\gamma_2 \to 0$, which gives the correct identification as in \eqref{Eq:Gamma2_limit}.
}

We now turn to the exact eigenstates and eigenvalues of the regularized fermion-phonon Hamiltonian.
A simple consequence of Proposition~\ref{Proposition:Diagonalization_of_fermion-phonon_model} is the following:

\corollary{}{
\label{Corollary:Hdiag}
The exact eigenstates of the Hamiltonian $H$ in \eqref{Eq:Fermion-phonon_Hamiltonian} are given by $\mathcal{U}\eta_{\mathbf{m}}^{BP}$ with the states $\eta_{\mathbf{m}}^{BP}$ in \eqref{Eq:etaBP}, and the corresponding eigenvalues are 
\begin{equation}
\label{Eq:eigenvals}
	\mathcal{E}_{\mathbf{m}}
	\define
	\mathcal{E}_0 + \frac{\pi v_F}{L}\left( q_{+}^2 + q_{-}^2 + 2\gamma_{1} q_{+} q_{-} \right)
		+ \omega_{0} m_P(0) + \sum_{X} \sum_{p \neq 0} \tilde{v}_{X}(p) \abs{p} m_{X}(p).
\end{equation}
Thus $H$ defines a self-adjoint operator with pure point spectrum.
}

As a special case of this corollary, $\tilde\Omega \define \mathcal{U}\Omega$ is the ground state of $H$ with corresponding eigenvalue $\mathcal{E}_0$, i.e.\ $\mathcal{E}_0$ has the physical interpretation as ground state energy.

\begin{proof}[Proof of Corollary~\ref{Corollary:Hdiag}] 
It is clear from \eqref{Eq:QRqRqOmega} and \eqref{Eq:Diagonalized_Hamiltonian} that $H\mathcal{U}\eta^{BP}_{\mathbf{m}} = \mathcal{E}_{\mathbf{m}}\mathcal{U}\eta^{BP}_{\mathbf{m}}$. 
Since the states in \eqref{Eq:etaBP} form a complete orthonormal basis in $\mathcal{F}$, the same holds true for these eigenstates.
This implies all other results (cf.\ Lemma~\ref{Lemma:Self_adjointness}). 
\end{proof} 

To compute fermion correlation functions we use the same notation as in \eqref{Eq:Psi_pm_operators} and the regularized fermion field operators in \eqref{Eq:Regularized_psi_explicit0}; see Section~\ref{SubSec:Mathematics_of_bosonization_II:Correlation_functions} for motivation and further explanations.
We thus define
\begin{multline}
\label{Eq:Fermion_correlation_functions1}
	\langle\tilde{\Omega},  \psi_{r_1}^{q_1}(x_1,t_1;0^+)
		\ldots \psi_{r_N}^{q_N}(x_N,t_N;0^+) \tilde{\Omega}\rangle \\
	\define
	\lim_{\epsilon_1,\ldots,\epsilon_N\to 0^+}
	\langle\tilde{\Omega}, \psi_{r_1}^{q_1}(x_1,t_1;\epsilon_1)
		\ldots \psi_{r_N}^{q_N}(x_N,t_N;\epsilon_N) \tilde{\Omega}\rangle,
\end{multline} 
where $\tilde{\Omega}$ is the ground state of the fermion-phonon Hamiltonian $H$ and the time evolution is given by $H$, i.e.\ $\psi_r^q(x,t;\epsilon) \define \exp(iHt)\psi_r^q(x;\epsilon)\exp(-iHt)$. 
The following result reduces the computation of any such correlation function to an exercise in normal ordering products of vertex operators:

\corollary{}{
\label{Corollary:FCF} 
The fermion correlation functions in \eqref{Eq:Fermion_correlation_functions1} are equal to 
\begin{equation}
\label{Eq:Fermion_correlation_functions2}
	\langle\Omega, \tilde\psi_{r_1}^{q_1}(x_1,t_1;0^+)
		\ldots \tilde \psi_{r_N}^{q_N}(x_N,t_N;0^+) \Omega\rangle 
\end{equation} 
with
\begin{multline}
\label{Eq:tildepsi} 
	\tilde\psi_r^q(x,t;\epsilon)
	= \frac{Z_{a,\epsilon} }{\sqrt{L}}
		\! \bosonnoord{ R_{r}^{qr}
			e^{-2\pi i q \left( (rx-v_{F} t) Q_{r} + \gamma_{1} v_{F}t Q_{-r} \right)/L} } \\
	\begin{aligned}
	& \times \prod_{X}
		\! \bosonnoord{ \exp \left( qr \sum_{p\neq 0} \frac{2\pi}{Lp}
			\rho_{X}(p) \hat{J}_{r,X}(-p)
			e^{-ip(x-r\tilde{v}_X(p) t)-\epsilon\abs{p}/2} \right) } \\
	& \times \prod_{X}
		\! \bosonnoord{ \exp \left( - qr \sum_{p\neq 0} \frac{2\pi}{Lp}
			\sigma_{X}(p) \hat{J}_{-r,X}(-p)
			e^{-ip(x+r\tilde{v}_X(p) t)-\epsilon\abs{p}/2} \right) }
	\end{aligned}
\end{multline} 
and
\begin{equation} 
\label{Eq:Zaepsilon}
	Z_{a,\epsilon}
	= \exp \left( - \sum_{0<p\leq \pi/a} \frac{2\pi}{Lp} (\sigma_F^2+\sigma_P^2)
		e^{-\epsilon p} \right).
\end{equation} 
}
\noindent (The proof can be found further below.)

By arguments similar to those in Section~\ref{SubSec:Mathematics_of_bosonization_II:Correlation_functions}, it follows from the corollary above that the correlation functions in \eqref{Eq:Fermion_correlation_functions1} are equal to
\begin{equation} 
	\langle\Omega,R_{r_1}^{q_1r_1} \ldots R_{r_N}^{q_Nr_N} \Omega\rangle
	\left(\frac{Z_{a,0} }{\sqrt{L}} \right)^{N/2}
	\prod_{1 \leq n < m \leq N} C^{q_n,q_m}_{r_n,r_m}(x_n-x_m, t_n-t_m; 0^+) 
\end{equation}
with building blocks $C_{r,r'}^{q,q'}(\cdot, \cdot; \cdot)$ which can be computed by normal ordering products of two operators of the form in \eqref{Eq:tildepsi}: 
\begin{equation} 
	\tilde{\psi}_{r\pp}^q(x,t;\epsilon) \tilde\psi_{r'}^{q'}(x',t';\epsilon')
	=  C^{q,q'}_{r,r'}(x-x',t-t';\epsilon+\epsilon')
	\! \bosonnoord{ \tilde\psi_{r\pp}^q(x,t;\epsilon) \tilde\psi_{r'}^{q'}(x',t';\epsilon') }.
\end{equation}
In Section~\ref{SubSec:Continuum_limit}, we give explicit formulas for these building blocks and for all fermion correlation functions in the continuum and thermodynamic limits $a \to 0^+$ and $L \to \infty$.
Fermion two-point correlation functions for $a > 0$ and $L < \infty$ can be found in \cite{MedenSchonhammerGunnarsson:1994}.

\begin{proof}[Proof of Corollary~\ref{Corollary:FCF}]
To obtain \eqref{Eq:Fermion_correlation_functions2} we insert $\tilde\Omega = \mathcal{U}\Omega$ and $I=\mathcal{U}\mathcal{U}^{\dagger}$ into \eqref{Eq:Fermion_correlation_functions1} and introduce
\begin{equation} 
\label{Eq:tildepsi_explicit1}
	\tilde{\psi}_r^q(x,t;\epsilon)
	\define
	\mathcal{U}^{\dagger} \psi_r^q(x,t;\epsilon) \mathcal{U}
		= e^{i\tilde H t} \tilde \psi_r^q(x;\epsilon) e^{-i\tilde Ht}
\end{equation} 
with $\tilde{H} \define \mathcal{U}^{\dagger} H \mathcal{U}$ in \eqref{Eq:Diagonalized_Hamiltonian}.
The explicit form of the operators in \eqref{Eq:tildepsi_explicit1} can be computed using \eqref{Eq:Regularized_psi_explicit}--\eqref{Eq:cNe} and Proposition~\ref{Proposition:Diagonalization_of_fermion-phonon_model} as follows.
Since $\mathcal{U}$ commutes with $R_r$ and $Q_r$,
\begin{equation}
\label{Eq:tildepsi_explicit2}
	\tilde\psi^q_{r}(x; \epsilon)
		= \frac{\mathcal{N}_{\epsilon}}{\sqrt{2\pi\epsilon}}
			\! \bosonnoord{R_{r}^{qr} e^{-2\pi iqrx Q_{r}/L}} \! 
		\exp \left( qr \sum_{p \neq 0} \frac{2\pi}{Lp}
			\mathcal{U}^{\dagger} \hat{J}_{r}(-p)\mathcal{U} e^{-ipx - \epsilon \abs{p}/2}
		\right),
\end{equation}
where it follows from \eqref{Eq:UJrU} and \eqref{Eq:JrXJrX} that the rightmost exponential is equal to 
\begin{multline}
\label{Eq:tildepsi_explicit3}
	\prod_{X} \exp \left( qr \sum_{p\neq 0} \frac{2\pi}{Lp}
		\rho_{X}(p) \hat{J}_{r,X}(-p) e^{-ipx -\epsilon\abs{p}/2} \right) \\
	\times \prod_{X} \exp \left( - qr \sum_{p\neq 0} \frac{2\pi}{Lp}
		\sigma_{X}(p) \hat{J}_{-r,X}(-p) e^{-ipx -\epsilon\abs{p}/2} \right).
\end{multline}
We are left to compute the time evolution given by $\tilde{H}$:
for the density operators,\footnote{This can be shown by differentiating the l.h.s.\ of \eqref{Eq:Time_evolution_of_Jr} with respect to $t$, computing the commutator of $\hat{J}_r(p)$ with $\tilde H \define \mathcal{U}^{\dagger} H \mathcal{U}$ by recalling \eqref{Eq:JrX}, and by integrating the resulting differential equation.}
\begin{equation}
\label{Eq:Time_evolution_of_Jr}
	e^{i\tilde Ht} \hat{J}_{r,X}(p) e^{-i\tilde Ht}
		= \hat{J}_{r,X}(p) e^{-irp\tilde{v}_X(p)t},
\end{equation}
and for the Klein factors,\footnote{This can be computed using \eqref{Eq:JrRr} for $p=0$; the computation is similar to that of the time evolution of the density operators.}
\begin{equation} 
	e^{i\tilde Ht}R_re^{-i\tilde Ht}
		= \bosonnoord{R_r e^{2i\pi r v_F(Q_r + \gamma_1 Q_{-r})t/L }}.
\end{equation}
Combining these with \eqref{Eq:tildepsi_explicit1}--\eqref{Eq:tildepsi_explicit3} yields the result in \eqref{Eq:tildepsi}; all that remains is to write this expression in a boson-normal-ordered form using \eqref{Eq:Bosonnoord1} and \eqref{Eq:cr}, which gives rise to the factor
\begin{equation} 
	\sqrt{1-e^{-2\pi\epsilon/L}}
	\exp \left( -\sum_{0 < p \leq \pi/a} \frac{\pi}{Lp}
		(\rho_F^2+\sigma_F^2+\rho_P^2+\sigma_P^2-1) e^{-\epsilon p} \right)
	= \frac{\sqrt{2\pi\epsilon/L}}{\mathcal{N}_\epsilon} Z_{a,\epsilon}
\end{equation}
(this follows from \eqref{Eq:cNe}, \eqref{Eq:Rho_sigma_identities}, and \eqref{Eq:Zaepsilon}).
\end{proof}

\subsection{Continuum limit}
\label{SubSec:Continuum_limit}
As mentioned after Corollary~\ref{Corollary:FCF}, all fermion correlation functions in the fermion-phonon model can be computed from the results in Section~\ref{SubSec:Exact_solution}. 
These formulas are in general complicated, but if one is interested in intermediate length scales they can be simplified.
One way to compute such simplified formulas is as limits $a \to 0^+$ and $L \to \infty$ of the fermion correlation functions in the regularized model; this is, however, demanding from a computational point of view. 
In this section we construct a renormalized model which corresponds to the continuum limit $a \to 0^+$ of the regularized fermion-phonon model.
Using this renormalized model, the continuum-limit results for the fermion correlation functions can be obtained with less computational effort, but, as will be discussed, the continuum limit requires additive and multiplicative renormalizations. 

To see that the limit $a\to 0^+$ is non-trivial, note that the ground state energy $\mathcal{E}_0$ in \eqref{Eq:Fermion_phonon_model_ground_state_energy} diverges like $O(L/a^2)$.
Moreover, for $\epsilon=0$, the constant in \eqref{Eq:Zaepsilon} is vanishing in this limit: 
\begin{equation} 
\label{Eq:Za0}
	Z_{a,0}
	= \left( e^{\gamma} \frac{L}{2a} \right)^{-(\sigma_F^2+\sigma_P^2)}
		\left( 1 + O \left( \frac{a}{L} \right) \right) 
\end{equation} 
with the Euler-Mascheroni constant $\gamma=0.5572\ldots$.\footnote{This follows from $\sum_{n=1}^{N} 1/n = \gamma + \ln(N) + O(1/N)$.} 
The reason for these difficulties is that the unitary operator $\mathcal{U}$ in Proposition~\ref{Proposition:Diagonalization_of_fermion-phonon_model} does not exist in this limit.
It is therefore remarkable that all these divergencies can be eliminated in a simple way after a similarity transformation with $\mathcal{U}$ using additive and multiplicative renormalizations as follows.
We define a \emph{renormalized Hamiltonian} 
\begin{equation} 
\label{Eq:Hren}
	\tilde{H}_{\text{ren}}
	\define
	\lim_{a\to 0^+} \left( \mathcal{U}^{\dagger} H \mathcal{U} - \mathcal{E}_0 \right) 
\end{equation} 
and \emph{renormalized fields}
\begin{equation}
\label{Eq:Psiren} 
	\Psi_{r}^{q}(x,t;\epsilon)
	\define
	\lim_{a\to 0^+} Z_{a,\epsilon}^{-1}
		\left( \frac{2\pi\ell}{L} \right)^{(\sigma_F^2 + \sigma_P^2)}
		\mathcal{U}^{\dagger} \psi_{r}^{q}(x,t;\epsilon) \mathcal{U},  
\end{equation} 
where $\ell>0$ is an arbitrary length parameter introduced for dimensional reasons.\footnote{Note that $Z_{a,\epsilon}^{-1} \left( {2\pi\ell}/{L} \right)^{(\sigma_F^2 + \sigma_P^2)} \to \left( e^{\gamma} {\pi\ell}/{a} \right)^{\sigma_F^2+\sigma_P^2} \left( 1 + O\left( {a}/{L} \right) \right)$ in the limit $\epsilon \to 0^{+}$.}
It follows from the results in Section~\ref{SubSec:Exact_solution} that these define a \emph{renormalized fermion-phonon model} in the following sense:   

\propositionlist{}{
\label{Proposition:Fermion_phonon_QFT}
\
\begin{enumerate}[label=\textnormal{(\alph*)}, leftmargin=0pt, itemindent = 10pt, labelwidth=0pt, align=left, nosep]

\item
\label{Item:Fermion_phonon_QFT_A}
The Hamiltonian in \eqref{Eq:Hren} and the fields in \eqref{Eq:Psiren} are equal to 
\begin{equation}
\label{Eq:Hren1}
	\tilde{H}_{\text{ren}}
	= \frac{\pi v_F}{L} \left( Q_{+}^2 + Q_{-}^2+ 2\gamma_1 Q_{+}Q_{-} \right)
		+ \omega_0 b_P^{\dagger}(0)b_P\pdag(0) \\
	+ \sum_{X} \sum_{p \neq 0} \omega_X(p) b^{\dagger}_{X}(p) b\pdag_{X}(p)
\end{equation}
and
\begin{multline}
\label{Eq:Psiren1}
	\Psi_r^q(x,t;\epsilon)
	= \frac{1}{\sqrt{L}}\left( \frac{2\pi\ell}{L}\right)^{\sigma_F^2+\sigma_P^2}
		\! \bosonnoord{ R_{r}^{qr}
			e^{-2\pi i q \left( (rx-v_{F} t) Q_{r} + \gamma_{1} v_{F}t Q_{-r} \right)/L} } \\
	\begin{aligned}
	& \times \prod_{X}
		\! \bosonnoord{ \exp \left( qr \sum_{p\neq 0} \frac{2\pi}{Lp}
			\rho_{X} \hat{J}_{r,X}(-p)
			e^{-ip(x-r\tilde{v}_X t)-\epsilon\abs{p}/2} \right) } \\
	& \times \prod_{X}
		\! \bosonnoord{ \exp \left( - qr \sum_{p\neq 0} \frac{2\pi}{Lp}
			\sigma_{X} \hat{J}_{-r,X}(-p)
			e^{-ip(x+r\tilde{v}_X t)-\epsilon\abs{p}/2} \right) },
	\end{aligned}
\end{multline} 
and they define self-adjoint and unitary operators on $\mathcal{F}$, respectively.

\item 
\label{Item:Fermion_phonon_QFT_B}
The renormalized fermion correlation functions 
\begin{equation} 
\label{Eq:FCFren}
	\lim_{L \to \infty} \lim_{a \to 0^+}
	\left( e^{\gamma}\frac{\pi\ell}{a} \right)^{N(\sigma_F^2+\sigma_P^2)}
		\langle\tilde{\Omega},  \psi_{r_1}^{q_1}(x_1,t_1;0^+)
		\ldots \psi_{r_N}^{q_N}(x_N,t_N;0^+) \tilde{\Omega}\rangle 
\end{equation} 
exist and are identical with
\begin{equation} 
\label{Eq:CF1}
	\lim_{L \to \infty}
	\langle\Omega, \Psi_{r_1}^{q_1}(x_1,t_1;0^+)
		\ldots \Psi_{r_N}^{q_N}(x_N,t_N;0^+) \Omega\rangle 
\end{equation}
as functions away from singularities.\footnote{This is the strongest statement of this identity which we prove in this paper; see Remark~\ref{Remark:Correlation_function_identity}.}

\end{enumerate} 
}
\noindent (The proof can be found at the end of this section.)

\remark{}{
\label{Remark:Correlation_function_identity}
We only prove that the correlation functions in \eqref{Eq:FCFren} and \eqref{Eq:CF1} are identical as functions away from the singularities.
To see why, we emphasize that the second expression is shown to equal
\begin{equation}
\label{Eq:CF1_with_limits}
	\lim_{L \to \infty} \lim_{\epsilon_1, \ldots, \epsilon_N \to 0^{+}} \lim_{a \to 0^+}
	\left( e^{\gamma}\frac{\pi\ell}{a} \right)^{N(\sigma_F^2+\sigma_P^2)}
		\langle\tilde{\Omega}, \psi_{r_1}^{q_1}(x_1,t_1;\epsilon_1)
		\ldots \psi_{r_N}^{q_N}(x_N,t_N;\epsilon_N) \tilde{\Omega}\rangle
\end{equation}
in the proof of Proposition~\ref{Proposition:Fermion_phonon_QFT}, and this differs from the first by the order in which the limits $\epsilon_j \to 0^{+}$ and $a \to 0^{+}$ are taken.
We cannot exclude the possibility that these limits cannot be interchanged, and, if that were the case, then the correlation functions in \eqref{Eq:FCFren} and \eqref{Eq:CF1} would define different distributions.\footnote{This is analogous to the situation for $1/(x \pm i0^{+})$ and $\mathcal{P} (1/x)$ which are identical as functions away from the singularity in $x = 0$ but define different distributions.}
We emphasize that, from a physical point of view, the correct order of limits is as in \eqref{Eq:FCFren}.
}

As for the regularized model, the renormalized Hamiltonian $\tilde{H}_{\text{ren}}$ determines the time evolution of the renormalized fields, i.e.\ $\Psi_r^q(x,t;\epsilon) = \exp(i\tilde{H}_{\text{ren}}t) \Psi_r^q(x;\epsilon) \exp(-i\tilde{H}_{\text{ren}}t)$. 
It would be interesting to extend this renormalized fermion-phonon model to allow for an efficient computation of boson correlation functions. 
Another interesting use would be to compute correlation functions at finite temperature; see \cite{deWoulLangmann:2012} for such results for a similar model. 
However, this is beyond the scope of the present paper. 

Let us briefly discuss the physical interpretation of the renormalized fermion-phonon model.
Our construction of the regularized fermion-phonon model is on the Fock space for the corresponding non-interacting model, and the Fock vacuum $\Omega$ has the physical interpretation as ground state of the latter model. 
The ground state $\tilde{\Omega}$ of the interacting model with \emph{local interactions} (which corresponds to the limit $a \to 0^+$) does not exist in this Fock space; this is clear since, for example, the ground state energy diverges. 
However, using a similarity transformation with the unitary operator $\mathcal{U}$, one can construct a new Hilbert space which is such that, by definition, it contains the model ground state: the definition is such that the transformation maps $\tilde{\Omega} = \mathcal{U} \Omega$ to the Fock vacuum $\Omega = \mathcal{U}^{\dagger} \tilde{\Omega}$.
Since the normal ordering was defined with reference to the original Fock vacuum, the additive and multiplicative renormalizations can be interpreted as changing the normal ordering such that it uses the interacting ground state as reference instead.
The difference between these two normal orderings diverges in the limit $a \to 0^{+}$, but this is to be expected since the similarity transformation maps the original Fock vacuum to the state $\mathcal{U}^{\dagger}\Omega$ which does not exist in this limit. 

The renormalized model can be used to efficiently compute the continuum and thermodynamic limits $a \to 0^+$ and $L \to \infty$ of the renormalized fermion correlation functions.
The strategy is the same as described for the regularized model at the end of Section~\ref{SubSec:Exact_solution}. 
The computational advantage of working with the renormalized model is that each of the renormalized fields $\Psi_{r}^{q}(x,t;\epsilon)$ in \eqref{Eq:Psiren1} is essentially a product of four vertex operators of the form
\begin{equation}
\label{Eq:Vnuru}
	\WW^{\nu}_{r,X}(x;\epsilon)
	\define
	\bosonnoord{ \exp \left( \nu \sum_{p\neq 0} \frac{2\pi}{Lp}\hat{J}_{r,X}(-p)
		e^{-ipx-\epsilon \abs{p}/2} \right) }
\end{equation} 
with different real parameters $\nu$ (cf.\ \eqref{Eq:Psiren1}).
We also have the following general formula for writing products of such vertex operators in a normal-ordered form:
\begin{equation}
\label{Eq:WW_normal_ordered}
	\WW^{\nu}_{r,X}(x;\epsilon) \WW^{\nu'}_{r,X}(x';\epsilon')
	= \left(
		\frac{ i\exp \left( -\frac{i\pi}{L}
			[r(x-x') + \tfrac{i}{2}(\epsilon+\epsilon')] \right) }
		{ 2 \sin \left( \frac{\pi}{L}
			[r(x-x') + \tfrac{i}{2}(\epsilon+\epsilon')] \right) }
	\right)^{-\nu\nu'}
	\! \bosonnoord{ \cdots }
\end{equation}
with $\bosonnoord{ \cdots }$ meaning boson normal ordering of the l.h.s.\ and
\begin{equation}
\label{Eq:W_C_factor}
	\frac{2\pi\ell}{L}
	\times
		\frac{ i\exp \left( -\frac{i\pi}{L}[rx + \tfrac{i}{2}\epsilon] \right) }
			{ 2 \sin \left( \frac{\pi}{L}[rx + \tfrac{i}{2}\epsilon] \right) }
	\xrightarrow[\epsilon \to 0^+]{L \to \infty}
	\frac{i\ell}{rx + i 0^+}
\end{equation}
(this follows from \eqref{Eq:Gamma_pm_Gamma_pm} using \eqref{Eq:Log_Taylor_expansion}).  
Moreover, it is useful to note that the zero-mode contributions when normal ordering products of the renormalized fields $\Psi_{r}^{q}(x,t;\epsilon)$ can be ignored in the limit $L \to \infty$ (since they always come with a factor $1/L$); see Remark~\ref{Remark:Include_zero_modes} for a discussion of the zero-mode contributions for $L < \infty$.
It thus follows that 
\begin{equation} 
\label{Eq:Psi_Psi_normal_ordered}
	\Psi_{r\pp}^{q}(x,t;\epsilon) \Psi_{r'}^{q'}(x',t';\epsilon')
	= C_{r,r'}(x-x',t-t';\epsilon+\epsilon')^{-qq'}
		\! \bosonnoord{ \Psi_{r\pp}^{q}(x,t;\epsilon) \Psi_{r'}^{q'}(x',t';\epsilon') }
\end{equation}
with
\begin{equation}
\label{Eq:Cpmpm}
\begin{aligned}
	\left( \frac{2\pi\ell}{L} \right)^{\rho_F^2 + \sigma_F^2 + \rho_P^2 + \sigma_P^2}
		C_{\pm,\pm}(x,t;\epsilon)
	& \xrightarrow[\epsilon \to 0^+]{L \to \infty}
		\prod_{r,X} \left(
			\frac{i\ell}{rx -\tilde{v}_Xt + i 0^{+}}
		\right)^{\rho_{X}^2 \delta_{r,\pm} + \sigma_{X}^2 \delta_{r,\mp}}, \\
	\left( \frac{2\pi\ell}{L} \right)^{2(\rho_F\sigma_F+\rho_P\sigma_P)}
		C_{\pm,\mp}(x,t;\epsilon)
	& \xrightarrow[\epsilon \to 0^+]{L \to \infty}
		\prod_{r,X} \left(
			\frac{i\ell}{rx -\tilde{v}_Xt + i 0^{+}}
		\right)^{\rho_{X}\sigma_{X}}.
\end{aligned}
\end{equation}
As we mentioned after Corollary~\ref{Corollary:FCF}, by arguments similar to those in Section~\ref{SubSec:Mathematics_of_bosonization_II:Correlation_functions}, this implies the following:

\corollary{}{
\label{Corollary:GreenN1} 
In the thermodynamic limit $L \to \infty$, the renormalized fermion correlation functions in \eqref{Eq:CF1} are
\begin{multline}
\label{Eq:Fermion_phonon_N-point_correl_fncs}
	\lim_{L \to \infty}
	\langle\Omega, \Psi_{r_1}^{q_1}(x_1,t_1;0^+)
		\ldots \Psi_{r_N}^{q_N}(x_N,t_N;0^+) \Omega\rangle
	= \langle\Omega, R_{r_1}^{q_1 r_1} \ldots R_{r_N}^{q_N r_N} \Omega\rangle \\
	\times \left( \frac{1}{2\pi\ell} \right)^{N/2}
		\prod_{0\leq n<m\leq N} \prod_{r,X}
		\left(
			\frac{i \ell}{r(x_n-x_m) - \tilde{v}_X(t_n-t_m)+i 0^+}
		\right)^{-q_n q_m c_{r,X;r_n,r_m}}
\end{multline}
with
\begin{equation}
	c_{r,X;\pm,\pm}
	\define
	\rho_X^2 \delta_{r,\pm} + \sigma_X^2 \delta_{r,\mp},
	\qquad
	c_{r,X;\pm,\mp}
	\define
	\rho_X \sigma_X,
\end{equation}
where the first factor on the r.h.s.\ of \eqref{Eq:Fermion_phonon_N-point_correl_fncs} is a sign factor \textnormal{(}cf.\ \eqref{Eq:OmRRROm}\textnormal{ff}\textnormal{)}.
}

\begin{proof} 
The formulas for $c_{r,X;r_n,r_m}$ follow from the exponents in \eqref{Eq:Cpmpm}, and we are thus left to prove that the factors $(2\pi\ell/L)^{\sigma_F^2+\sigma_P^2}$ from the renormalized fields in \eqref{Eq:Psiren1} exactly match the factors needed for the limit $L \to \infty$ to exist.
This is equivalent to showing that
\begin{equation} 
	\left( \frac{1}{\sqrt{L}}
		\left( \frac{2\pi\ell}{L} \right)^{\sigma_F^2+\sigma_P^2} \right)^N
	= \left( \frac{1}{2\pi\ell} \right)^{N/2} \prod_{1\leq n<m\leq N}
		\left( \frac{2\pi\ell}{L} \right)^{-q_n q_m \sum_{r,X} c_{r,X;r_n,r_m}}
\end{equation} 
if $\langle\Omega, R_{r_1}^{q_1 r_1} \ldots R_{r_N}^{q_N r_N} \Omega\rangle \neq 0$,
which is true by Corollary~\ref{Lemma:qqrr}.
\end{proof}

The explicit expressions for all fermion correlation functions can be found from Corollary~\ref{Corollary:GreenN1}.
For example, the special case $N = 2$ gives the two-point correlation functions stated in the introduction (cf.\ \eqref{Eq:Green2}).
Similarly, as an example of a four-point correlation function ($N=4$), we find
\begin{multline}
\label{Eq:4pointfunction}
	\lim_{L \to \infty}
		\langle\Omega, \Psi_{+}\pdag(x_1,t_1;0^+) \Psi_{-}^{\dagger}(x_2,t_2;0^+)
			\Psi_{-}\pdag(x_3,t_3;0^+) \Psi_{+}^{\dagger}(x_4,t_4;0^+) \Omega\rangle \\
	= \left( \frac{1}{2\pi \ell} \right)^2 \prod_{r,X}
		\left(
			\frac{rx_{13} - \tilde{v}_{X}t_{13} + i0^{+}}
				{rx_{12} - \tilde{v}_{X}t_{12} + i0^{+}}
		\right)^{\rho_X \sigma_X}
		\left(
			\frac{rx_{24} - \tilde{v}_{X}t_{24} + i0^{+}}
				{rx_{34} - \tilde{v}_{X}t_{34} + i0^{+}}
		\right)^{\rho_X \sigma_X} \\
	\begin{aligned}
	& \times \prod_{X}
		\left(
			\frac{i \ell}{x_{14} - \tilde{v}_{X}t_{14} + i0^{+}}
		\right)^{\rho_X^2}
		\left(
			\frac{i \ell}{-x_{14} - \tilde{v}_{X}t_{14} + i0^{+}}
		\right)^{\sigma_X^2} \\
	& \times \prod_{X}
		\left(
			\frac{i \ell}{x_{23} - \tilde{v}_{X}t_{23} + i0^{+}}
		\right)^{\sigma_X^2}
		\left(
			\frac{i \ell}{-x_{23} - \tilde{v}_{X}t_{23} + i0^{+}}
		\right)^{\rho_X^2}
	\end{aligned}
\end{multline}
using the shorthand notation $x_{nm} \define x_{n} - x_{m}$ and $t_{nm} \define t_{n} - t_{m}$.

\remark{}{
\label{Remark:Include_zero_modes}
One can get simple formulas even without passing to the thermodynamic limit: for the fermion two-point correlation functions we checked that the zero-mode contributions exactly cancel the $(x \pm r\tilde{v}_{X}t)$-dependence from the numerator in \eqref{Eq:WW_normal_ordered} by using the identity
\begin{equation}
	\left( \rho_{F}^2 + \sigma_{F}^2 \right) \tilde{v}_{F}
		+ \left( \rho_{P}^2 + \sigma_{P}^2 \right) \tilde{v}_{P} = v_{F},
\end{equation}
which can be derived from \eqref{Eq:CS_matrices_explicit}.
We expect that a similar cancelation takes place for all fermion correlation functions.
} 

\begin{proof}[Proof of Proposition~\ref{Proposition:Fermion_phonon_QFT}]
The states in \eqref{Eq:etaBP} form a complete set of orthonormal eigenstates of the renormalized Hamiltonian in \eqref{Eq:Hren1}, and the corresponding eigenvalues are $\lim_{a \to 0^{+}}(\mathcal{E}_{\mathbf{m}}-\mathcal{E}_0)$ with $\mathcal{E}_{\mathbf{m}}$ in \eqref{Eq:eigenvals}.   
It follows that $\tilde{H}_{\text{ren}}$ defines a self-adjoint operator on $\mathcal{F}$ (cf.\ Lemma~\ref{Lemma:Self_adjointness}). 
This also shows that the limit in \eqref{Eq:Hren} is in the strong sense on the domain of finite linear combinations of the states in \eqref{Eq:etaBP}. 

Equation \eqref{Eq:Psiren} implies \eqref{Eq:Psiren1} for the renormalized fields.
They are thus proportional to a product of unitary operators (this follows from Proposition~\ref{Proposition:Vertex_operators}) and are therefore also unitary.
This completes the proof of Part~\ref{Item:Fermion_phonon_QFT_A}.

The limit in \eqref{Eq:Psiren} is in the weak sense on $\mathcal{F}$, and it thus follows that the correlation functions in \eqref{Eq:CF1} and \eqref{Eq:CF1_with_limits} are equal.
This proves Part~\ref{Item:Fermion_phonon_QFT_B} (see Remark~\ref{Remark:Correlation_function_identity} for further explanations).
\end{proof}

\subsection{Proof of Proposition~\ref{Proposition:Diagonalization_of_fermion-phonon_model}}
\label{Sec:Proof_of_diagonalization_proposition}
One important feature allowing to solve the fermion-phonon model exactly is that bosons with different magnitude of momentum $p$ decouple.
To make this manifest we write the Hamiltonian in \eqref{Eq:Bosonized_Hamiltonian_precise} as 
\begin{equation}
\label{Eq:Fermion_phonon_Hamiltonian_sum}
	H = \sum_{p \geq  0} h_p
\end{equation}
(this follows since $(h_{p} + h_{-p})/2 = h_{p}$) with the zero-mode part
\begin{equation} 
	h_0
	= \frac{\pi}{L} \left( v_{F} \left( Q_{+}^2 + Q_{-}^2 + 2\gamma_1 Q_{+}Q_{-} \right)
	+ \hat{\Pi}_{P}(0)^2+ \omega_{0}^{2} \hat{\Phi}_{P}(0)^2 \right)  -\frac{1}{2}\omega_0
\end{equation} 
and the parts with momenta $p>0$ written as 
\begin{equation} 
\label{Eq:hp}
	h_{p}
	= \frac{2\pi}{L} \left(
		\hat{\boldsymbol{\Pi}}^{\dagger}(p) \mathbf{A}(p)\hat{\boldsymbol{\Pi}}(p)
		+ \hat{\boldsymbol{\Phi}}^{\dagger}(p) \mathbf{B}(p)\hat{\boldsymbol{\Phi}}(p)
	\right) - \left( \omega^0_F(p) + \omega^0_P(p) \right)
\end{equation}
using the following matrix notation:
\begin{align}
	\hat{\boldsymbol{\Pi}}(p)
	& \define
	\big( \hat{\Pi}_{F}(p), \hat{\Pi}_{P}(p) \big)^{T},
	& \hat{\boldsymbol{\Phi}}(p)
	& \define
	\big( \hat{\Phi}_{F}(p), \hat{\Phi}_{P}(p) \big)^{T},
	\label{Eq:vPi_vPhi_definition} \\
	\mathbf{A}(p)
	& \define
	\left( \begin{matrix}
		1-\gamma_1(p)	& 0 \\
		0				& 1
	\end{matrix} \right),
	& \mathbf{B}(p)
	& \define
	p^2 \left( \begin{matrix}
		v_F^2[1+\gamma_1(p)]	& v_Fv_P\gamma_2(p) \\
		v_Fv_P\gamma_2(p)	& v_P^2
	\end{matrix} \right), 
\end{align} 
with the shorthand notation $\gamma_{1,2}(p) \define \gamma_{1,2}\Theta(\pi/a - \abs{p})$.
To ensure that this describes a stable system, we must require that both matrices $\mathbf{A}(p)$ and $\mathbf{B}(p)$ are positive definite.
This is true if and only if $\gamma_1 < 1$ and $\gamma_{2}^2 < 1 + \gamma_{1}$, which, as we mentioned, are equivalent to  the conditions in \eqref{Eq:Restrictions}.

The parts with $p=0$ and $p>\pi/a$ are already diagonal: 
\begin{align} 
	h_0
	& = \frac{\pi v_{F}}{L} (Q_{+}^2 + Q_{-}^2 + 2\gamma_1 Q_{+}Q_{-})
		+ \omega_0b^{\dagger}_P(0) b\pdag_P(0), \\
	h_{p>\pi/a}
	& = \sum_X \omega_X^0(p) \left( b^{\dagger}_X(p)b\pdag_X(p) + b^{\dagger}_X(-p)b\pdag_X(-p) \right). 
\end{align} 
The task of diagonalizing the parts $h_{0<p\leq\pi/a}$ can be reduced to diagonalizing the matrix
\begin{align} 
\label{Eq:Cmatrix}
	\mathbf{C}(p)
	\define
	\mathbf{A}(p)^{1/2}\mathbf{B}(p) \mathbf{A}(p)^{1/2}
	= p^2 \left( \begin{matrix}
		v_F^2(1-\gamma_1^2)				& v_Fv_P\gamma_2\sqrt{1-\gamma_1} \\
		v_Fv_P\gamma_2\sqrt{1-\gamma_1}	& v_P^2
	\end{matrix} \right), 
\end{align}
which is also positive definite if and only if $\gamma_1 < 1$ and $\gamma_{2}^2 < 1 + \gamma_{1}$.
It follows that there exists a unitary matrix $\mathbf{U}$ and a diagonal matrix $\boldsymbol{\omega}(p) = \diag(\omega_F(p),\omega_P(p))$ such that 
\begin{equation} 
\label{Eq:Cmatrix1}
	\mathbf{C}(p) = \mathbf{U}\boldsymbol{\omega}(p)^2\mathbf{U}^{\dagger}.
\end{equation}
Indeed, one can write 
\begin{equation} 
	h_{0<p\leq\pi/a}
	= \frac{2\pi}{L} \left(\tilde{\boldsymbol{\Pi}}^{\dagger}(p) \tilde{\boldsymbol{\Pi}}(p)
		+ \tilde{\boldsymbol{\Phi}}^{\dagger}(p) \boldsymbol{\omega}(p)^2
		\tilde{\boldsymbol{\Phi}}(p)\right) - \left( \omega^0_F(p)+\omega^0_P(p) \right)
\end{equation} 
with 
\begin{equation} 
\label{Eq:CT}
	\tilde{\boldsymbol{\Pi}}(p)
		= \mathbf{U}\mathbf{A}^{1/2}\hat{\boldsymbol{\Pi}}(p),
	\qquad 
	\tilde{\boldsymbol{\Phi}}(p)
		= \mathbf{U}^{\dagger}\mathbf{A}^{-1/2}\hat{\boldsymbol{\Phi}}(p)
\end{equation}
defined in the same way as in \eqref{Eq:vPi_vPhi_definition}.
This can be shown to be a canonical transformation, i.e.\ one can show that the operators $\tilde{\Pi}_X(p)$ and $\tilde{\Phi}_X(p)$ satisfy CCR similar to \eqref{Eq:Fermion-phonon_CCR}. 
By introducing boson operators $\BB_{X}^{(\dagger)}(p)$ corresponding to the ones in \eqref{Eqs:Phonon_Pi_Phi_representation} and \eqref{Eq:Fermion_Phi_Pi_operators}:
\begin{equation}
\label{Eqs:Pi_phi_tilde_operators_relation_to_B_operators}
\begin{aligned}
	\tilde{\Pi}\pdag_{X}(p)
		& \define -i \sqrt{ \frac{L}{2\pi} } \sqrt{\frac{\omega_{X}(p)}{2}}
			\left( \BB\pdag_{X}(p) - \BB^{\dagger}_{X}(-p) \right),\\
	\tilde{\Phi}\pdag_{X}(p)
		& \define \sqrt{ \frac{L}{2\pi} } \frac{1}{\sqrt{2\omega_{X}(p)}}
			\left( \BB\pdag_{X}(p) + \BB^{\dagger}_{X}(-p) \right), 
\end{aligned}
\end{equation}
one obtains
\begin{equation} 
	h_{0<p\leq\pi/a}
	= \sum_X \left( \omega_{X}(p) \left( \BB^{\dagger}_X(p)\BB\pdag_X(p)
		+ \BB^{\dagger}_X(-p)\BB\pdag_X(-p) \right)
		+ \left( \omega_X(p) - \omega_X^0(p) \right) \right) . 
\end{equation} 
It is known that there exist unitary operators $\mathcal{U}_p$ such that 
\begin{equation} 
\label{Eq:BB_from_b}
	\mathcal{U}^{\dagger}_p \BB^{(\dagger)}_{X}(\pm p)\mathcal{U}\pdag_p
		= b^{(\dagger)}_{X}(\pm p),
\end{equation}
which can be constructed using the operators $b^{(\dagger)}_X(\pm p)$; see e.g.\ Appendix~C.1 in \cite{deWoulLangmann:2012} for an explicit construction of these unitary operators.
It follows that the unitary operator   
\begin{equation}
\label{Eq:U_from_Up} 
	\mathcal{U}
	\define
	\prod_{0<p\leq \pi/a} \mathcal{U}_p
\end{equation} 
satisfies \eqref{Eq:Diagonalized_Hamiltonian}--\eqref{Eq:Fermion_phonon_model_ground_state_energy} (note that this is a \emph{finite} product and thus well-defined). 
Moreover, one can easily verify \eqref{Eq:Fermion-phonon_renormalized_velocities_general}--\eqref{Eq:Fermion-phonon_renormalized_velocities} by computing the eigenvalues of the matrix in \eqref{Eq:Cmatrix} (cf.\ \eqref{Eq:Fermion-phonon_dispersion_relations} and \eqref{Eq:Cmatrix1}).
This completes the proof of Part~\ref{Item:Diagonalization_of_fermion-phonon_model_A} of Proposition~\ref{Proposition:Diagonalization_of_fermion-phonon_model}.

To prove Part~\ref{Item:Diagonalization_of_fermion-phonon_model_B}, invert \eqref{Eq:CT} and write the resulting equations in component form: 
\begin{equation}
	\hat{\Pi}_{X}(p) =  \sum_{X'} \left( M_{\Pi} \right)_{X,X'} \tilde{\Pi}_{X'}(p),
	\qquad
	\hat{\Phi}_{X}(p) =  \sum_{X'} \left( M_{\Phi} \right)_{X,X'} \tilde{\Phi}_{X'}(p)
\end{equation}
using matrices $\mathbf{M}_{\Pi} \define \mathbf{A}^{-1/2}\mathbf{U}^{\dagger}$ and $\mathbf{M}_{\Phi} \define  \mathbf{A}^{1/2}\mathbf{U}$ (as before, we restrict our discussion to the non-trivial case $0 < \abs{p} \leq \pi/a$). 
Using \eqref{Eqs:Phonon_Pi_Phi_representation}, \eqref{Eq:Fermion_Phi_Pi_operators}, and \eqref{Eqs:Pi_phi_tilde_operators_relation_to_B_operators} one finds
\begin{equation}
	b\pdag_{X}(p)
	= \sum_{X'} \left( \mathcal{C}\pdag_{X,X'} \BB\pdag_{X'}(p)
		+ \mathcal{S}\pdag_{X,X'} \BB^{\dagger}_{X'}(-p) \right)
\end{equation}
with
\begin{equation}
\begin{aligned}
\label{Eqs:CS_matrices}
	\mathcal{C}_{X,X'}
		& \define \frac{1}{2} \left(
			\sqrt{ \frac{v_{X}}{\tilde{v}_{X'}} }
				\big( M_{\Phi} \big)_{X,X'}
			+ \sqrt{ \frac{\tilde{v}_{X'}}{v_{X}} }
				\big( M_{\Pi} \big)_{X,X'} \right), \\
	\mathcal{S}_{X,X'}
		& \define \frac{1}{2} \left(
			\sqrt{ \frac{v_{X}}{\tilde{v}_{X'}} }
				\big( M_{\Phi} \big)_{X,X'}
			- \sqrt{ \frac{\tilde{v}_{X'}}{v_{X}} }
				\big( M_{\Pi} \big)_{X,X'} \right).
\end{aligned}
\end{equation}
It thus follows from \eqref{Eq:BB_from_b}--\eqref{Eq:U_from_Up} that 
\begin{equation}
\label{Eq:Transformation_of_b_for_fermion-phonon_model_explicit}
	\mathcal{U}^{\dagger} b\pdag_{X}(p) \mathcal{U}
		= \sum_{X'} \left( \mathcal{C}\pdag_{X,X'} b\pdag_{X'}(p)
			+ \mathcal{S}\pdag_{X,X'} b^{\dagger}_{X'}(-p) \right),
\end{equation}
and recalling the relation between  $b^{(\dagger)}_X(p)$ and $\hat{J}_{r,X}(p)$ in \eqref{Eq:JrX} we obtain 
\begin{equation}
\label{Eq:Transformation_of_J_for_fermion-phonon_model_explicit}
	\mathcal{U}^{\dagger} \hat{J}_{r,X}(p) \mathcal{U}
		= \sum_{X'} \left( \mathcal{C}_{X,X'} \hat{J}_{r,X'}(p)
			+ \mathcal{S}_{X,X'}\hat{J}_{-r,X'}(p) \right),
\end{equation}
which, for $X=F$, implies \eqref{Eq:UJrU}--\eqref{Eq:rhoXp_sigmaXp} with $\rho_{X'} \define \mathcal{C}_{F,X'}$ and $\sigma_{X'} \define -\mathcal{S}_{F,X'}$. 
The explicit expressions in \eqref{Eq:CS_matrices_explicit} follow from \eqref{Eqs:CS_matrices} by straightforward computations which we omit.
Lastly, the consistency of \eqref{Eq:Transformation_of_b_for_fermion-phonon_model_explicit} with the CCR in \eqref{Eq:CCR_b} implies
\begin{equation}
	\sum_{X'} \Bigl( \abs{\mathcal{C}_{X,X'}}^2 - \abs{\mathcal{S}_{X,X'}}^2 \Bigr) = 1
\end{equation} 
which, for $X=F$, is equivalent to the identity in \eqref{Eq:Rho_sigma_identities}.


\section{Concluding remarks}
\label{Sec:Concluding_remarks}
We constructed and solved a one-dimensional fermion-phonon model using bosonization.
We also gave a self-contained account of the mathematical basis of bosonization together with complete proofs.
Our main results include a renormalized model in the continuum limit, which was obtained by additive and multiplicative renormalizations (cf.\ Proposition~\ref{Proposition:Fermion_phonon_QFT}), and explicit formulas for all fermion correlation functions in the continuum and thermodynamic limits using this renormalized model (cf.\ Corollary~\ref{Corollary:GreenN1}).
We emphasize that positivity of our correlation functions is manifest from the method by which they were computed: the renormalized model we derived is analogous to the one used by Carey, Ruijsenaars, and Wright \cite{CareyRuijsenaarsWright:1985} to prove positivity of Klaiber's $N$-point correlation functions for the massless Thirring model \cite{Klaiber:LectTheoPhys:1968}, and their argument also applies to our model.

As an application of the tools developed in this paper, we address one point which was raised in the introduction: we mentioned that certain fermion four-point correlation functions can be used to study physical properties of the fermion-phonon model related to charge-density-wave (CDW) and superconducting (SC) instabilities \cite{ChenLeeLuchini:1988, VoitSchulz:1985}.
The correlation functions of interest are $\lim_{L \to \infty} \langle\Omega, O\pdag_{A}(x,t) O^{\dagger}_{A}(0,0) \Omega\rangle$ for $A \in \left\{ \text{CDW},\text{SC} \right\}$ with $O_{\text{CDW}}(x,t) \define \bosonnoord{ \Psi_{+}\pdag(x,t;0^+) \Psi_{-}^{\dagger}(x,t;0^+) }$ and $O_{\text{SC}}(x,t) \define \bosonnoord{ \Psi_{-}\pdag(x,t;0^+) \Psi_{+}\pdag(x,t;0^+) }$.
We emphasize that these operators must be in a boson-normal-ordered form for the correlation functions to exist, and this corresponds to an additional multiplicative renormalization; this is seen by comparing the equations below with \eqref{Eq:4pointfunction}. 
Using Corollary~\ref{Corollary:GreenN1} we find
\begin{equation}
	\lim_{L \to \infty}
		\langle\Omega, O\pdag_{\text{CDW}}(x,t) O^{\dagger}_{\text{CDW}}(0,0) \Omega\rangle
	= \left( \frac{1}{2\pi \ell} \right)^2 \prod_{X}
	\left( \frac{\ell^2}{x^2  - (\tilde{v}_{X}t - i0^{+})^2 } \right)^{(\rho_X -\sigma_X)^2}
\end{equation}
and
\begin{equation}
	\lim_{L \to \infty}
		\langle\Omega, O\pdag_{\text{SC}}(x,t) O^{\dagger}_{\text{SC}}(0,0) \Omega\rangle
	= \left( \frac{1}{2\pi \ell} \right)^2 \prod_{X}
	\left( \frac{\ell^2}{x^2  - (\tilde{v}_{X}t - i0^{+})^2 } \right)^{(\rho_X +\sigma_X)^2},
\end{equation}
which are of interest in physics since they allow to construct phase diagrams \cite{ChenLeeLuchini:1988, VoitSchulz:1985}.

We derived our results using a particular regularization, but, as we mentioned, there exist a vast number of different regularizations. 
Physical arguments suggest that the correlation functions in the continuum and thermodynamic limits are universal in the sense that they are independent of the details of the regularization. 
It would be interesting to prove this universality conjecture. 
A related question is if the correlation functions obtained by passing to the continuum limit are identical as distributions to the continuum limit of the correlation functions for the regularized model.
(We only proved the equality of these functions in a weaker sense.) 
As for the previous question, physical arguments suggest that this is the case, but to prove this conjecture one would have to show that the correlation functions in \eqref{Eq:FCFren} and \eqref{Eq:CF1} are identical as distributions.
To our knowledge, both of these questions are open even for the massless Thirring model.

\paragraph*{Acknowledgments:}
We would like to thank Alan Carey, Jan Derezi{\'n}ski, Luca Fresta, and Jonas de Woul for valuable suggestions on the manuscript, and Vieri Mastropietro for encouraging us to write the paper.
Financial support by the G\"oran Gustafsson Foundation (GGS 1221) is acknowledged.


\begin{appendices}


\section{Notation and conventions}
\label{Appendix:Notation}
The symbols $\mathbb{C}$, $\mathbb{R}$, and $\mathbb{Z}$ denote the sets of complex, real, and integer numbers, and $\mathbb{N}$ and $\mathbb{N}_0$ denote the positive and non-negative integers, respectively.
The symbol $\partial_x$ is short for $\partial/\partial x$.
We write ``$\define$'' to emphasize a definition. 
The complex conjugate of  $c\in \mathbb{C}$ is $\overline{c}$.
The Heaviside function $\Theta(\cdot)$ is defined such that $\Theta(x)=1$ for $x\geq 0$ and $\Theta(x)=0$ for $x<0$.
We use the symbols $[a,b] \define ab - ba$ and $\{a,b\} \define ab + ba$ for the commutator and the anticommutator, respectively.
We write $I$ for the identity operator, but often identify $c \in \mathbb{C}$ with $cI$.
Two symbols $\dagger$ and $*$ are used for Hermitian conjugation of operators: the first is used in cases where operator domains are well-defined or of no concern, and the second in a weaker sense such that, if $A$ is an operator defined on a dense domain $\mathcal{D}$, then ``$A^{\dagger}$" means ``$A^*$ restricted to $\mathcal{D}$."

Unless otherwise stated, and when there is no risk of confusion, a free variable or index in an identity can take any allowed value.
For example, the relations in \eqref{Eq:Fermion_field_anticommutation_relations} hold true for all $k,k'\in(2\pi/L)(\mathbb{Z}+1/2)$ and $r,r'=\pm$, whereas \eqref{Eq:Fermion_field_action_on_Omega} holds true for $r=\pm$ but only those $k\in(2\pi/L)(\mathbb{Z}+1/2)$ which satisfy the given condition. 

\paragraph*{Parameters:}
The following parameters are used:
\begin{equation*}
\begin{array}{cll}
	\hline\hline
	\multicolumn{1}{c}{\text{Parameter}}	& \multicolumn{1}{c}{\text{Physical significance}}\\
	\hline
	L		& \text{infrared cutoff (length of space)}		\\
	a		& \text{ultraviolet cutoff (interaction range)}	\\
	v_{F}	& \text{Fermi velocity}							\\
	v_{P}	& \text{phonon velocity}							\\
	\gF		& \text{fermion-fermion coupling constant}		\\
	\gP		& \text{fermion-phonon coupling constant}			\\
	\hline\hline
\end{array}
\end{equation*} 

\paragraph*{Variables:}
The following variable names are reserved for particular sets: 
\begin{equation*}
\begin{array}{rcl}
	\hline\hline
	\multicolumn{1}{c}{\text{Variables}}	& \multicolumn{1}{c}{\text{Set}}
		& \multicolumn{1}{c}{\text{Physical interpretation}}								\\
	\hline
	r, r', r_j				& \{+,-\} 					& \text{chiralities}				\\
	x, x', y, x_j			& [-L/2,L/2]					& \text{positions}				\\
	k, k', k_j				& (2\pi/L)(\mathbb{Z}+1/2)	&\text{fermion momenta}			\\
	p, p', p_j				& (2\pi/L)\mathbb{Z}			& \text{boson momenta} 			\\
	X, X'					& \{F,P\}					& \text{boson flavor indices}	\\
	\hline\hline
\end{array}
\end{equation*}

\paragraph*{Summation conventions:}
The following abbreviations are used:
\begin{equation*}
\begin{array}{cc}
	\hline\hline
	\multicolumn{1}{c}{\text{Abbreviation}}	& \multicolumn{1}{c}{\text{Meaning}}	\\
	\hline
	\displaystyle{\sum_{r}}
		& \displaystyle{\sum_{r = \pm}}											\\
	\displaystyle{\sum_{k}}
		& \displaystyle{\sum_{k \in \frac{2\pi}{L}(\mathbb{Z}+\frac12)}}			\\
	\displaystyle{\sum_{p\neq 0}}
		& \displaystyle{\sum_{p \in \frac{2\pi}{L}\mathbb{Z}\setminus\{0\}}}		\\
	\multicolumn{2}{c}{\text{(etc.)}}												\\
	\hline\hline
\end{array}
\end{equation*}

\paragraph*{Useful formula:}
The following formula is used repeatedly:
\begin{equation}
\label{Eq:Log_Taylor_expansion}
	\sum_{p>0} \frac{2\pi}{Lp} e^{-\xi p}
		= - \ln \left( 1 - e^{-2\pi \xi/L} \right)
	\qquad
	(\Re(\xi) > 0).
\end{equation}


\section{Bosonization in a nutshell}
\label{Appendix:Bosonization_nutshell}
In this appendix we give a survey of the results developed in this paper using a notation common in the physics literature.
It serves to connect the results in this paper to the latter and as a summary.
We also include remarks on the history of bosonization.

\subsection{Bosonization}
The collection of mathematical results known as bosonization concern the quantum field theory model defined by the Hamiltonian
\begin{equation}
\label{Eq:Free_fermion_Hamiltonian_position_space}
	H_0
	\define
	\int_{-L/2}^{L/2} \dop{}{x} v_F \sum_{r}
		\noord{ \psi^{\dagger}_{r}(x) r(-i\partial_{x}) \psi\pdag_{r}(x) }
\end{equation}
with fermion fields $\psi^{(\dagger)}_{r}(x)$ satisfying antiperiodic boundary conditions and the CAR
\begin{equation}
\label{Eq:CAR_formal_nutshell}
	\bigl\{ \psi\pdag_{r\pp}(x), \psi^{\dagger}_{r'}(y) \bigr\} = \delta_{r,r'} \delta(x-y),
	\qquad
	\bigl\{ \psi\pdag_{r\pp}(x), \psi\pdag_{r'}(y) \bigr\} = 0.
\end{equation}
The fermion densities are defined as
\begin{equation} 
\label{Eq:Jr_nutshell}
	J_{r}(x)
	\define
	\noord{ \psi^{\dagger}_{r}(x)\psi\pdag_{r}(x) }.
\end{equation} 
By \emph{bosonization} we mean the following three identities:
\begin{gather}
	\text{\emph{The fermion densities obey non-trivial commutation relations:}} \nonumber \\
	\left[ J_{r}(x), J_{r'}(y) \right]
		= r \frac{1}{2\pi i}\delta_{r,r'} \partial_{x} \delta(x-y).
		\vphantom{\sum^{x}} \label{Eq:Jordan}
\end{gather}
\begin{gather}
	\text{\emph{The Hamiltonian can be expressed in terms of fermion densities:}} \nonumber \\
	H_0 = \pi v_F \sum_r \int_{-L/2}^{L/2} \dop{}{x} \! \noord{ J_{r}(x)^2 } \!.
	\vphantom{\sum^{x}} \label{Eq:Kronig}
\end{gather}
\begin{gather}
	\text{\emph{The fermion fields can be expressed in terms of fermion densities:}} \nonumber \\
	\psi_{r}(x) = \exp \left( r2\pi i \partial_{x}^{-1} J_{r}(x) \right).
	\vphantom{\sum^{x}} \label{Eq:Schotte}
\end{gather}
The r.h.s.\ of \eqref{Eq:Jordan} is an example of a quantum field theory anomaly known as a \emph{Schwinger term}, and the result in \eqref{Eq:Kronig} is known by the name \emph{Kronig's identity}.\footnote{Actually, what is usually called Kronig's identity is
\begin{equation*}
	\int_{-L/2}^{L/2} \dop{}{x}
		\! \noord{ \psi^{\dagger}_{r}(x) r (-i\partial_{x}) \psi\pdag_{r}(x) }
	= \pi \int_{-L/2}^{L/2} \dop{}{x} \! \noord{ J_{r}(x)^2 } \!.
\end{equation*}
However, we only need the somewhat weaker form stated in \eqref{Eq:Kronig}.}
Moreover, the inverse differential operator $\partial_x^{-1}$ in \eqref{Eq:Schotte} hides an important integration constant, which includes an infinite normalization constant and a so-called Klein factor (cf.\ \eqref{Eq:Regularized_psi_explicit}).
Note that all three identities are consequences of mathematical subtleties related to normal ordering. 

To explain the name \emph{bosonization} we note that
\begin{equation}
\label{Eq:Fermion_phi_pi_operators_position_space}
	\Pi_{F}(x)
	\define
	- \sqrt{v_{F} \pi} \big[ J_{+}(x) - J_{-}(x) \big],
	\qquad
	\Phi_{F}(x)
	\define
	\partial_{x}^{-1} \sqrt{\frac{\pi}{v_{F}}} \big[ J_{+}(x) + J_{-}(x) \big]
\end{equation}
are boson field operators since they are Hermitian and satisfy CCR:
\begin{equation}
	\bigl[ \Phi_{F}(x), \Pi_{F}(y) \bigr] = i \delta(x-y),
	\qquad
	\bigl[ \Phi_{F}(x), \Phi_{F}(y) \bigr] = \bigl[ \Pi_{F}(x), \Pi_{F}(y) \bigr] = 0,
\end{equation}
and it follows from \eqref{Eq:Kronig} that the fermion Hamiltonian in \eqref{Eq:Free_fermion_Hamiltonian_position_space} is identical with a boson Hamiltonian: 
\begin{equation}
\label{Eq:Kronigs_identity_applied_to_fermion-phonon_model}
	H_0 
	= \frac{1}{2} \int_{-L/2}^{L/2} \dop{}{x} \! \noord{ \Big( \Pi_{F}(x)^2
		+ v_{F}^2 \left[ \partial_{x}\Phi_{F}(x) \right]^2 \Big) } \!.
\end{equation}
Finally, the identity in \eqref{Eq:Schotte} can be used to express all fermion correlation functions in terms of boson correlation functions, and the latter can be computed exactly.

\remark{}{
\label{Remark:Dirac_fermions_KG_bosons}
By interpreting $v_F$ as the vacuum speed of light, the Hamiltonians in  \eqref{Eq:Free_fermion_Hamiltonian_position_space} and \eqref{Eq:Kronigs_identity_applied_to_fermion-phonon_model} define important models in particle physics: the former corresponds to massless Dirac fermions, and the latter to massless Klein-Gordon bosons, both in 1+1 dimensions.
The inverse differential operator in \eqref{Eq:Fermion_phi_pi_operators_position_space}, however, hides certain complications related to zero modes; see the closing discussion in Section~\ref{Sec:Mathematics_of_bosonization_I} for details.
}

\remark{}{
From the point of view of particle physics described in the previous remark, the fermion densities in \eqref{Eq:Jr_nutshell} correspond to fermion currents in light-cone coordinates.
This explains why we use the symbol $J_r(x)$ for these operators.
}

\subsection{Formal solution of the fermion-phonon model}
The formal bosonization results in \eqref{Eq:Free_fermion_Hamiltonian_position_space}--\eqref{Eq:Kronigs_identity_applied_to_fermion-phonon_model} can be used to bosonize the fermion-phonon model in \eqref{Eq:Fermion-phonon_formal_Hamiltonian}--\eqref{Eq:Restrictions} as follows.
By restating the model with an infrared cutoff $L > 0$, we note that the free-fermion Hamiltonian in \eqref{Eq:Free_fermion_Hamiltonian_position_space} is part of the fermion-phonon Hamiltonian with the cutoff in place.
As a result, this entire Hamiltonian can be cast in the form
\begin{multline}
\label{Eq:Bosonized_formal_Hamiltonian}
	H = \int_{-L/2}^{L/2} \dop{}{x} \Biggl( \frac{1}{2}
			\! \noord{ \left(\Pi_{F}(x)^2 + v_{F}^2 [\partial_x\Phi_{F}(x)]^2\right) }
		+ \frac{1}{2}
			\! \noord{ \left( \Pi_{P}(x)^2 + v_{P}^2 [\partial_{x} \Phi_{P}(x)]^2 \right) } \\
		+ \frac{\gF}{2\pi v_F}
			\left( -\Pi_{F}(x)^2 + v_{F}^2 [\partial_x\Phi_{F}(x)]^2 \right)
		+ \gP{}
			\sqrt{\frac{v_F}{\pi}} [\partial_{x} \Phi_{F}(x)]
				[\partial_{x} \Phi_{P}(x)] \Biggr)
\end{multline}
consisting only of bosons field operators $\Phi_{X}(x)$ and $\Pi_{X}(x)$ in the sense that the latter are Hermitian and satisfy CCR:
\begin{equation}
	\Big[ \Phi_{X}(x), \Pi_{X'}(y) \Big] = i \delta(x-y),
	\qquad
	\Big[ \Phi_{X}(x), \Phi_{X'}(y) \Big] = \Big[ \Pi_{X}(x), \Pi_{X'}(y) \Big] = 0.
\end{equation}
A Hamiltonian of this type can be diagonalized by a Bogoliubov transformation.
Moreover, as in the free-fermion case, all fermion correlation functions can be computed using the identity in \eqref{Eq:Schotte}.
In the formal discussion given here, we have ignored serious mathematical difficulties.
For instance, it is standard practice to construct a Hilbert space for the non-interacting model, i.e.\ the Hamiltonian in \eqref{Eq:Fermion-phonon_formal_Hamiltonian} with $\gF{} = \gP{} = 0$.
However, this Hilbert space does not contain the ground state of the interacting model.
Furthermore, one can show that the exact fermion two-point correlation functions for the interacting model are inconsistent with the CAR in \eqref{Eq:CAR_formal_nutshell}.
This was discussed by Wilson in \cite{Wilson:1970} for the massless Thirring model, which corresponds to the special case $\gP{} = 0$, and it is equally true for $\gP{} \neq 0$.

\subsection{From formal to precise results}
To avoid difficulties with the distributional nature of quantum fields, we work with field operators in Fourier space:
\begin{equation} 
\label{Eq:Psi_Jr_Fourier_transforms}
	\hat{\psi}\pdag_{r}(k)
	= \frac{1}{\sqrt{2\pi}} \int_{-L/2}^{L/2}\dop{}{x} \psi\pdag_{r}(x) e^{-ikx},
	\qquad 
	\hat{J}_{r}(p)
	= \int_{-L/2}^{L/2}\dop{}{x} J_{r}(x) e^{-ipx},
\end{equation}
where $k \in (2\pi/L)(\mathbb{Z}+1/2)$ and $p \in (2\pi/L)\mathbb{Z}$ are fermion and boson momenta, respectively.
(Note that one can regard $\hat{\psi}_r^{(\dagger)}(k)$ and $\hat{J}_r(p)$ as fermion field operators and density operators, respectively, smeared out by test functions, the latter being plane waves.)
This can be used to write down and prove precise versions of the formal bosonization results above; see Sections~\ref{Sec:Mathematics_of_bosonization_I}~and~\ref{Sec:Mathematics_of_bosonization_II}.
Similarly, as in the case of free fermions, our strategy to make the fermion-phonon model precise is also to work in Fourier space:
\begin{align}
	\label{Eq:Phonon_field_Fourier_transform_nutshell}
	\hat{\Pi}_{P}(p)
		& = \frac{1}{\sqrt{2\pi}} \int_{-L/2}^{L/2}\dop{}{x} \Pi_{P}(x) e^{-ipx},
	& \hat{\Phi}_{P}(p)
		& = \frac{1}{\sqrt{2\pi}} \int_{-L/2}^{L/2}\dop{}{x} \Phi_{P}(x) e^{-ipx}, \\
	\label{Eq:Interaction_potentials_Fourier_transform_nutshell}
	\hat{\gF{}}_a(p)
		& = \int_{-L/2}^{L/2}\dop{}{x} \gF{}_a(x) e^{-ipx},
	& \hat{\gP{}}_a(p)
		& = \int_{-L/2}^{L/2}\dop{}{x} \gP{}_a(x) e^{-ipx}.
\end{align}
This and \eqref{Eq:Psi_Jr_Fourier_transforms} can be used to obtain the precise formulation of the fermion-phonon model in Section~\ref{Sec:Fermion-phonon_model} from the formal definition in the introduction.


\section{Operators and quadratic forms}
\label{Appendix:Quadratic_forms}
Many non-trivial mathematical questions in quantum physics are due to unbounded (Hilbert space) operators.
For instance, products and commutators of unbounded operators need not be defined, formally symmetric operators need not be self-adjoint, etc.; see e.g.\ \cite{ReedSimon:1972}. 
To keep this paper self-contained, we collect a number of results which allow us to settle such questions for the operators encountered in the main text.

\subsection{Quadratic forms} 
Consider a complex Hilbert space $\mathcal{F}$ with inner product $\langle\cdot,\cdot\rangle$, and let $\mathcal{D}$ be a dense subset of $\mathcal{F}$.
A \emph{quadratic form} $q$ on $\mathcal{D}$ is a map $\mathcal{D}\times\mathcal{D}\to\mathbb{C}$, $(f,g) \mapsto q(f,g)$ such that $q(\cdot,g)$ is conjugate linear and $q(f,\cdot)$ is linear; see e.g.\ Section~VIII.6 in \cite{ReedSimon:1972}. 

Choose some fixed complete orthonormal basis in $\mathcal{F}$ labeled by a countable set of quantum numbers; we write this basis as $\left\{ \eta_n \right\}_{n \in \mathbb{N}}$, without loss of generality. 
We are mainly interested in the case where $\mathcal{F}$ is a fermion Fock space and $\left\{ \eta_n \right\}_{n \in \mathbb{N}}$ are either the fermion states in \eqref{Eq:Fermion_states} or the boson states in \eqref{Eq:Boson_states}.

Let $\mathcal{D}$ be the set of all finite linear combinations of the basis states $\eta_n$,\footnote{To be precise, $\mathcal{D}$ is the set of all states $\eta = \sum_{n=1}^N c_n\eta_n$ with $c_n \in \mathbb{C}$ and \emph{finite} $N \in \mathbb{N}$.} and let 
\begin{equation} 
\label{Eq:A} 
	A
	\define
	\left( A_{n,m} \right)_{n,m \in \mathbb{N}}
\end{equation} 
be some infinite matrix with complex entries. 
Such a matrix can always be identified with a well-defined quadratic form on $\mathcal{D}$ as follows: 
\begin{equation} 
\label{Eq:qA} 
	q_A(f,g) = \sum_{n,m=1}^\infty A_{n,m} \langle f,\eta_n \rangle \langle \eta_m,g \rangle
	\qquad
	(f,g \in \mathcal{D})
\end{equation} 
(note that this sum is always finite). 
Moreover, the adjoint $A^{\dagger}$ of $A$, given by
\begin{equation} 
	(A^{\dagger})_{m,n}
	\define
	\overline{A_{n,m}}, 
\end{equation} 
is always well-defined as a quadratic form. 
However, to be able to identify this quadratic form with an operator on $\mathcal{F}$:
\begin{equation} 
\label{Eq:Af} 
	A f
	\define
	\sum_{n,m=1}^\infty A_{n,m} \eta_n \langle \eta_m, f \rangle
	\qquad
	(f \in \mathcal{F}), 
\end{equation} 
one has to impose certain restrictions on the matrix elements $A_{n,m}$.
Note that we abuse notation and use the same symbol $A$ for the matrix in \eqref{Eq:A}, the corresponding quadratic form in \eqref{Eq:qA}, and, if it exists, the corresponding operator on $\mathcal{F}$. 
If a quadratic form $A$ can be identified with an operator, then the matrix elements correspond to expectation values of this operator: 
\begin{equation} 
\label{Eq:Amn} 
	A_{n,m} = \langle \eta_{n}, A\eta_{m} \rangle . 
\end{equation} 
It should be noted that the product of two quadratic forms $A$ and $B$ need not be a well-defined operator even if both forms can be identified with operators.
However, under certain restrictions on the matrix elements, it is true.
In such cases, the operator product corresponds to the matrix product:  
\begin{equation} 
\label{Eq:AB} 
	(AB)_{n,m} = \sum_{k=1}^\infty A_{n,k}B_{k,m}. 
\end{equation}

\subsection{Four lemmas}
We now define a set of quadratic forms as in \eqref{Eq:qA} satisfying conditions which rule out many of the possible difficulties mentioned.  

\definition{}{
\label{Def:Oinfty}
Let $\mathcal{O}_{f}$ be the set of all matrices $A$ as in \eqref{Eq:A} such that there exists an $M < \infty$ such that, for all $n \in \mathbb{N}$, 
\begin{equation}
\label{Eq:Of}
	A_{n,m} = A_{m,n} = 0
	\qquad
	\forall m>M. 
\end{equation}
} 

From this definition it follows that $A \in \mathcal{O}_{f}$ if and only if $A\eta, A^{\dagger}\eta \in \mathcal{D}$ for all $\eta \in \mathcal{D}$.
The following result is therefore obvious:

\lemma{}{
The set $\mathcal{O}_{f}$ can be identified with a $*$-algebra of operators on $\mathcal{F}$ with involution $\dagger$ and identity $I$ \textnormal{(}i.e.\ $I_{n,m}=\delta_{n,m}$\textnormal{)}.
}

The above lemma makes precise in which sense many of the products and commutators of operators in the main text are well-defined.

It should be noted that an operator $A \in \mathcal{O}_{f}$ satisfying $A=A^{\dagger}$ does not necessarily define a self-adjoint operator; see e.g.\ \cite{GrosseLangmann:1992} for the stronger conditions needed to ensure this.
However, there is one important special case where this is trivially true: 

\lemma{}{
\label{Lemma:Self_adjointness} 
Given $A \in \mathcal{O}_{f}$ such that
\begin{equation} 
	A\eta_{n} = a_{n}\eta_{n}
	\qquad
	\text{with} \;\,	a_{n} \in \mathbb{R} \;\, \forall n \in \mathbb{N}.
\end{equation} 
Then $A$ is the restriction of a unique self-adjoint operator on $\mathcal{F}$.
} 
\noindent (This is a simple consequence of the definition of self-adjoint operators; see e.g.\  \cite{ReedSimon:1972}.)

We often abuse notation and use the same symbol for the self-adjoint extension of such an operator $A$. 

In the main text we introduce vertex operators as quadratic forms of the kind discussed above, and we show that the product in \eqref{Eq:AB} of two such quadratic forms is well-defined. 
There is an important case when one can conclude from this that a quadratic form can be identified with a unitary operator: 

\lemma{}{
\label{Lemma:Unitarity} 
Let $A$ be a quadratic form as defined in \eqref{Eq:A}\textnormal{--}\eqref{Eq:qA} and such that $A^{\dagger}A$ and $AA^{\dagger}$ both are well-defined and equal to the identity $I$.
Then $A$ can be identified with a unique unitary operator on $\mathcal{F}$. 
}

If this is the case we abuse notation and use the same symbol for the unitary operator extending $A$.

\begin{proof}
Let $f \in \mathcal{F}$ and $f^{(n)}$, $n \in \mathbb{N}$, a Cauchy sequence in $\mathcal{D}$ converging to $f$. Then $A^{\dagger} A =I$ implies that $A f^{(n)}$ is also a Cauchy sequence in $\mathcal{D}$ with a unique limit $\lim_{n \to \infty} Af^{(n)} \define Af$ depending only on $f$. In a similar manner, $A A^{\dagger}=I$ implies that $A^{\dagger}$ has a unique extension from $\mathcal{D}$ to $\mathcal{F}$. It is easy to see that these extensions satisfy $A^{\dagger} Af=A A^{\dagger} f=f$ for all $f \in \mathcal{F}$. 
\end{proof}  

In the main text we use a simple strategy to establish commutation relations of unbounded operators with a common dense invariant domain $\mathcal{D}$: we introduce a cutoff $\Lambda$ to approximate the unbounded operators by bounded ones (for which domain questions can be ignored), and we derive the result by taking the limit where the cutoff is removed; see also Remark~\ref{Remark:Commutators}. For the convenience of the reader we state a definition and a lemma underlying this strategy.

\definition{Operator convergence}{
\label{Definition:Operator_convergence}
Suppose $A \in \mathcal{O}_{f}$ and let $\{ A^{\Lambda} \}_{\Lambda > 0}$ be a family of bounded operators.
Then
\begin{equation} 
\label{Eq:AtoA}
	A = \lim_{\Lambda \to \infty} A^{\Lambda}
	\;\, \text{on} \;\, \mathcal{D}
\end{equation}
means the following: for fixed $f \in \mathcal{D}$, the identity $A^\Lambda f= Af$ holds independent of $\Lambda$ for sufficiently large values \textnormal{(}i.e.
$\exists \Lambda_0<\infty$ such that $A^\Lambda f = Af$ $\forall \Lambda > \Lambda_0 $\textnormal{)}. 
}  

\lemma{}{
\label{Lemma:Commutators}
Let $\{ A^{\Lambda} \}_{\Lambda > 0}$ be a family of bounded operators and suppose $A \in \mathcal{O}_{f}$ such that \eqref{Eq:AtoA} holds true, and similarly for $\{ B^{\Lambda} \}_{\Lambda > 0}$ and $B$ \textnormal{(}if $B$ is bounded one can set $B^\Lambda=B$\textnormal{)}. 
For fixed $f \in \mathcal{D}$, assume there exists some $C \in \mathcal{O}_{f}$ such that
\begin{equation} 
	\left[ A^\Lambda, B^{\Lambda'} \right] f = Cf
\end{equation} 
holds true independent of $\Lambda$ and $\Lambda'$ for sufficiently large values.
Then $[A,B] = C$ on $\mathcal{D}$. 
}
\noindent (The proof is straightforward and thus omitted.)


\end{appendices}





\end{document}